\documentclass[useAMS,usenatbib]{mn2e} 
\usepackage{epsf}
\usepackage{epsfig}
\usepackage{times}
\usepackage{amssymb, amsmath}
\usepackage{url} 

\title[21cm Redshift Space Distortion: Methodology Re-examined]
  {Redshift Space Distortion of the 21cm Background from the Epoch of 
    Reionization I: Methodology Re-examined}
\author[Y.~Mao et al.]
  {Yi Mao,$^1$\thanks{Email: ymao@astro.as.utexas.edu}
  Paul R. Shapiro,$^1$\thanks{Email: shapiro@astro.as.utexas.edu}
  Garrelt Mellema,$^2$ Ilian T. Iliev,$^3$ 
  Jun Koda,$^1$ 
  \newauthor 
  and Kyungjin Ahn $^4$ \\   
  $^1$ Department of Astronomy and Texas Cosmology Center, University of Texas, Austin, TX 78712, USA \\
  $^2$ Department of Astronomy \& Oskar Klein Centre, AlbaNova, Stockholm University, SE-106 91 Stockholm, Sweden \\
  $^3$ Astronomy Centre, Department of Physics \& Astronomy, Pevensey II Building, University of Sussex, Falmer, Brighton BN1 9QH, UK \\
  $^4$ Department of Earth Science Education, Chosun University, Gwangju 501-759, Korea 
  } 
\date{Accepted 2012 January 2.  Received 2011 December 22; in original form 2011 April 11}

\pagerange{\pageref{firstpage}--\pageref{lastpage}} \pubyear{2011}

\def\LaTeX{L\kern-.36em\raise.3ex\hbox{a}\kern-.15em
    T\kern-.1667em\lower.7ex\hbox{E}\kern-.125emX}

\begin{document}

\label{firstpage}

\maketitle

\begin{abstract}

The peculiar velocity of the intergalactic gas responsible for the cosmic 21cm background from the epoch of reionization (EOR) and beyond introduces an anisotropy in the three-dimensional power spectrum of brightness temperature fluctuations. Measurement of this anisotropy by future 21cm surveys is a promising tool for separating cosmology from 21cm astrophysics. 
However, previous attempts to model the signal have often neglected peculiar velocity or only approximated it crudely. 
This paper re-examines the effects of peculiar velocity on the 21cm signal in detail, improving upon past treatment and addressing several issues for the first time. 
(1) We show that even the {\it angle-averaged} power spectrum, $P(k)$, is affected significantly by the peculiar velocity. 
(2) We re-derive the brightness temperature dependence on atomic hydrogen density, spin temperature, peculiar velocity and its gradient, and redshift, to clarify the roles of thermal vs. velocity broadening and finite optical depth. 
(3) We show that properly accounting for finite optical depth eliminates the unphysical divergence of the 21cm brightness temperature in overdense regions of the IGM found by previous work that employed the usual optically-thin approximation. 
(4) We find that the approximation made previously to circumvent the diverging brightness temperature problem by capping the velocity-gradient 
can misestimate the power spectrum on all scales. 
(5) We further show that the observed power spectrum in redshift-space remains finite {\it even} in the optically-thin approximation if one properly accounts for the redshift-space distortion. 
However, results that take full account of finite optical depth show that 
this approximation is only accurate in the limit of high spin temperature.
(6) We also show that the linear theory for redshift-space distortion widely employed to predict the 21cm power spectrum results in a $\sim 30\%$ error in the observationally relevant wavenumber range $k \sim 0.1 - 1\,h/$Mpc, when strong ionization fluctuations exist (e.g.\ at the 50\% ionized epoch). We derive an alternative, quasi-linear formulation which improves upon the accuracy of the linear theory. 
(7) We describe and test two numerical schemes to calculate the 21cm signal from reionization simulations to incorporate peculiar velocity effects in the optically-thin approximation accurately, by real- to redshift-space re-mapping of the H~I density. One is particle-based, the other grid-based, and while the former is most accurate, we demonstrate that the latter is computationally more efficient and can be optimized so as to achieve sufficient accuracy.

\end{abstract}



\section{Introduction}

Neutral hydrogen atoms in the intergalactic medium (IGM) at high redshift produce a diffuse background of redshifted 21cm radiation which encodes information about the physical conditions in the early universe during and before the epoch of reionization (EOR, $z > 6$).  
Three-dimensional mapping of this 21cm background (a.k.a.\ 21cm tomography) has recently been proposed as a promising cosmological probe. In principle, it has greater potential than the cosmic microwave background (CMB) since it can map most of our horizon volume, thus providing unprecedented cosmological information \citep{Mao:2008ug}. 

The next few decades promise to become a golden age for 21cm tomography, with about a half-dozen experiments already proposed or underway for measuring the 21cm background from the EOR, including the upcoming first generation such as 
21CMA\footnote{\url{http://21cma.bao.ac.cn/}}, 
MWA\footnote{\url{http://www.haystack.mit.edu/ast/arrays/mwa/} \label{foot:mwa}}, 
LOFAR\footnote{\url{http://www.lofar.org} \label{foot:lofar}}, 
GMRT\footnote{\url{http://gmrt.ncra.tifr.res.in}}, 
and PAPER\footnote{\url{http://astro.berkeley.edu/~dbacker/eor/}}, 
and the next generation such as 
SKA\footnote{\url{http://www.skatelescope.org}}, 
and the {\it Omniscope}
\footnote{http://en.wikipedia.org/wiki/Fast\_Fourier\_Transform\_Telescope, \mbox{formerly} termed {\it Fast Fourier Transform Telescope}.} 
(\citealt{Tegmark09,Tegmark10}). 
These telescopes will measure the 21cm signal either statistically (first generation telescopes) or by precise imaging and map making (next generation telescopes).

Observations will measure the power spectra of 21cm brightness temperature fluctuations from the EOR. The information that 21cm power spectra encode is twofold. First, cosmic reionization leaves its imprint, such as the size distribution of the H~II region, on 21cm power spectra. Since the topology and geometry of ionized bubbles is sensitive to the properties of the ionizing sources (see, e.g., \citealt{Friedrich11}), we can learn about the ionizing sources from 21cm power spectra. For example, we can distinguish models with only high-mass atomic cooling sources from models with both high-mass and self-regulated low-mass atomic cooling sources \citep{Iliev11}. Second, 21cm power spectra are also sensitive to cosmological parameters because the latter determine the matter density fluctuations at high redshifts. 
The precision with which 21cm tomography can constrain cosmological parameters has been forecast in several studies. Some of these consider mapping diffuse hydrogen in the IGM before and during the EOR 
\citep{McQuinn:2005hk,Bowman07,Santos06,Mao:2008ug,Barger09,Adshead11}, 
others mapping neutral hydrogen in galactic halos after reionization
\citep{Wyithe08,Visbal09}. 
These studies show that cosmological constraints based on CMB measurements can be significantly improved if combined with 21cm measurements. In addition, it has been demonstrated in the literature that 21cm power spectra can also constrain many cosmological models beyond the vanilla $\Lambda$CDM model, e.g., spatial curvature and the running of the spectra of primordial scalar density perturbations (\citealt{Mao:2008ug,Barger09}), neutrino masses (\citealt{Mao:2008ug,Pritchard08}), compensated isocurvature perturbations (\citealt{Gordon09}), primordial non-Gaussian density perturbations (\citealt{Joudaki11}), cosmic string wakes (\citealt{Brandenberger10}), and anisotropic matter density fluctuations (\citealt{Hernandez11}). 

In view of this promise which observations of 21cm power spectra hold for testing and constraining cosmological and astrophysical models, further progress is required to ensure that predictions are accurate enough to fulfill this promise. This accuracy depends not only on the realistic astrophysical modeling of reionization and the H~I spin temperature, but also on the methods used to extract the 21cm signal from simulations of the EOR. We focus here on this 21cm methodology issue, and leave aside the issue of the accuracy of the underlying reionization models and simulations. For this purpose, we will make use of the results of a recent reionization simulation of our own, based upon a radiative transfer calculation combined with a high-resolution N-body simulation of $\Lambda$CDM. While this simulation represents the current state-of-the-art in large-scale reionization simulations, it will serve here only as our illustrative testbed. The accuracy and realism of the simulation, itself, is not our concern here, as we focus, instead, on the accuracy of our method for extracting the 21cm signal from such simulations. 

All observations will give the redshifted 21cm signal in observer redshift-space, where the frequency not only depends on the cosmological redshift, but also on the peculiar velocity of the IGM. 
However, most theoretical endeavors, both in analytical modelling (e.g., \citealt{Furlanetto04,Iliev02}), 
semi-numerical (e.g., \citealt{Alvarez09,Zahn07,Zahn11}) and in numerical simulations 
(e.g., \citealt{Shapiro06,2008AIPC.1035...68S}, \citealt{Mellema06b},  \citealt{2008MNRAS.384..863I}; \citealt{McQuinn07}; \citealt{Trac07}), 
have focused on predicting the statistics of the  21cm signal (e.g. the power spectrum of brightness temperature fluctuations) without taking peculiar velocities into account. 
On the other hand, peculiar velocities will influence the 21cm brightness temperature significantly as was for example shown by \cite{Mellema06b}. 

In the linear regime, the effects of peculiar velocities have been studied
analytically by \cite{Bharadwaj01}, \cite{Bharadwaj04}, \cite{Barkana05} and
\cite{2006ApJ...643..585W}.  
It has been shown that, in this regime, it is possible to separate the contributions to the brightness temperature fluctuation statistics from the patchiness of reionization and the cosmological density fluctuations, respectively (\citealt{Barkana05}). This, it is hoped, would make it possible to use 21cm measurements to solve for cosmological parameters. 
However, the effects of {\it nonlinearity} remain largely unexplored\footnote{\cite{2008PhRvD..78j3512S} presented a nonlinear analysis of the redshift-space distortion. However, they assumed that the 21cm brightness temperature fluctuations are Gaussian, which may not be valid for the EOR (see, e.g. Fig.~14 of \citealt{Mellema06b}).}. There are two kinds of nonlinearity that may contribute, one associated with the gravitational growth of matter density and velocity perturbations, the other due to ionization patchiness. 
The linear theory formula for the 21cm redshift-space power spectrum \citep{Barkana05} widely employed in the literature was derived under the assumption that not only the matter density and velocity fluctuations are linear, but so are the ionization fluctuations. The latter assumption clearly breaks down on the scale of the size of the H~II region. 
We shall investigate here the accuracy of this linear theory formula, particularly for the wavenumbers that are expected to be probed by current and future 21cm surveys of the EOR. 
For this purpose, it is important to develop schemes that can calculate the fully nonlinear 21cm background.

Given the rapid progress of observations (e.g., GMRT has placed an upper bound on the 21cm power spectrum at $z\approx 8.6$ in their first result release [\citealt{Paciga11}], and MWA and LOFAR are close to their data collection stage), we urgently need a thorough understanding of how peculiar velocities enter into predictions of the 21cm signal in observer redshift-space from results of modelling or simulations in real space.

Along these lines, 
\citet[ their Figs.~4, 9 and 10]{Mellema06b} were the first to consider 
the effect of peculiar velocities on the 21cm brightness temperature fluctuations in observer redshift-space when making spectra and maps along the line of sight (LOS). 
They found significant differences between maps and spectra of brightness temperature with and without the effects of peculiar velocities.
However, they did not account for this effect when calculating statistical properties such as the power spectra of the brightness temperature fluctuations, nor did they explain in detail how the effects of peculiar velocities were implemented. 
\cite{Lidz07} claimed to compute the ``full redshift-space'' 21cm power spectrum, but gave no details of their calculation. 
\citet{Thomas09} claimed to include the effects of peculiar velocities in making 21cm maps with their 1D radiative transfer simulation, 
without presenting any details on how these were calculated nor any analysis of the effects on statistical quantities.  

The 21cm brightness temperature can diverge in the overdense regions of the IGM when corrected for peculiar velocity in the optically-thin approximation, because the nonlinear velocity gradient may cancel the Hubble flow in these regions. 
In this paper we shall investigate the origin of this divergence and how this unphysical effect can be avoided. 
Recently, \citet{Santos10} proposed an approximate scheme to circumvent this divergence when computing the 21cm power spectrum in semi-numerical models of the evolving IGM, also adopted by \citet{Mesinger11}. In this scheme a numerical cap on the value of the velocity gradient is imposed. 
The accuracy of their approximation, referred to henceforth as the ``$\nabla v$-limited'' prescription, has not yet been determined. We shall investigate this below.

Our paper is the first in a series which sets out to build a solid and self-consistent computational scheme to predict the {\it fully nonlinear} 21cm background accurately in observer redshift space, given density, velocity and ionization fraction information in real space. This paper will focus on the methodology for incorporating the effects of peculiar velocity in a nonlinear way. We leave the second paper of this series (\citealt{Shapiro11}) to focus on the additional nonlinear effects of inhomogeneous reionization coupled to peculiar velocity and to test the validity of using the anisotropy of the 21cm background fluctuations to separate the astrophysical effects of reionization from these of the background cosmology. Some of our results were previously summarized by us in \cite{Mao10}. 

This paper is organized as follows. In \S~\ref{sec:lesson}, we will demonstrate how important the effects of peculiar velocity are by comparing the angle-averaged power spectra $P(k)$ of brightness temperature fluctuations when peculiar velocity is neglected, calculated from reionization simulations, with an approximate scheme that takes peculiar velocity into account, motivated by linear theory. 
We then clarify our terminology in \S~\ref{sec:term}.  
In \S~\ref{sec:21cm-limit}, we use a heuristic derivation to present a simple picture of redshift-space distortions of the 21cm background in the limit of low optical depth and high spin temperature, and clarify the similarities and differences with galaxy redshift surveys. 
To properly take into account peculiar velocity, including the effects of finite spin temperature and optical depth and the distinction between thermal- and velocity-broadening of the line profile, we present in \S~\ref{sec:PVeffects} the 21cm brightness temperature derived from the equation of transfer in an expanding universe. 
We then derive the 21cm power spectrum as measured in redshift-space in a hierarchy of approximations, from the exact nonlinear power spectrum with finite optical depth to the linear theory in the limit of low optical depth. 
Since the standard linear theory formula for 21cm redshift-space distortion (e.g.\ \citealt{Barkana05}) assumes that all departures from the cosmic mean values (matter density, peculiar velocity, and ionization fraction) are of linear amplitude, while ionization fluctuations are not small for scales comparable to the size of the H~II region, we present here an improved version which takes account of ionization fluctuations to higher order. In \S~\ref{sec:schemes}, we propose two computational schemes, 
one based on particle data 
and one based on grid data. 
We test and compare the accuracy and efficiency of these two schemes. 
In \S~\ref{sec:finite-opt-comparison}, we investigate the accuracy of the optically-thin approximation with regard to the 21cm power spectrum. 
In \S~\ref{sec:BL05}, we test the accuracy of the linear theory formula of \cite{Barkana05} for redshift-space distortion, widely employed to predict the 21cm power spectrum, and the new quasi-linear $\mu_{\bf k}$-decomposition presented in \S~\ref{sec:PVeffects}. 
In \S~\ref{sec:vecgrad-lim}, we discuss the origin of the divergence of the brightness temperature found in previous works and how it can be avoided. We also compare the results of the ``$\nabla v$-limited'' prescription for dealing with this problem to the results from our new schemes. 
We conclude in \S~\ref{sec:conclusion}. We include some technical details of post-processing massive numerical particle data in Appendix~\ref{app:sph}.

\begin{figure*}
 \begin{center}
   \includegraphics[height=0.35\textheight]{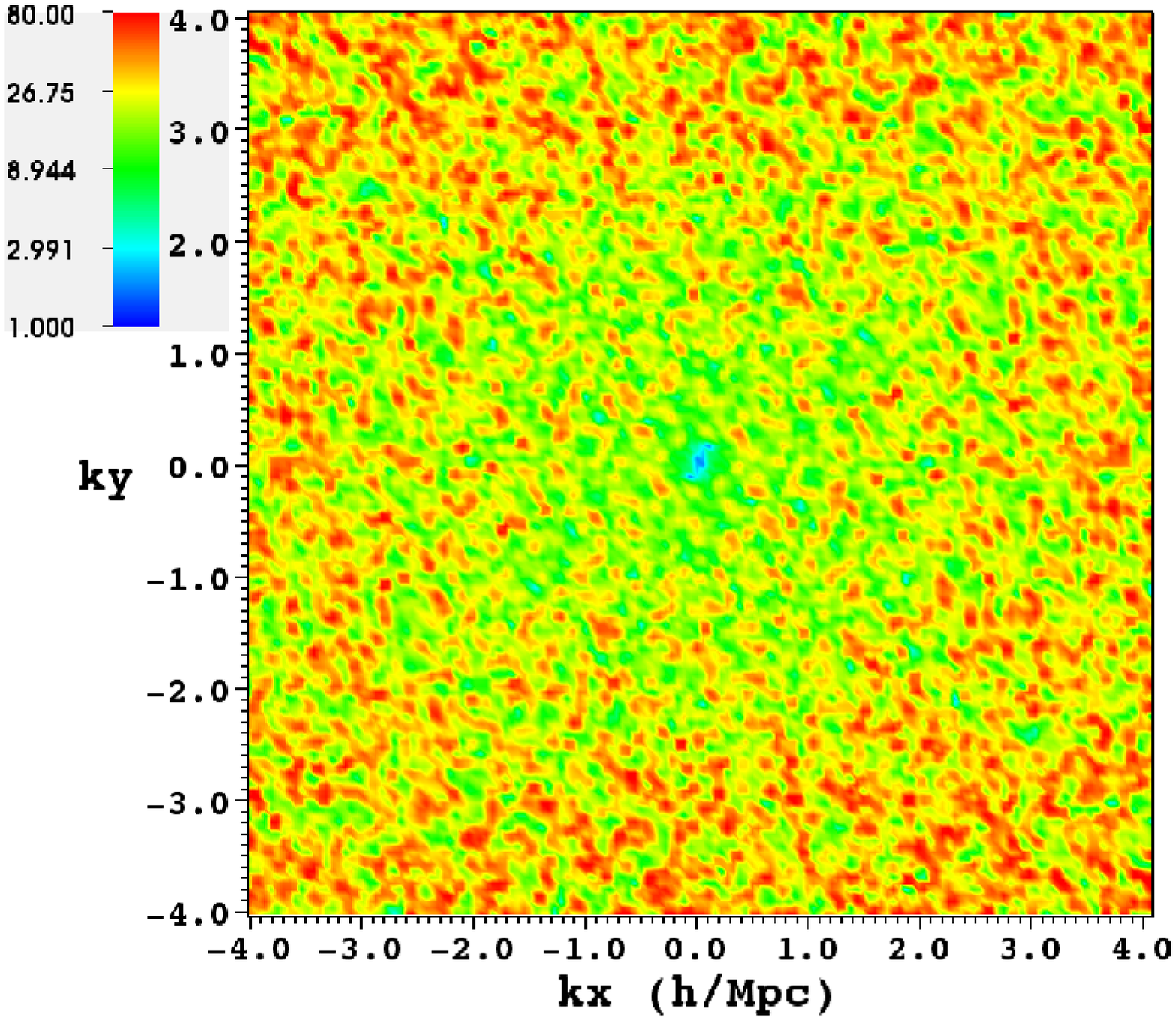} 
   \includegraphics[height=0.35\textheight]{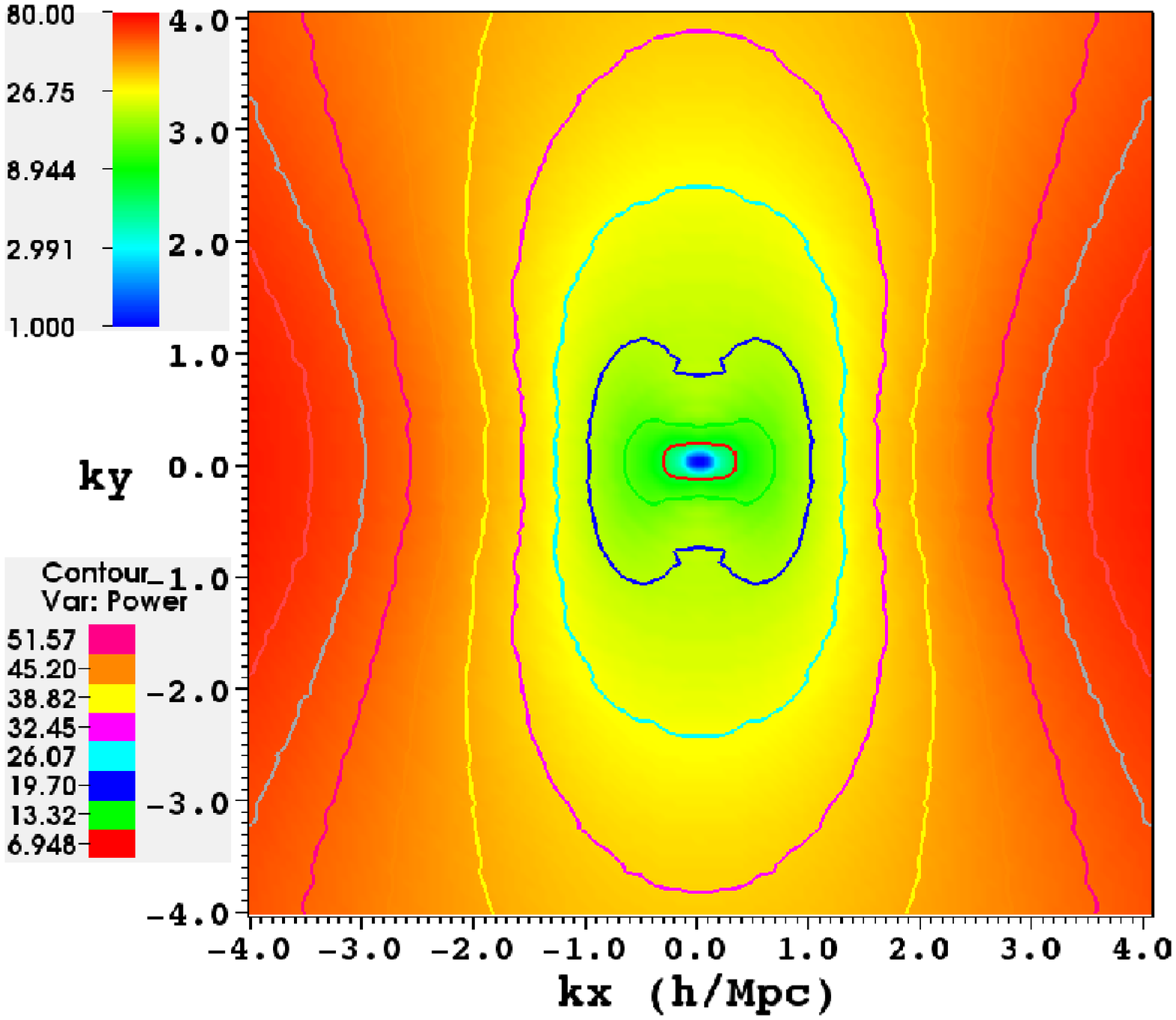}  
   \includegraphics[height=0.35\textheight]{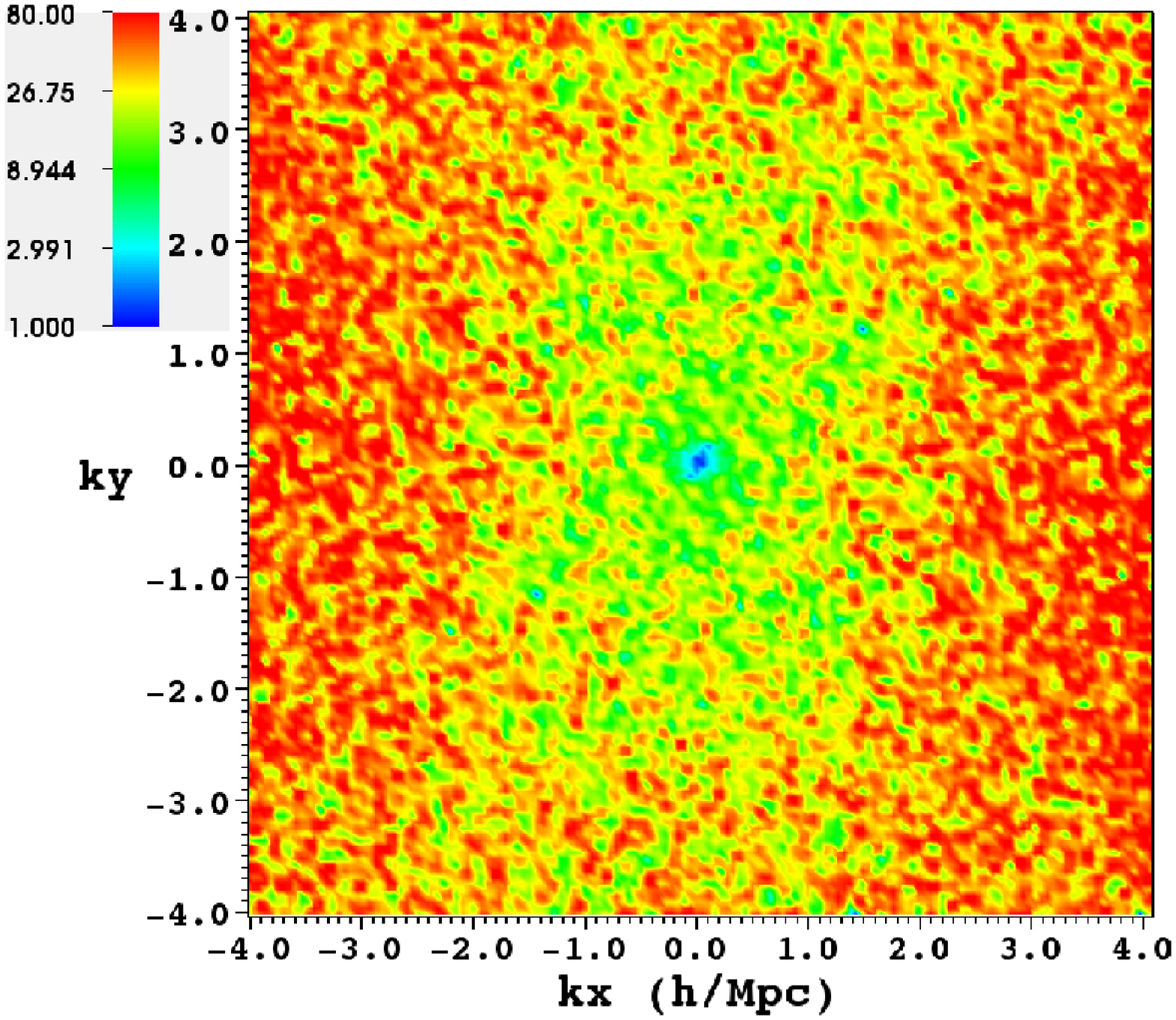}
 \end{center}
\caption{3D power spectra $\Delta^2({\bf k})\equiv k^3 P_{21}({\bf k})/2\pi^2$ (in units of mK$^2$) of 21cm brightness temperature fluctuations. The panels show a slice through the $k_x$-$k_y$ plane, with the LOS along the $x$-axis, calculated from our numerical simulation at the 50\% ionized epoch ($z=9.457$). Top left: UPV scheme; top right: quasi-linear $\mu_{\bf k}$-decomposition scheme; bottom: the fully nonlinear PPM-RRM scheme.}
\label{fig:visual}
\end{figure*}

\section{How Important Is Peculiar Velocity?} 
\label{sec:lesson}

Before developing our methodologies we will illustrate the effects of peculiar velocities on the 21cm power spectrum as measured
in redshift-space, to show their importance. 
For this purpose, we focus on the limiting case in which the spin temperature greatly exceeds the CMB temperature and the optical depth is small, so we can write the differential brightness temperature, $\delta T_b \equiv T_b-T_{\rm CMB}$, as follows: 
\begin{equation}
\delta T_b(\nu_{\rm obs}) = \widehat{\delta T}_b (z_{\rm cos}) \, \frac{1+\delta^r_{\rho_{\rm HI}}({\bf r})}{\left|1+\delta_{\partial_r v}({\bf r})\right|}\,,
\end{equation}
where the pre-factor $\widehat{\delta T}_b$ is the cosmic mean value in this limit, to be defined in equation~(\ref{eqn:dtbhat}). Here $z_{\rm cos}$ is the cosmological redshift, $\bf r$ is the comoving real-space coordinates, 
and $\delta^r_{\rho_{\rm HI}}$ and $\delta^r_{\rho_{\rm H}}$ are the fluctuations of neutral and total hydrogen density in real space, respectively; i.e. $\rho_{\rm HI} = \rho_{\rm H}\,x_{\rm HI}$, 
and $\delta^r_{\rho_{\rm HI}} = \delta^r_{\rho_{\rm H}} + \delta^r_{x_{\rm HI}} + \delta^r_{\rho_{\rm H}} \, \delta^r_{x_{\rm HI}}$, where $x_{\rm HI}$ is 
the neutral hydrogen fraction. Also, we define the quantity
\begin{equation}
\delta_{\partial_r v}({\bf r}) \equiv \frac{1+z_{\rm cos}}{H(z_{\rm cos})} \frac{dv_\parallel}{dr_\parallel }({\bf r})\,,
\label{eqn:dvdr-def}
\end{equation}
the gradient of the proper radial peculiar velocity along the LOS, normalized by the conformal Hubble constant $H/(1+z_{\rm cos})$. 
The power spectrum of brightness temperature fluctuations in observer redshift-space can then be written as 
$ \left< \widetilde{\delta T_b^*} ({\bf k}) \widetilde{\delta T_b} ({\bf k}') \right> \equiv (2\pi)^3 P_{\Delta T}^{\rm 3D} ({\bf k}) \delta^{(3)}({\bf k} - {\bf k}') $, where $\widetilde{\delta T_b} ({\bf k})$ is the Fourier transform of $\delta T_b$. Hereafter $P_{x,x}$ is the auto-power spectrum of the field $x$, and $P_{x,y}$ is the cross-power spectrum of the fields $x$ and $y$. 

In Figure~\ref{fig:visual} we present slices for
three versions of the three-dimensional power spectrum, the first being the one without including any effects of peculiar velocities (hereafter dubbed the ``uncorrected for peculiar velocity'', or UPV, scheme), given by
\begin{equation}
P_{\Delta T}^{\rm UPV,3D}({\bf k}) = \widehat{\delta T}_b^2(z_{\rm cos})  P_{\delta^r_{\rho_{\rm HI}},\delta^r_{\rho_{\rm HI}}} ({\bf k}) \,.
\label{eqn:power_upv}
\end{equation}

The second version is calculated according to the ``quasi-linear $\mu_{\bf k}$-decomposition scheme'' (a generalization of linear theory in \citealt{Barkana05}; but see the exact definition and derivation in \S~\ref{sec:linearRSD} below),
\begin{eqnarray}
P_{\Delta T}^{s,\rm qlin, 3D} ({\bf k}) &=& \widehat{\delta T}_b^2(z_{\rm cos}) \left[ P_{\delta^r_{\rho_{\rm HI}},\delta^r_{\rho_{\rm HI}}} (k)
  \right.\nonumber \\
& &  \left. + 2\, P_{\delta^r_{\rho_{\rm H}},\delta^r_{\rho_{\rm HI}}} (k)\,\mu_{\bf k}^2 + P_{\delta^r_{\rho_{\rm H}},\delta^r_{\rho_{\rm H}}} (k)\, \mu_{\bf k}^4 \right]\,.
\label{eqn:Kaiser3D}
\end{eqnarray}
On large scales, according to linear theory, the second and fourth moments of the $\mu_{\bf k}$-decomposition in equation (\ref{eqn:Kaiser3D}) come from the cross-correlation of the peculiar velocity gradient with neutral hydrogen density fluctuations and the auto-correlation of the peculiar velocity gradient, respectively.
Here $\mu_{\bf k} \equiv k_\parallel/|{\bf k}|$ where $k_\parallel$ is the LOS component of $\bf k$. 
The moments in the RHS of equation~(\ref{eqn:Kaiser3D}) are angle-averaged in a spherical ${\bf k}$-space shell with $k=|{\bf k}|$, i.e., $P_{\delta^r_{\rho_{\rm HI}},\delta^r_{\rho_{\rm HI}}} (k) = \langle P_{\delta^r_{\rho_{\rm HI}},\delta^r_{\rho_{\rm HI}}} ({\bf k})  \rangle$, etc. 
Note that in equation~(\ref{eqn:Kaiser3D}), the quasi-linear $\mu_{\bf k}$-decomposition power spectrum can be computed directly from the real-space data, avoiding the need to specify a computational scheme for calculating the redshift-space-distorted 21cm signal data cube.

The third version of the three-dimensional power spectrum of 21cm brightness temperature fluctuations shown in Figure~\ref{fig:visual} is calculated using a numerical scheme that finds the fully nonlinear redshift-space-distorted 21cm brightness temperature signal as a function of position and frequency (the ``PPM-RRM'' scheme, see \S~\ref{sec:PPM-RRM}). This last version of $P_{\Delta T}^{\rm 3D}({\bf k})$ will be derived in the sections which follow, based on the results of numerical reionization simulations. 

The simulation data used for Figure~\ref{fig:visual} are taken from a radiative transfer (RT) simulation of a $114\,h^{-1}$ Mpc box with $256^3$ RT resolution (more fully presented in \S~\ref{sec:sim}). For the UPV scheme (top left), the power spectrum is seen to be numerically fluctuating in equal-$|{\bf k}|$ shells, but otherwise to be isotropic in the sense that it does not show any directional preference. For the quasi-linear $\mu_{\bf k}$-decomposition scheme (top right), the power spectrum is perfectly distorted along the LOS direction, i.e.\ elongated for small $k$ and squeezed for large $k$. 
For the PPM-RRM scheme (bottom center), it is hard to see the distortion for the small-$k$ modes due to the small number of modes, but the compressed nature of the large-$k$ modes is clearly visible, albeit with some numerical noise. Clearly, peculiar velocities introduce noticeable anisotropies in the 21cm power spectra.

\begin{figure}
\begin{center}
  \includegraphics[height=0.36\textheight]{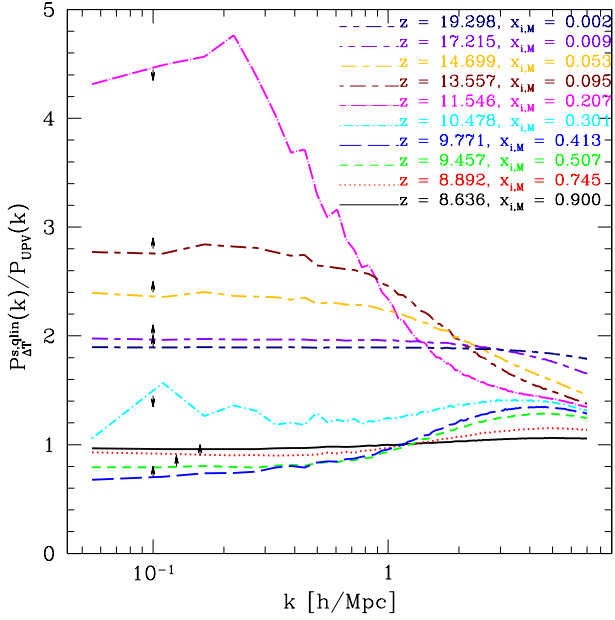} 
\end{center}
\caption{Ratio of 21cm redshift-space-distorted power spectrum in the quasi-linear $\mu_{\bf k}$-decomposition scheme and 21cm power spectrum in the UPV scheme as a function of comoving wavenumber $k$, at a series of redshift $z$ and mass-averaged ionization fraction $x_{i,M}$. Arrows indicate the direction of the evolution of the curves at low $k$ as reionization proceeds. Starting from the curve at $x_{i,M}=0.002$ (dark blue, short dash - long dash) near the ratio $=1.87$ limit, the ratio at low $k$ moves up through the curves, in sequence, at $x_{i,M}=0.009$ (purple, short dash - long dash), $x_{i,M}\approx 0.05$ (orange, short dash - long dash),  $x_{i,M}\approx 0.1$ (dark red, short dash - long dash), flips the direction at $x_{i,M}\approx 0.2$ (magenta, dot - long dash), then moves down through the curves at $x_{i,M}\approx 0.3$ (cyan, dot - short dash), flips the direction again at $x_{i,M}\approx 0.4$ (blue, long dash), moves up through the curves at $x_{i,M}\approx 0.5$ (green, short dash), $x_{i,M}\approx 0.75$ (red, dot), and approaches the curve at $x_{i,M}\approx 0.9$ (black, solid) near the ratio $=1$ limit.
}
\label{fig:redvsreal}
\end{figure}

To make a more quantitative comparison, we compute the angle-averaged power spectrum $P_{\Delta T}^{s,\rm qlin, 1D}(k) = \langle P_{\Delta T}^{s,\rm qlin, 3D}({\bf k})\rangle$ for the quasi-linear $\mu_{\bf k}$-decomposition scheme, 
\begin{eqnarray}
P_{\Delta T}^{s,\rm qlin, 1D}  (k) &=& \widehat{\delta T}_b^2(z_{\rm cos}) \left[ P_{\delta^r_{\rho_{\rm HI}},\delta^r_{\rho_{\rm HI}}} (k)
  \right.\nonumber \\
& &  \left. + \frac{2}{3} P_{\delta^r_{\rho_{\rm H}},\delta^r_{\rho_{\rm HI}}} (k)+ \frac{1}{5} P_{\delta^r_{\rho_{\rm H}},\delta^r_{\rho_{\rm H}}} (k) \right]\,,\label{eqn:Kaiser-ave.}
\end{eqnarray}
as a function of $k=|{\bf k}|$,
and the same for the UPV scheme, $P_{\Delta T}^{\rm UPV,1D} (k)=\widehat{\delta T}_b^2(z_{\rm cos})  P_{\delta^r_{\rho_{\rm HI}},\delta^r_{\rho_{\rm HI}}} (k)$. Figure~\ref{fig:redvsreal} shows the ratio of these two, $P_{\Delta T}^{s,\rm qlin, 1D}(k)/P_{\Delta T}^{\rm UPV,1D}(k)$, for ten different phases of reionization. Two limiting cases are obvious: for the early phases of reionization, the ratio approaches an almost constant
value of 1.87; for the late phases the ratio tends to 1.0. These limits hold best at low $k$. They can be understood as follows. At early times, the neutral fraction fluctuations $\delta^r_{x_{\rm HI}}$ are negligible, i.e.\ the neutral hydrogen density traces the total hydrogen density almost exactly, and, therefore, the 21cm power spectrum in the quasi-linear $\mu_{\bf k}$-decomposition scheme differs from $P_{\Delta T}^{\rm UPV,1D} \approx \widehat{\delta T}_b^2(z_{\rm cos})  P_{\delta^r_{\rho_{\rm H}} \delta^r_{\rho_{\rm H}}}$ by a factor of $1+\frac{2}{3}+\frac{1}{5} = 1.87$. At late times, neutral fraction fluctuations dominate over density fluctuations
\footnote{The variance in neutral fraction can be estimated as $(\Delta x_{\rm HI})^2=\left< (x_{\rm HI} - \bar{x}_{\rm HI})^2 \right> \approx \bar{x}_{\rm HI}(1-\bar{x}_{\rm HI})$, so the rms neutral fraction {\it fluctuation} is $\delta^{\rm rms}_{x_{\rm HI}} = \Delta x_{\rm HI}/\bar{x}_{\rm HI} \approx \sqrt{(1-\bar{x}_{\rm HI})/\bar{x}_{\rm HI}}$. Thus the neutral fraction fluctuations grow as reionization proceeds, even though the variance in neutral fraction decreases near the end of reionization.}
, so its auto-power $P_{\delta^r_{x_{\rm HI}},\delta^r_{x_{\rm HI}}}$ becomes the dominant term in both versions of the
power spectra, making their ratio approach unity.   
As pointed out above, the density fluctuation terms in equation (\ref{eqn:Kaiser3D}) reflect the redshift-space distortion caused by peculiar velocity. Hence, the effect of peculiar velocity on the power spectrum of 21cm brightness temperature fluctuations becomes subdominant towards the end of reionization, as noted also by \cite{McQuinn:2005hk} and \cite{Mesinger07}.

Between these two limits, Figure~\ref{fig:redvsreal} shows that as reionization proceeds the ratio evolves rather nonlinearly, changing both amplitude and shape non-monotonically. Reionization proceeds ``inside-out'' in our simulation, i.e.\ overdense regions ionize earlier than underdense regions, so the cross-power $P_{\delta^r_{\rho_{\rm H}},\delta^r_{x_{\rm HI}}}$ between density fluctuation and neutral fraction fluctuation is negative at large scales. Shortly after the onset of reionization ($x_i \lesssim 0.2$), 
the total density power spectrum $P_{\delta^r_{\rho_{\rm H}},\delta^r_{\rho_{\rm H}}}$ still dominates over the other terms, but 
the cross-power $P_{\delta^r_{\rho_{\rm H}},\delta^r_{x_{\rm HI}}}$ also contributes significantly and is the next most important term, 
so $P_{\Delta T}^{s,\rm qlin, 1D}/P_{\Delta T}^{\rm UPV,1D} \approx (1.87 P_{\delta^r_{\rho_{\rm H}},\delta^r_{\rho_{\rm H}}} + 2.67 P_{\delta^r_{\rho_{\rm H}},\delta^r_{x_{\rm HI}}})/(P_{\delta^r_{\rho_{\rm H}},\delta^r_{\rho_{\rm H}}} + 2 P_{\delta^r_{\rho_{\rm H}},\delta^r_{x_{\rm HI}}}) \approx 1.87 - 1.07 (P_{\delta^r_{\rho_{\rm H}},\delta^r_{x_{\rm HI}}}/P_{\delta^r_{\rho_{\rm H}},\delta^r_{\rho_{\rm H}}})$, moving the ratio up since the neutral fraction fluctuations increase as reionization proceeds. When reionization reaches the midway point ($x_i \gtrsim 0.4$) and large ionized bubbles have formed, the neutral fraction auto-power $P_{\delta^r_{x_{\rm HI}},\delta^r_{x_{\rm HI}}}$ starts dominating over other powers and the cross-power $P_{\delta^r_{\rho_{\rm H}},\delta^r_{x_{\rm HI}}}$ becomes subleading, so $P_{\Delta T}^{s,\rm qlin, 1D}/P_{\Delta T}^{\rm UPV,1D} \approx 1+ (2/3)(P_{\delta^r_{\rho_{\rm H}},\delta^r_{x_{\rm HI}}}/P_{\delta^r_{\rho_{\rm HI}},\delta^r_{\rho_{\rm HI}}}) \approx 1+(2/3)(P_{\delta^r_{\rho_{\rm H}},\delta^r_{x_{\rm HI}}}/P_{\delta^r_{x_{\rm HI}},\delta^r_{x_{\rm HI}}})$. Since the cross power $P_{\delta^r_{\rho_{\rm H}},\delta^r_{x_{\rm HI}}}$ is negative, the ratio is less than unity. As reionization proceeds towards its final stages, the neutral fraction fluctuations continue to grow, pushing the ratio closer and closer to unity. Between $x_i \approx 0.2$ and $0.4$, the competition between neutral fraction fluctuations and density fluctuations makes the ratio at large scales first turn around at a large value $\sim 4-5$ ($x_i \approx 0.2$), then move all the way down to less than 1 ($x_i \approx 0.4$), then turn around again and begin to approach the limit of 1. 

These comparisons illustrate the nontrivial effects when applying redshift space distortions using the quasi-linear $\mu_{\bf k}$-decomposition scheme. However, the fully nonlinearly distorted 21cm power spectrum may well show a more complicated behavior, which has not been previously explored. In order to calculate the fully-nonlinear redshift-space-distorted 21cm power spectrum, we need a robust scheme to compute it from simulation results. The aim of this paper is to develop such a scheme, taking into account all peculiar velocity effects. In a subsequent paper we will use this scheme to study the nonlinear distortion in the 21cm power spectrum and test the validity of quasi-linear $\mu_{\bf k}$-decomposition scheme upon which the 21cm cosmology is based.

\section{Terminology}
\label{sec:term}

Before we proceed to the main content, we summarize in this section our terminology which otherwise may be confusing. 

\subsection{Reference Frames}\label{sec:term-ref-frame}
We distinguish between different reference frames. These are
\begin{itemize}

\item {\bf Emitter space:} the local rest-frame of the emitting atoms. 

\item {\bf FRW space:} the cosmic reference frame in which space is uniformly expanding, as described by the Friedmann-Robertson-Walker metric. 

The emitter space and the FRW space are related by the local Lorentz transformation at the position of emitting atoms, and the relative motion of these two frames is the peculiar velocity of atoms. 

From the observer's point of view, the coordinates of source $(t,{\bf r})$ in FRW space can be relabeled by $t_{\rm arrival}$ (arrival time of radiation emitted at time $t$ by source located at comoving location ${\bf r}$), $z_{\rm cos}$ ({\it cosmological} redshift experienced by photons from time $t$ of their emission to the time $t_{\rm arrival}$ at which they reach the observer), and ${\bf \Theta}$ (angular coordinates on the sky). 

\item {\bf Observer real space:} for fixed $t_{\rm arrival} = t_{\rm present}$ (present time), the observer can reconstruct a part of the FRW space theoretically -- those $(t,{\bf r})$ on the  light-cone that can be determined by $z_{\rm cos}$. In particular, $r=r(z_{\rm cos})|_{t_{\rm present}}=\int_0^{z_{\rm cos}} c\,dz'/H(z')$. We call this the observer real space. In the rest of this paper, quantities measured in real space are superscripted with $r$, so for example $n^r$ is a number density in real space.

\item {\bf Observer redshift space:} in practice, observers can only measure the {\rm observed} redshift of radiation, since the wavelength is redshifted both cosmologically and by the Doppler shift associated with peculiar velocity, $\nu_{\rm obs} = \nu_0 /(1+z_{\rm obs})$ and $1+z_{\rm obs} = (1+z_{\rm cos})(1- \frac{v_\parallel}{c})^{-1}$. Observers can set up a ``distorted'' comoving coordinate system, known as observer {\it redshift} space, in which the position of the emitter is the {\it apparent} comoving position
if the redshift is interpreted as only cosmological, i.e.\ $ s \equiv r(z_{\rm obs})|_{t_{\rm present}}= \int_0^{z_{\rm obs}} c\,dz'/H(z')$, 
which shifts the {\it real} comoving coordinate $r$ along the LOS ($\hat{r}$) to 
\begin{equation}
{\bf s} = {\bf r} + \frac{(1+z_{\rm obs})}{H(z_{\rm obs})} v_\parallel(t,{\bf r}) \, \hat{r}\,.
\label{eqn:red-to-real-coord}
\end{equation} 
Note that the transformation between observer real and redshift spaces is not covariant (in a general relativistic sense) or even Galilean invariant, since it does not preserve spatial intervals at fixed time. In the rest of this paper, quantities measured in observer redshift space are superscripted with $s$, so for example, $n^s$ is a number density in redshift space.

\end{itemize}

\subsection{3D Mapping Distortion}
One can distinguish between several types of distortions, namely
\begin{itemize}

\item {\bf Apparent location distortion in redshift-space:} 
when the observed frequency of a spectral line from a distant source is used to locate the source along the LOS, the answer depends upon solving equation~(\ref{eqn:red-to-real-coord}), which requires a knowledge of the LOS peculiar velocity of the source at the time of emission. The term ``redshift-space distortion'' usually refers to the error one makes in locating the source by assuming the peculiar velocity to be zero. 

\item {\bf Brightness temperature distortion in real-space:}
Radiative transfer effects can result in a modification of the observed 21cm brightness temperature due to gradients in the velocity field, as shown in \S~\ref{sec:rederive}. This effect is independent of the adoption of either real- or redshift-space. In other words, 
even if an observer could construct a 3D mapping of 21cm brightness temperature in observer real space by knowing the peculiar velocities along the LOS, gradients in the peculiar velocity field can still modify the magnitude of brightness temperature.

\item {\bf 21cm redshift-space distortion:} 
This is the combination of the previous two distortions, namely the apparent location distortion in redshift-space and the brightness temperature distortion in real-space. The observed 21cm signal is modified by the presence of peculiar velocities according to this combination.

\end{itemize}

\subsection{Power Spectra}
Power spectra can be calculated in different dimensions in $\mathbf{k}$-space and with
different methods for applying the effects of peculiar velocities. We use
the following terminology:
\begin{itemize}

\item {\bf 3D power spectrum $P_{\rm 3D}({\bf k})$:} The power spectrum in three-dimensional ${\bf k}$-space.

\item {\bf 1D power spectrum $P_{\rm 1D}(k)$:} The power spectrum in one-dimensional $|{\bf k}|$-space (or simply $k$-space), obtained by averaging the 3D power spectrum over modes in spherical shells in ${\bf k}$-space: $P_{\rm 1D}(k) \equiv \left< P_{\rm 3D}({\bf k})\right>_{\rm shell}$ with $k = |{\bf k}|$. 

\item {\bf 21cm power spectrum:} An abbreviation of ``power spectrum of 21cm brightness temperature fluctuations''. 

\item {\bf 21cm redshift-space-distorted power spectrum:} The 21cm power spectrum in observer redshift space, i.e.\ taking the {\it 21cm redshift-space distortion} into account. 

\item {\bf 21cm real-space power spectrum:} The 21cm power spectrum evaluated {\it with} velocity gradient corrections and yet in real space, i.e.\ the power spectrum which results from the Fourier transform of the scalar field corresponding to the true (i.e.\ peculiar-velocity-corrected) 21cm brightness temperature at each point in real-space 
at a single cosmic time. 
This power spectrum so-defined is not the power spectrum of the observed 21cm brightness temperature field evaluated in {\it redshift-space} in which each plane transverse to the line-of-sight corresponds to a single observed frequency. Instead, this ``real-space power spectrum'' represents the brightness temperature at different observed frequencies for different locations in real space, as a result of Doppler shifts caused by peculiar velocity. 

\item {\bf 21cm UPV power spectrum:} The 21cm power spectrum evaluated {\it without} any velocity gradient corrections and in real space, i.e.\ taking into account neither the {\it brightness temperature distortion in real-space} nor the {\it apparent location distortion in redshift-space}; ``UPV'' stands for ``uncorrected for peculiar velocity''. 

\item {\bf 21cm quasi-linear $\bmath{\mu}_{\bf k}$-decomposition power spectrum:} an abbreviation of 21cm power spectrum calculated with the ``quasi-linear $\mu_{\bf k}$-decomposition scheme'' (see \S~\ref{sec:comp-sche-summ}).

\end{itemize}

\subsection{Computational Schemes}
\label{sec:comp-sche-summ}

Below we develop different schemes for applying the effects of peculiar velocities. We summarize these here.
\begin{itemize}

\item {\bf Linear theory:} A scheme to compute 21cm power spectrum in redshift-space, where all fields, density, velocity and ionization fraction, are linearized; introduced by \cite{Barkana05}. 

\item {\bf Quasi-linear $\bmath{\mu}_{\bf k}$-decomposition scheme:} A scheme to compute 21cm power spectrum in redshift-space, assuming the density and velocity fields to be linear, but without constraints on the ionization fraction field; introduced in \S~\ref{eqn:lin-muk-decomp}.

\item {\bf PPM-RRM scheme} (``Particle-to-Particle-to-Mesh Real-to-Redshift-Space-Mapping''): A {\it particle-based} numerical scheme to construct the 21cm data cube in observer redshift space, using the real- to redshift-space re-mapping of density, velocity and ionization fraction data. Introduced in \S~\ref{sec:PPM-RRM}.

\item {\bf MM-RRM scheme} (``Mesh-to-Mesh Real-to-Redshift-Space-Mapping''): Same as the PPM-RRM scheme, but {\it grid-based}; introduced in \S~\ref{sec:MM-RRM}.

\item {\bf DEMRF scheme} (``Direct Evaluation by Multiple Real-space FFTs''): A scheme to compute the 21cm power spectrum in redshift-space by a direct integration technique; introduced in \S~\ref{sec:DEMRF}. 

\end{itemize}

\section{21cm Redshift Space Distortion: Optically Thin and High $\bmath{T_s}$ Limit}
\label{sec:21cm-limit}

In this section we consider the simplest scenario, namely in the limit of small optical depth and high spin temperature $T_s \gg T_{\rm CMB}$, and show that in this limit, peculiar velocities affect the 21cm brightness temperature in an analogous way to the redshift-space distortion in galaxy redshift surveys. 

Recall that galaxy redshift surveys can distinguish individual galaxies. In other words, galaxies can be counted directly. Peculiar velocities move galaxies to their apparent locations, thereby affecting the number density of galaxies in redshift-space. For 21cm surveys, however, individual 21cm-line emitters --- each neutral hydrogen atom --- cannot be resolved and the H~I number density can only be inferred from the observed brightness temperature of 21cm emission. This fundamental difference from galaxy redshift surveys implies that radiative transfer effects associated with peculiar velocities must be taken into account when calculating the redshift-space distortion of the 21cm background.

In the optically thin limit, the emission from each individual H~I atom can be regarded as independently transferred along the LOS. In the high spin temperature ($T_s \gg T_{\rm CMB}$) limit, the stimulated emission/absorption is negligible compared to the spontaneous emission. Therefore when both limits apply, each H~I atom can be thought of as an independently shining 21cm-line source with the intrinsic luminosity $L_{\nu_0} = h\nu_0 A_{10}$, where $\nu_0 = 21\,{\rm cm}/c = 1420.4057$~MHz, and $A_{10} = 2.85\times 10^{-15} {\rm s}^{-1}$ is the Einstein spontaneous emission coefficient of the 21cm transition. Then the emissivity at frequency $\nu'_{\rm RF}$ in the emitter space is 
\begin{equation}
j_\nu^{\rm RF} = \frac{1}{4\pi} L_{\nu_0} n_1^r \phi(\nu'_{\rm RF})\,,
\end{equation}
where $n_1^r\approx (3/4)n_{\rm HI}$ is the number density of H~I atoms in the upper hyperfine state in real-space. The function $\phi(\nu'_{\rm RF})$ is the line profile and satisfies the normalization condition $\int_{-\infty}^\infty \phi(\nu)\, d\nu =1 $. The radiative transfer equation in FRW space then becomes
\begin{equation}
\frac{dI_\nu}{d\xi} = j_\nu\,,
\end{equation}
where $I_\nu$ is the comoving specific intensity of a light ray. The ray path can be labeled by the {\it proper} distance along it, $d\xi = c\,dt$, where $t$ is the physical time. Since the emissivity transforms as $\nu^2$ (see, e.g., \citealt{Mihalas78}), $j_\nu = (1-\frac{v_\parallel}{c})^2\, j_\nu^{\rm RF}$ in FRW space. 
The observed specific intensity then is
\begin{equation}
I_{\nu_{\rm obs}} = \frac{1}{4\pi} L_{\nu_0} \int a^3 (1-\frac{v_\parallel}{c})^2\, n_1^r \phi(\nu'_{\rm RF})\,d\xi\,.
\end{equation}

In the idealized case of no thermal broadening, the line profile is a $\delta$-function peaked at the transition frequency seen from the emitter space, i.e., 
\begin{equation}\label{eqn:delta-line-prof}
\phi(\nu'_{\rm RF}) = \delta(\nu'_{\rm RF} - \nu_0)\,.
\end{equation}
Therefore, the integration picks up the integrand evaluated at the location of emission. Using an identity, whose derivation will be described in \S~\ref{sec:PVeffects}, 
\begin{equation}
\left|\frac{d\nu'_{\rm RF}}{d\xi} \right| = \frac{1}{c}\nu_0\,H(a)\, \left| 1+\frac{1}{aH(a)} \frac{dv_\parallel}{dr_\parallel} \right|\,,
\end{equation}
where $r_\parallel$ is the comoving LOS distance, 
we find that the observed specific intensity is 
\begin{equation}
I_{\nu_{\rm obs}} = \frac{ c L_{\nu_0} a \nu_{\rm obs}^2}{4\pi \nu_0^3 H(a)} \frac{n_1^r}{\left|1+\frac{1}{aH(a)} \frac{dv_\parallel}{dr_\parallel} \right|}\,.
\end{equation}

Now we consider the distortion of apparent location. The number density in redshift-space $n_1^s$ satisfies 
\begin{equation}
n_1^s = \frac{n_1^r}{\left| 1+\frac{1}{aH(a)} \frac{dv_\parallel}{dr_\parallel} \right|}\,,
\end{equation}
since the number of H~I atoms are preserved between real- and redshift-space, and the volume element in redshift-space is distorted as $\delta V^s = \delta V^r \left|1+\frac{1}{aH(a)} \frac{dv_\parallel}{dr_\parallel}\right|$. Therefore we find that
\begin{equation}
I_{\nu_{\rm obs}} = \frac{ c L_{\nu_0} a \nu_{\rm obs}^2}{4\pi \nu_0^3 H(a)} n_1^s\,.
\label{eqn:spe-int-in-limit}
\end{equation}
(We will express eq.~\ref{eqn:spe-int-in-limit} in terms of the familiar brightness temperature in \S~\ref{sec:PVeffects}.) This is to say, 
the simple proportionality relation between the specific intensity (or,  brightness temperature) and the neutral hydrogen density is preserved 
with and without peculiar velocities. 

There is a simple explanation for equation~(\ref{eqn:spe-int-in-limit}). 
In the limit of optically thin and high spin temperature, 21cm radiation from each neutral atom is emitted and then transferred independently. Therefore, the radiative transfer effects of peculiar velocity on a pocket of gas is simply equivalent to the simple picture of having all emitters shine from their apparent locations. Note that this net effect combines the peculiar velocity effects on the radiative transfer and on the distortion of apparent locations of sources.

Equation~(\ref{eqn:spe-int-in-limit}) establishes that, in this limit, there is an analogy between 21cm brightness temperature measurements and galaxy number density measurements, in that the neutral hydrogen atom number in 21cm tomography corresponds to the galaxy number in galaxy surveys. Therefore the 21cm power spectrum should be affected by peculiar velocities in a form similar to the linear redshift space distortion on large scales (first shown by \citealt{Barkana05}), similar to the galaxy matter power spectrum \citep{Kaiser87}. In both cases, the effects of peculiar velocities can be thought of as the distortion due to displacing sources to their apparent LOS locations. 

In the more general case in which optical depth is not small and/or $T_s \lesssim T_{\rm CMB}$, however, the analogy between the redshift-space distortion of the 21cm background signal and that in galaxy redshift surveys breaks down. Since galaxy redshift surveys can resolve and count discrete galaxies, they do not depend upon measuring the unresolved intensity of galactic emission to deduce the number density of galaxies. For the 21cm background, however, we cannot resolve individual sources, (i.e. individual atoms), so we must use the specific intensity (or brightness temperature) to infer the source density, e.g. in the optically-thin/high $T_s$ limit, according to equation~(\ref{eqn:spe-int-in-limit}) above. If the conditions of low optical depth and high spin temperature are not satisfied, however, equation~(\ref{eqn:spe-int-in-limit}) no longer applies. In that case, the luminosity emitted per atom then depends upon the unknown spin temperature, and the received intensity is also no longer linear in the optical depth. In order to interpret redshift-space-distorted 21cm maps, in general, therefore, we cannot simply borrow the analogy of the galaxy redshift surveys.  We discuss the details of this in \S~\ref{sec:PVeffects}. 

\section{Effects of Peculiar Velocity on the Observed 21cm Background}
\label{sec:PVeffects}

Given density, velocity, and ionization information in real space, peculiar velocities can affect the observed 21cm signal through two effects: (1) the observed 21cm brightness temperature can be modified by the gradient of radial peculiar velocity of the gas along the LOS, and (2) the {\it apparent} location of the gas can be shifted from its {\it real}-space location because of the Doppler shift due to its peculiar velocity. We will address the first effect in \S~\ref{sec:rederive}, and then combine both effects to form a self-consistent picture of 21cm redshift space distortion in \S~\ref{sec:doppler} and \S~\ref{sec:linearRSD}. 

\subsection{The Transfer of 21cm Radiation Through the Intergalactic Medium}
\label{sec:rederive}

The effect of peculiar velocity gradients on {\it observed} 21cm brightness temperature was first addressed in \cite{Bharadwaj01} and subsequently in \cite{Bharadwaj04}, and \citet{Barkana05}. 
\cite{Bharadwaj01} and \cite{Bharadwaj04} only explored the simpler limit of high spin temperature ($T_s \gg T_{\rm CMB}$) and optically thin radiative transfer, and implicitly assumed that the velocity gradient is small so that the factor $1/\left(1+\frac{1}{aH(a)} \frac{dv_\parallel}{dr_\parallel} \right)$ can be linearized. 
\citet{Barkana05} attributed this velocity gradient correction to the effect of the fixed thermal width of the 21cm scattering cross section, without showing the details of the derivation. Since we aim to understand peculiar velocity thoroughly, it is worthwhile to re-derive this effect from first principle, i.e., solving the radiative transfer equation, and keeping all contributions of peculiar velocity to linear order $v/c$. 
In this section we show that it is the peculiar velocity of the {\it bulk motion}, not the {\it thermal} broadening, that is responsible for making its correction in 21cm brightness temperature. We check the validity of the optically thin approximation and show that it can break down in certain conditions, although it is mostly valid in the IGM.  
We find also that, in addition to the well-known velocity gradient correction, the contribution of spin temperature to 21cm brightness temperature can be modified by a term of order $\mathcal{O}(v/c)$. 

\subsubsection{The Formal Solution}

Consider a light ray with {\it comoving} specific intensity $I_\nu$ \footnote{ 
It is sometimes customary to use the {\it proper} specific intensity $I_\nu^{(p)}$, which is related to the comoving specific intensity by $I_\nu=I_\nu^{(p)} a^3$. }
passing through a gas element. In an expanding universe, in which $\nu \propto 1/a$, the radiative transfer equation reads (\citealt{Gnedin97,Wise11,Zhang07})
\begin{equation}
\frac{\partial I_\nu}{c\,a\,\partial \eta} + \frac{\hat{n}}{a}\cdot\nabla I_\nu - \frac{H(a)}{c}\frac{\partial I_\nu}{\partial \ln\nu} = - \kappa_\nu I_\nu + j_\nu\,,
\end{equation}
where $I_\nu$ is a function of conformal time $\eta$, comoving coordinates $\bf r$, frequency $\nu$ and direction $\hat{n}$. Here $a$ is the cosmic scale factor and $H(a)$ is the Hubble constant at $a$. The ray path can be labeled by the {\it proper} distance along it, $d\xi = c\,dt$, where $t$ is the physical time. The radiative transfer equation can be rewritten in terms of the Lagrangian total derivative 
\begin{equation}
\frac{dI_\nu}{d\xi} = - \kappa_\nu I_\nu + j_\nu\,.
\end{equation}
Here $\kappa_\nu$ and $j_\nu$ are the absorption coefficient and the {\it comoving} spontaneous emission coefficient at the frequency $\nu$ in FRW space, respectively. 

We label $\nu_{\rm obs}$ to be the frequency observed today, $\nu'= \nu_{\rm obs}/a$ the frequency at some proper distance $\xi'$ along the ray path in FRW space, and $\nu'_{\rm RF} = \nu'(1-\frac{v_\parallel}{c})^{-1}$ the frequency in emitter space, where $v_\parallel$ is the radial proper peculiar velocity of the gas. Hereafter, the subscript or superscript ``RF'' stands for ``rest-frame''.

By defining the optical depth $\tau_\nu$ forward along the ray path as
\begin{equation}
d\tau'_{\nu'} \equiv \kappa_{\nu'} d\xi'\,,
\end{equation}
the radiative transfer equation has the formal solution for the specific intensity observed today at frequency $\nu_{\rm obs}$
\begin{equation}
\label{eqn:radiativetransfer}
I_{\nu_{\rm obs}} = I_{\nu_{\rm obs}}^{\rm CMB}\, e^{-\tau_{\nu_{\rm obs}}} + \int_0^{\tau_{\nu_{\rm obs}}} S_{\nu'}(\xi') 
 e^{-(\tau_{\nu_{\rm obs}}-\tau'_{\nu'})} d\tau'_{\nu'}\,.
\end{equation}
Here we assume that the ray has the same comoving specific intensity as the CMB ($I_{\nu_{\rm obs}}^{\rm CMB}$) when the ray was on the far side of the gas element from the observer. 
$S_{\nu'}(\xi')= j_{\nu'}/\kappa_{\nu'}$ is the {\it comoving} source function at the frequency $\nu'$ seen in FRW space at the proper distance $\xi'$ on the ray path.  $\tau_{\nu_{\rm obs}}$ is the integrated optical depth through the gas. 

\subsubsection{Optical Depth}

In emitter space, the absorption coefficient is 
\begin{equation}
\kappa_\nu^{\rm RF} = \frac{1}{c}h\nu_0 (n_0 B_{01} - n_1 B_{10} ) \phi(\nu'_{\rm RF})\,,
\end{equation} 
where $B_{01}$ and $B_{10}$ are the Einstein probability coefficients for induced upward and downward transitions, respectively, between the lower state with density $n_0$ and higher state with density $n_1$. 
The spin temperature is defined to be the excitation temperature between the hyperfine states, i.e. 
\begin{equation}
\frac{n_1}{n_0} \equiv \frac{g_1}{g_0} e^{-T_\star/T_s} = 3 e^{-T_\star/T_s}\,,
\end{equation}
where $g_0 = 1$ and $g_1 =3$ are the statistical weights. $T_\star \equiv h \nu_0/k_B = 0.068$ K is the temperature corresponding to the rest-frame frequency $\nu_0$. 
For 21cm transitions, all astrophysical applications satisfy $T_s \gg T_\star$, so $n_0 = n_{\rm HI}/4$ where $n_{\rm HI}$ is the {\it proper} number density of neutral hydrogen. 
\footnote{Note that strictly speaking, $n_{\rm HI}$ is the number density in emitter space. But the number densities in emitter space and in FRW space only differ in the relativistic limit, i.e., $n^{\rm RF}/n^{\rm cos} = dV_{\rm cos}/dV_{\rm RF}=dt_{\rm RF}/dt = 1/\sqrt{1-v^2/c^2}$, so we can ignore the difference to linear order $v/c$.} 
It is straightforward to show that 
\begin{equation}
\kappa_\nu^{\rm RF} = \frac{3c^2 A_{10} T_\star n_{\rm HI}\phi(\nu'_{\rm RF})}{32\pi \nu_0^2 T_s} \,,
\end{equation} 
using the identities $g_1 B_{10} = g_0 B_{01} = c^3 g_1 A_{10}/8\pi h\nu_0^3$. 

Now we transform our calculation to FRW space. The absorption coefficient transforms as $\nu^{-1}$ (see, e.g., \citealt{Mihalas78}), so in FRW space 
\begin{equation}
\kappa_{\nu'} = \kappa_\nu^{\rm RF} (\nu'_{\rm RF}/\nu') = \kappa_\nu^{\rm RF}\,(1-\frac{v_\parallel}{c})^{-1}\,.
\end{equation}
Therefore the optical depth is 
\begin{equation}\label{eqn:opt-dep-gen}
\tau_{\nu_{\rm obs}} = \int \kappa_{\nu'} d\xi' 
= \int \frac{3c^2 A_{10} T_\star n_{\rm HI} }{32\pi \nu_0^2 T_s (1-\frac{v_\parallel}{c})} \phi(\nu'_{\rm RF})\,d\xi'\,.
\end{equation}

For the 21cm line transition, in the idealized case of no thermal broadening, the line profile is $\phi(\nu'_{\rm RF}) = \delta(\nu'_{\rm RF} - \nu_0)$. 
Integrating a $\delta$-function picks up the integrand evaluated at the peak which physically corresponds to the location on the ray path where the transition actually takes place, its proper distance labeled as $\xi_r$ (hereafter in this section, the {\it subscript} $r$ stands for ``radiation''). 
We assume that each ray with a given observed frequency only experiences one 21cm transition event along the ray path. (We discuss the multi-transition case in \S~\ref{sec:multievent}.) 
The line profile can be rewritten as 
\begin{equation}
\phi(\nu'_{\rm RF}) = \frac{\delta(\xi'-\xi_r)}{\left| d\nu'_{\rm RF}/d\xi' \right|_{\xi_r}}\,.
\end{equation}
Here we assume the non-singular case, i.e., $\left(d\nu'_{\rm RF}/d\xi' \right)_{\xi_r} \ne 0$. (We discuss the singular case in \S~\ref{sec:singular-revis}.) 
We use the relation $\nu'_{\rm RF} = \nu_{\rm obs} a^{-1}(1-\frac{v_\parallel}{c})^{-1}$ to take the derivative $d\nu'_{\rm RF}/d\xi'$, and then evaluate it at $\xi_r$ where $\nu'_{\rm RF} = \nu_0$. It is straightforward to show that 
\begin{equation}
\left(\frac{d\nu'_{\rm RF}}{d\xi'}\right)_{\xi_r} = \frac{\nu_0}{a_r\,c} \frac{\partial V_\parallel}{\partial r_\parallel}\,,
\end{equation}
where $V_\parallel = a\,r_\parallel H(a) + v_\parallel$ is the proper velocity along the LOS and 
\begin{equation}
\frac{\partial V_\parallel}{\partial r_\parallel} = a H(a) + \frac{dv_\parallel}{dr_\parallel}\,,
\end{equation}
with $r_\parallel$ the comoving LOS distance. 
Therefore the optical depth is 
\begin{equation} \label{eqn:opt-depth}
\tau_{\nu_{\rm obs}}= \frac{3c^3 A_{10}T_{\star} a_r n_{\rm HI}(\xi_r) }{32\pi \nu_0^3 T_s (\xi_r) \left|\partial V_\parallel/\partial r_\parallel\right|_{\xi_r} (1-\frac{v_\parallel (\xi_r)}{c})}\,.
\end{equation}

\subsubsection{Observed Brightness Temperature}
\label{sec:brightT}

Now we simplify the formal solution of radiative transfer equation. 
Since $d\tau'_{\nu'} \propto \delta(\xi'-\xi_r) d\xi'$, the integral in equation~(\ref{eqn:radiativetransfer}) takes non-zero contribution only from $\xi'=\xi_r$, therefore $\int_0^{\tau_{\nu_{\rm obs}}} S_{\nu'}(\xi') 
 e^{-(\tau_{\nu_{\rm obs}}-\tau'_{\nu'})} d\tau'_{\nu'} 
 = S_{\nu'}(\xi_r) \int_0^{\tau_{\nu_{\rm obs}}} 
 e^{-(\tau_{\nu_{\rm obs}}-\tau'_{\nu'})} d\tau'_{\nu'}
 = S_{\nu'}(\xi_r) (1-e^{-\tau_{\nu_{\rm obs}}})$. 
\footnote{The factor $e^{-(\tau_{\nu_{\rm obs}}-\tau'_{\nu'})}$ is a step function at $\xi'=\xi_r$, so more rigorously, the integral yields $\int_0^{\tau_{\nu_{\rm obs}}} S_{\nu'}(\xi') e^{-(\tau_{\nu_{\rm obs}}-\tau'_{\nu'})} d\tau'_{\nu'} = S_{\nu'}(\xi_r) \tau_{\nu_{\rm obs}} [1-(1-e^{-\tau_{\nu_{\rm obs}}})\eta(0)]$.  The unit step function $\eta(x)$ at $x=0$ is undefined in general, but using an identity intrinsic in this problem $ 1 - e^{-\tau_{\nu_{\rm obs}}} = \int_0^{\tau_{\nu_{\rm obs}}} e^{-(\tau_{\nu_{\rm obs}}-\tau'_{\nu'})} d\tau'_{\nu'} = \tau_{\nu_{\rm obs}} [1-(1-e^{-\tau_{\nu_{\rm obs}}})\eta(0)]$, we can regulate $\eta(0)$ and obtain the same result. 
} 

In emitter space, $S_{\nu_0}^{\rm RF}= 2 k_B \nu_0^2 T_s(\xi_r)/c^2$, i.e.\ the Planck function evaluated with the spin temperature $T_s$ at $\xi_r$.  The source function transforms as $\nu^{3}$ (see, e.g., \citealt{Mihalas78}), so the {\it comoving} source function in FRW space is 
\begin{equation}
S_{\nu'} (\xi_r) = a_r^3 \left(\frac{\nu'}{\nu_0}\right)^3 S_{\nu_0}^{\rm RF} = \frac{2 k_B \nu_{\rm obs}^2}{c^2} T_s(\xi_r) a_r (1-\frac{v_\parallel}{c})\,,
\end{equation}
where $a_r^3$ accounts for the comoving factor. 

Suppose the ray has the frequency $\nu_p = \nu_{\rm obs}/a_p$ with some scale factor $a_p < a_r$, (i.e., when it is on the far side of the gas element from the observer,) and is in equilibrium with the CMB of temperature $T_{{\rm CMB},p} = T_{{\rm CMB},0}/a_p$. In the absence of intervening atoms, the comoving specific intensity observed today would be $I_{\nu_{\rm obs}}^{\rm CMB} = a_p^3 2 k_B \nu_p^2 T_{{\rm CMB},p}/c^2 = 2 k_B \nu_{\rm obs}^2 T_{{\rm CMB},0}/c^2 $. 

The 21cm brightness temperature at the observed frequency $\nu_{\rm obs}$ is defined by 
\begin{equation}
I_{\nu_{\rm obs}} \equiv \frac{2 k_B \nu_{\rm obs}^2}{c^2} T_{b}(\nu_{\rm obs}). 
\end{equation}
From equation~(\ref{eqn:radiativetransfer}) it is straightforward to show that 
\begin{equation}\label{eqn:optical-thin}
T_{b}(\nu_{\rm obs}) = T_{{\rm CMB},0}\, e^{-\tau_{\nu_{\rm obs}}} + T_s(\xi_r) a_r (1-\frac{v_\parallel}{c}) (1-e^{-\tau_{\nu_{\rm obs}}})\,.
\end{equation}
The 21cm line is generally optically thin to the IGM, i.e.\ $\tau_{\nu_{\rm obs}} \ll 1$. (We discuss the validity of this approximation in \S~\ref{sec:opt-thin-revis}.) In this limit, the differential brightness temperature is 
\begin{eqnarray}
\delta T_b(\nu_{\rm obs}) &\equiv & T_b(\nu_{\rm obs}) - T_{{\rm CMB},0} \\
&=& a_r\tau_{\nu_{\rm obs}} \left[ T_s(\xi_r)(1-v_\parallel/c) - T_{\rm CMB}(a_r) \right] \,,
\end{eqnarray}
or 
\begin{eqnarray}
\delta T_b(\nu_{\rm obs}) &=& \frac{3c^3 A_{10} T_\star n_{\rm HI}({\bf r}) a_r }{32\pi \nu_0^3 H(a_r)\left|1+(aH)^{-1}\frac{dv_\parallel}{dr_\parallel}({\bf r})\right|} \nonumber\\
& & \times \left[1 -\frac{ T_{\rm CMB}(a_r)}{T_s^{\rm eff}({\bf r})}\right]\,, 
\label{eqn:brightnessT}
\end{eqnarray}
where $\bf r$ is the real-space location of 21cm transition corresponding to the proper distance $\xi_r$ on the ray path. 
$T_{\rm CMB}(a_r) = T_{{\rm CMB},0}/a_r$ is the CMB temperature at the time of 21cm transition. 
Here we define the {\it effective} spin temperature 
\begin{equation}\label{eqn:dist-spin}
T_s^{\rm eff}({\bf r}) \equiv T_s({\bf r}) \left[1-\frac{v_\parallel({\bf r})}{c}\right]\,.
\end{equation}
Note that equation~(\ref{eqn:dist-spin}) only infers that the spin temperature manifests itself to 21cm brightness temperature and optical depth in a manner modified by the peculiar velocity, but this effect does not modify the level population of hydrogen hyperfine states, nor the spin temperature. The level population can in fact be modified by peculiar velocity through an effect pointed out by \cite{Chuzhoy06}. This is however a different effect from the one in equation~(\ref{eqn:dist-spin}) which is based on a {\it given} spin temperature. 

Equation~(\ref{eqn:brightnessT}) is in agreement with the well-known equation in \citet{Barkana05} except for the appearance of effective spin temperature $T_s^{\rm eff}$. However, this modification is actually not important for two reasons. First, it is of order $\mathcal{O}(v/c)$ and the bulk motion of gas is mostly non-relativistic. Second, many research papers focus on the epoch during reionization when $T_{\rm CMB}/T_s \ll 1$, when the spin temperature has a negligible effect on
the brightness temperature.

For convenience, we define the mean\footnote{This is not a volume-weighted mean, but essentially a mean in the redshift-space. See footnote \ref{ftnt:redshift-mean}.}
 brightness temperature in the limit $T_s\gg T_{\rm CMB}$ as 
\begin{eqnarray}
& & \widehat{\delta T}_b (z_{\rm cos}) \equiv  \frac{3c^3 A_{10} T_\star \bar{n}_{\rm HI}(z_{\rm cos})  }{32\pi \nu_0^3 (1+z_{\rm cos}) H(z_{\rm cos})} \nonumber\\
 &=& 23.88 \left(\frac{\Omega_{\rm b}h^2}{0.02}\right)\sqrt{\frac{0.15}{\Omega_{\rm M} h^2}\frac{1+z_{\rm cos}}{10}}\bar{x}_{\rm HI,m}(z_{\rm cos})\,{\rm mK}\,,
\label{eqn:dtbhat}
\end{eqnarray}
where the cosmological redshift is defined by $1+z_{\rm cos} \equiv 1/a_r$, $\bar{n}_{\rm HI}(z_{\rm cos})$ is the mean neutral hydrogen number density at $z_{\rm cos}$, $\bar{x}_{\rm HI,m}(z_{\rm cos})$ is the mean mass-weighted neutral fraction at $z_{\rm cos}$.  
Then we can rewrite equation~(\ref{eqn:brightnessT}) in terms of fluctuations, 
\begin{equation}
\delta T_b(\nu_{\rm obs}) = \widehat{\delta T}_b (z_{\rm cos}) \, \frac{1+\delta_{\rho_{\rm HI}}({\bf r})}{\left|1+\delta_{\partial_r v}({\bf r})\right|}\,\left[1 -\frac{ T_{\rm CMB}(a_r)}{T_s^{\rm eff}({\bf r})}\right]\,,
\label{eqn:dtb1}
\end{equation}
where $\delta_{\rho_{\rm HI}}({\bf r}) = [n_{\rm HI}({\bf r}) - \bar{n}_{\rm HI}(z_{\rm cos})]/\bar{n}_{\rm HI}(z_{\rm cos})$ is the neutral hydrogen density
fluctuation, and $\delta_{\partial_r v}({\bf r})$ is defined in equation~(\ref{eqn:dvdr-def}). 

\subsubsection{Line Profile Revisited: Velocity vs. Thermal Broadening}
\label{sec:line-prof-revis}

In general, the line profile can include thermal broadening, as well as velocity broadening due to bulk motion. The velocity broadening is naturally included by taking the $\delta$-function-shaped line profile peaked at the rest-frame frequency which is shifted both cosmologically and by Doppler effect. Our calculation (eqs.~\ref{eqn:opt-depth} and \ref{eqn:dtb1}) shows that the velocity gradient correction is due to the bulk motion of neutral atoms. However, in their original paper, \cite{Barkana05} explained the inclusion of the velocity gradient compactly, without showing details, as ``The velocity gradient term arises because the 21 cm scattering cross section has a fixed thermal width, which translates through the redshift factor $(1+ v_r /c)$ to a fixed interval in velocity''. This seems to mean that the thermal broadening is responsible for the velocity gradient correction. In this subsection, we will clarify that in the non-singular case, the contribution of thermal broadening is always subdominant to the velocity broadening. 

Basically, the thermal velocity of hydrogen atoms can contribute an additional Doppler shift of the line frequency. For a given $\nu_{\rm obs}$, neutral atoms can in principle have a possibility, given by the Maxwellian distribution, of seeing the radiation in the 21cm rest-frame frequency $\nu_0$, even if $\nu'_{\rm RF}$ (with Doppler shifted due to bulk motion) $\ne \nu_0$ . This is described by the Gaussian line profile, replacing equation~(\ref{eqn:delta-line-prof}),
\begin{equation}\label{eqn:therm-line-prof}
\phi(\nu'_{\rm RF}) = \frac{1}{\Delta\nu_{\rm th} \sqrt{\pi}} \exp{\left[ -\frac{(\nu'_{\rm RF} - \nu_0)^2}{\Delta\nu_{\rm th}^2} \right]}\,,
\end{equation}
where 
\begin{equation}
\Delta\nu_{\rm th}=\frac{\nu_0}{c}\sqrt{\frac{2k_B T_k}{m_H}}\,,
\end{equation}
is the thermal Doppler shift corresponding to a gas kinetic temperature $T_k$. 

In the non-singular case, i.e.\ $\left(d\nu'_{\rm RF}/d\xi' \right)_{\xi_r} \ne 0$, we can change the integration variable in equation~(\ref{eqn:opt-dep-gen}), 
\begin{equation}\label{eqn:reg-var-cha}
d\xi'=\frac{d\nu'_{\rm RF}}{ \left| d\nu'_{\rm RF}/d\xi' \right|}\,,
\end{equation}
and rewrite the optical depth with thermal broadening as 
\begin{equation}
\tau_{\nu_{\rm obs}}^{\rm T} = \int_{-\infty}^\infty \mathfrak{T}\Bigl(\xi'(\nu'_{\rm RF})\Bigr) \phi(\nu'_{\rm RF})\,d\nu'_{\rm RF}\,.
\end{equation}
where $\mathfrak{T}(\xi') $ is the function in equation~(\ref{eqn:opt-depth}) with $\xi_r$ replaced by $\xi'$ corresponding to $\nu'_{\rm RF}$, so by definition 
$\mathfrak{T}(\xi_r) \equiv \tau_{\nu_{\rm obs}}^{\rm NT}$ is the optical depth without thermal broadening for the observed frequency $\nu_{\rm obs}$.

Suppose the rest-frame frequency finds $\nu'_{\rm RF} = \nu_0$ at $\xi'=\xi_r$. 
The thermal width is small compared to $\nu_0$, since $\Delta\nu_{\rm th}/\nu_0\sim 10^{-5}$ if $T_k\sim 10^4$~K. Therefore the integrand is nonzero only near $\nu'_{\rm RF} = \nu_0$. So we can Taylor expand the integrand at $\nu_0$ to sub-leading order in $\mathcal{O}(\nu'_{\rm RF} - \nu_0)^2$, since the first order $\propto \int_{-\infty}^\infty d\nu'_{\rm RF} (\nu'_{\rm RF} - \nu_0) \exp{\left[ -\frac{(\nu'_{\rm RF} - \nu_0)^2}{\Delta\nu_{\rm th}^2} \right]} = 0 $. It is straightforward to show that the result is 
\begin{equation}
\tau_{\nu_{\rm obs}}^{\rm T} = \tau_{\nu_{\rm obs}}^{\rm NT} \left[1+ \Delta \tau_{\nu_{\rm obs}}^{\rm T}\right]\,,
\end{equation}
where the fractional correction due to thermal broadening is 
\begin{equation}\label{eqn:thermal}
\Delta \tau_{\nu_{\rm obs}}^{\rm T}  =  \frac{1}{4 \tau_{\nu_{\rm obs}}^{\rm NT}} 
 \frac{d^2 \mathfrak{T}\Bigl(\xi'(\nu'_{\rm RF})\Bigr)}{d{\nu'_{\rm RF}}^2}\biggr|_{\nu_0} \Delta\nu_{\rm th}^2 
 \sim \mathcal{O}\left(\frac{\Delta\nu_{\rm th}}{\nu_0}\right)^2 
 \sim 10^{-9}\,.
\end{equation}
Here we assume the gas temperature is about $10^4$~K, which is close to the maximum temperature attainable by neutral hydrogen before collisional ionization becomes important. Therefore, the contribution of thermal broadening is always negligible compared to the bulk motion.

\subsubsection{Observed Brightness Temperature: Optically Thick Limit}
\label{sec:singular-revis}

Our results for optical depth and brightness temperature seem to diverge for $\delta_{\partial_r v}=-1$ (see eqs.~\ref{eqn:opt-depth} and \ref{eqn:dtb1}). We discuss this singularity behavior in this subsection, and find that the divergence in optical depth can be relaxed by including thermal broadening, and the divergence in brightness temperature can be removed by dropping the optically thin approximation. 

We should first note that the singularity at $\delta_{\partial_r v}=-1$ corresponds to $\left(d\nu'_{\rm RF}/d\xi' \right)_{\xi_r} = 0$. In this case, the regular changing variable technique (eq.~\ref{eqn:reg-var-cha}) is invalid. Instead, one should Taylor expand $\nu'_{\rm RF}(\xi')$ near $\xi_r$ to second order, 
$\nu'_{\rm RF} = \nu_0 + \frac{1}{2} \beta (\xi'-\xi_r)^2 $, 
where $\beta = \frac{d^2 \nu'_{\rm RF}}{d\xi^{'\,2}}\biggr|_{\xi_r}$, and find that 
\begin{equation}
d\xi' = {\rm sgn}(\xi'-\xi_r) \,{\rm sgn}(\beta)\, \frac{d\nu'_{\rm RF}}{\sqrt{2\beta (\nu'_{\rm RF} -\nu_0)}}\,.
\end{equation}
Then the optical depth becomes
\begin{eqnarray}
\tau_{\nu_{\rm obs}} &=& 2\int_{\nu_0}^{{\rm sgn}(\beta)\cdot\infty} \,d\nu'_{\rm RF}\,{\rm sgn}(\beta)\,\frac{3c^2 A_{10} T_\star n_{\rm HI} }{32\pi \nu_0^2 T_s (1-\frac{v_\parallel}{c})} \nonumber \\
& & \times \frac{\phi(\nu'_{\rm RF})}{\sqrt{2\beta (\nu'_{\rm RF} -\nu_0)}} \,.
\end{eqnarray}

If there is no thermal broadening, the line profile is a $\delta$-function peaked at $\nu'_{\rm RF} = \nu_0$, and the optical depth is still divergent due to the $1/\sqrt{\nu'_{\rm RF} -\nu_0}$ factor. However, thermal broadening can remove this divergence. To see this, we can move the $\xi'$-dependent factors ($n_{\rm HI}$, $T_s$, and $v_\parallel$) out of the integral, evaluated at $\xi_r$, since the evaluation is concentrated near $\xi_r$. When applying the thermal line profile (eq.~\ref{eqn:therm-line-prof}), we find that the optical depth in the singular case ($\delta_{\partial_r v}=-1$) becomes
\begin{equation}\label{eqn:opt-sing}
\tau_{\nu_{\rm obs}} =\frac{3c^2 A_{10} T_\star n_{\rm HI}(\xi_r) }{32\pi \nu_0^2 T_s(\xi_r) (1-\frac{v_\parallel(\xi_r)}{c})} \times \frac{1.446}{\sqrt{|\beta| \Delta\nu_{\rm th}}}\,. 
\end{equation}
Here the factor $1.446$ is an approximation of $\sqrt{2/\pi}\times 2\,\Gamma(5/4)$. Since $\tau_{\nu_{\rm obs}} \propto 1/\sqrt{\Delta\nu_{\rm th}}$, the optical depth can be large when $\delta_{\partial_r v}$ is close to $-1$. We evaluate $\beta$ when $\delta_{\partial_r v} = -1$, 
\begin{equation}\label{eqn:opt-beta}
\beta = -\frac{\nu_0 H(a_r)^2}{c^2} \left[ 2 + \frac{a_r H'(a_r)}{H(a_r)} - \frac{c}{(a_r H(a_r))^2} \frac{d^2 v_\parallel}{d\,r_\parallel^2}\biggr|_{\xi_r} \right]\,,
\end{equation}
where $H'(a) = dH/da$. 

There are two astrophysical cases that can generate $\delta_{\partial_r v}=-1$. One is the virialized halo and the other is the spherical collapse at the turn-around point (pre-virialization). In both cases, the proper velocity is $V_\parallel = a\,r_\parallel H(a) + v_\parallel = 0 $ (seen from the halo center), so $dv_\parallel/dr_\parallel = - aH(a)$. Near the singular point, the optical depth can become large, thus invalidating the optically thin approximation. As a result, 
one cannot apply the popular equation (eq.~\ref{eqn:brightnessT}) to evaluate the brightness temperature, but should instead use the exact solution 
\begin{equation}
\label{eqn:opt-thick-formula}
\delta T_b(\nu_{\rm obs}) = a_r \left[ T_s(\xi_r)(1-\frac{v_\parallel}{c}) - T_{\rm CMB}(a_r) \right] \Bigl[ 1- e^{-\tau_{\nu_{\rm obs}}} \Bigr]
\end{equation}
where $\tau_{\nu_{\rm obs}}$ is given by equation~(\ref{eqn:opt-sing}) when $\tau_{\nu_{\rm obs}} \gtrsim 1$ or equation~(\ref{eqn:opt-depth}) when $\tau_{\nu_{\rm obs}}<1$ but not too small. The exact value of $\tau_{\nu_{\rm obs}}$ is not important as long as $\tau_{\nu_{\rm obs}} \gg 1$, since the $\tau$-dependent term should saturate $1-\exp{(-\tau_{\nu_{\rm obs}})} \approx 1$ for large $\tau_{\nu_{\rm obs}}$, in which case the brightness temperature is still finite, i.e.\ 
\begin{equation}\label{eqn:opt-thick-formula2}
\delta T_b(\nu_{\rm obs}) \approx a_r \left[ T_s(\xi_r)(1-\frac{v_\parallel}{c}) - T_{\rm CMB}(a_r) \right]
\end{equation}
instead of infinite as it would be using the popular equation~(\ref{eqn:brightnessT}). 

\begin{figure*}
\begin{center}
  \includegraphics[height=0.35\textheight]{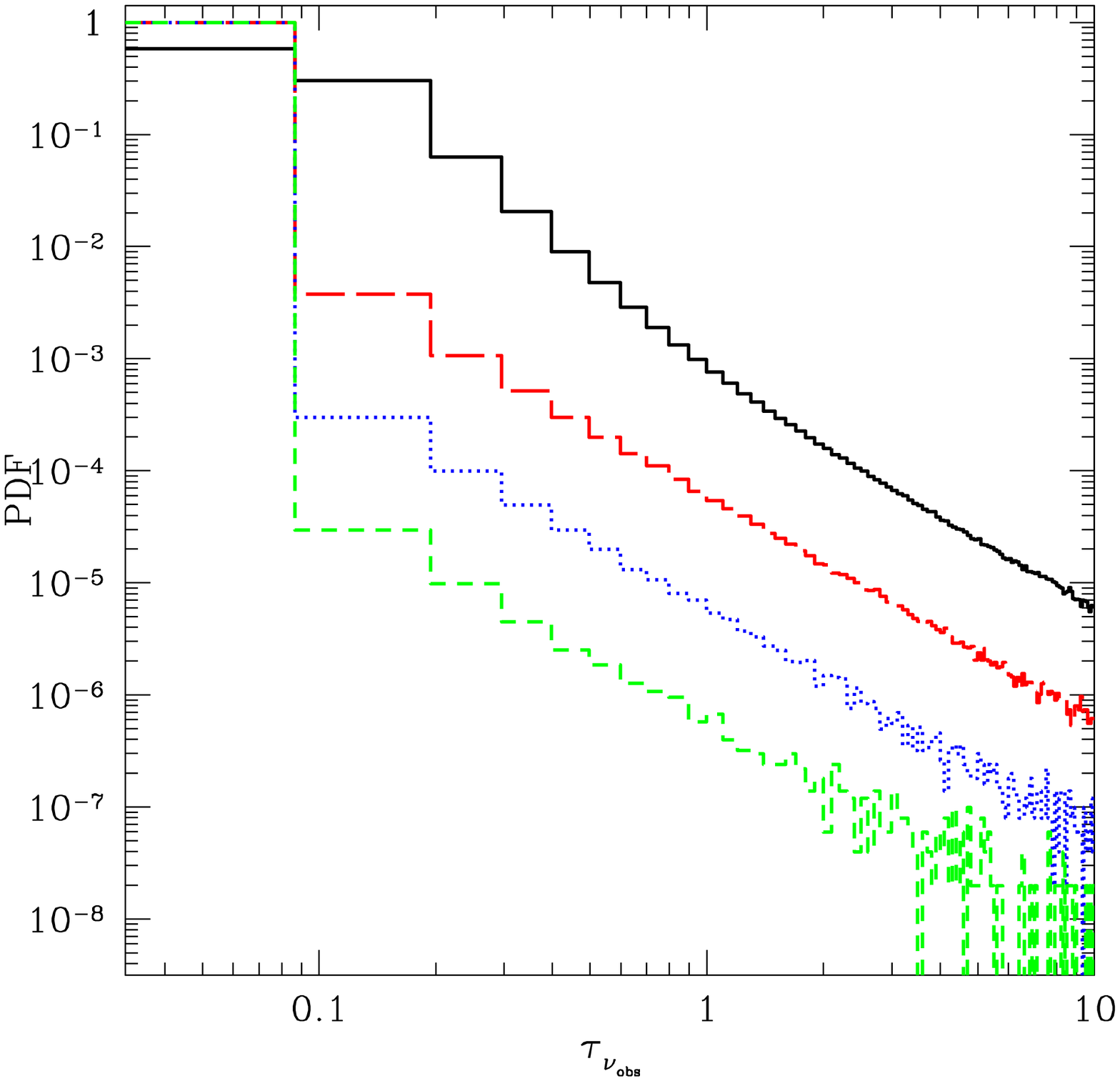}
  \includegraphics[height=0.35\textheight]{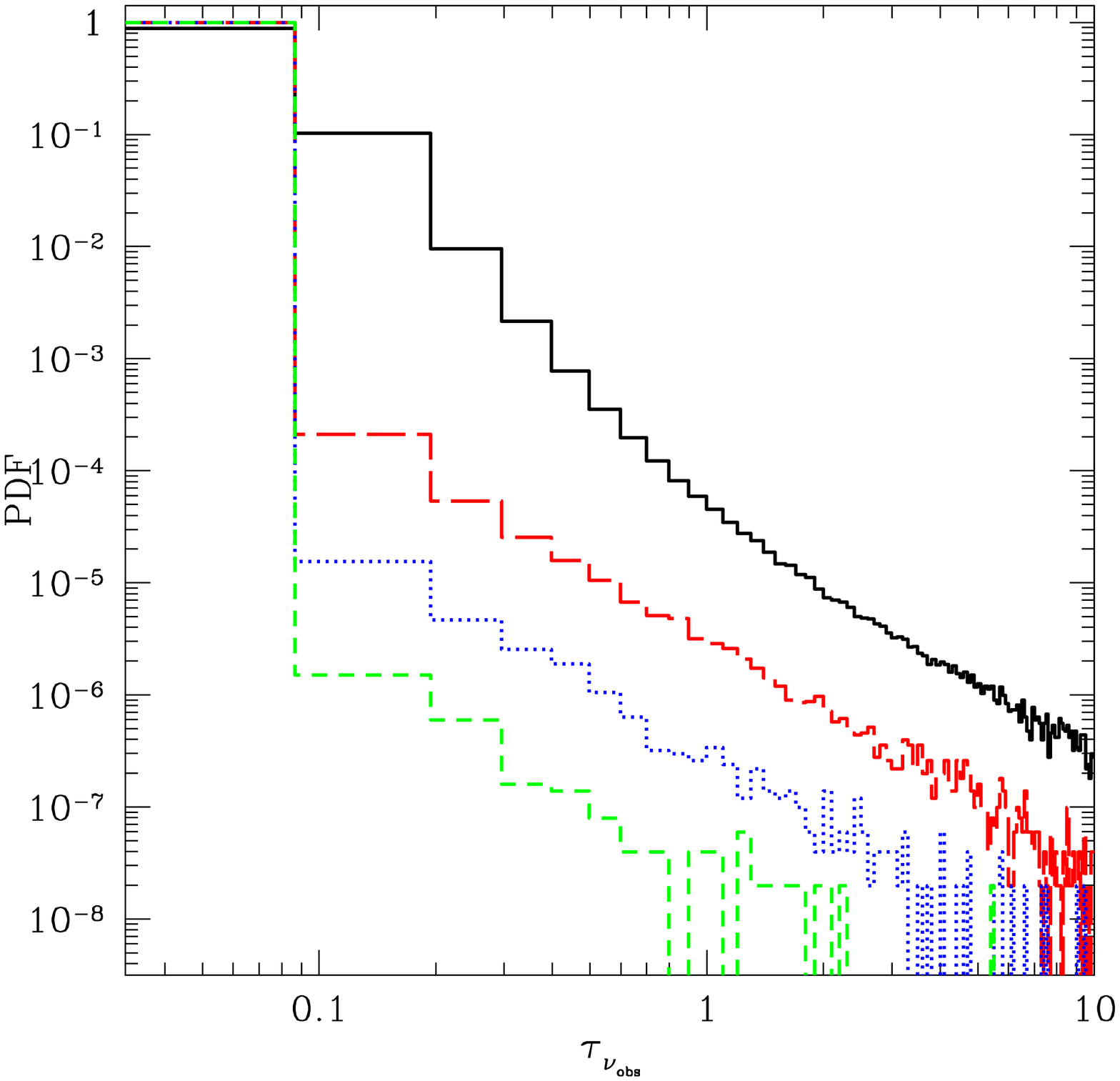} 
\end{center}
\caption{PDF of 21cm optical depth $\tau_{\nu_{\rm obs}}$ from our simulation data at $z=9.457$ (50\% ionized). The simulation is in a 114 Mpc/h box with IGM particle data smoothed onto a $256^3$ grid. PDF shows the probability of finding $\tau_{\nu_{\rm obs}}$ in intervals of $\Delta\tau_{\nu_{\rm obs}}=0.1$. We assume $T_s/T_{\rm CMB} = 0.1$ (solid, black), 1 (long-dashed, red), 10 (dotted, blue), and 100 (short-dashed, green). Left panel: assuming a fully neutral Universe ($x_{\rm HI} = 1$). Right panel: using the actual ionization pattern from the simulation.}
\label{fig:PDFtau}
\end{figure*}

\subsubsection{How Good is the Optically Thin Approximation during the EOR?}
\label{sec:opt-thin-revis}

It is often assumed that 21cm line is optically thin, fundamentally because 21cm hyperfine transition is highly forbidden with an extremely small probability of $A_{10} = 2.85\times 10^{-15} {\rm s}^{-1}$. However, peculiar velocity gradients can enhance the optical depth in overdense regions
\footnote{\cite{Iliev02} showed that the 21cm line can become optically thick inside dense mini-halos, but this is a different effect from the enhancement due to velocity gradients we consider here.
}. To see this, we rewrite the optical depth in equation~(\ref{eqn:opt-depth}) as 
\begin{eqnarray}
\tau_{\nu_{\rm obs}} &=& \frac{ \widehat{\delta T}_b (z_{\rm cos})}{T_{\rm CMB,0}} \frac{1+\delta_{\rho_{\rm HI}}}{\alpha  
\left|1+\delta_{\partial_r v}\right| (1-\frac{v_\parallel}{c})}\nonumber \\
& =& 0.00438 \left(\frac{\Omega_{\rm b}h^2}{0.02}\right)\sqrt{\frac{0.15}{\Omega_{\rm M} h^2}\frac{1+z_{\rm cos}}{10}}\frac{\bar{x}_{\rm HI,m}(z_{\rm cos})}{0.5} \nonumber \\
& &  \times \frac{1+\delta_{\rho_{\rm HI}}}{\alpha  
\left|1+\delta_{\partial_r v}\right| (1-\frac{v_\parallel}{c})}\,, 
\label{eqn:opt-dep-rewrit}
\end{eqnarray}
where $\alpha \equiv T_s({\bf r})/T_{\rm CMB}(z_{\rm cos})$. For example in our reionization simulation at $z\sim 9$ when $\bar{x}_{\rm HI,m}\sim 0.5$, and with reasonable assumptions such as non-relativistic bulk motion $v\ll c$ and small fluctuation $1+\delta_{\rho_{\rm HI}} \sim 1$, the optical depth can become of order unity when the velocity gradient is very negative, such that $\left|1+\delta_{\partial_r v}\right| \lesssim \frac{0.004}{\alpha}$, as can happen in some overdense regions. This condition on the velocity gradient widens when 21cm occurs in absorption ($T_s < T_{\rm CMB}$) and becomes narrower when it occurs in emission ($T_s > T_{\rm CMB}$). 

Figure~\ref{fig:PDFtau} shows the PDF of the $\tau_{\nu_{\rm obs}}$ distribution of the IGM from our simulation data (see simulation details in \S~\ref{sec:sim}). We smooth the N-body particle mass in the IGM onto a regular $256^3$ grid,  compute the cell's velocity gradient using the SPH-like smoothing method described in Appendix~\ref{sec:dvdrmethod}, and compute the optical depth of the IGM. For simplicity, we drop the $1-v/c$ factor in the optical depth calculation by assuming  non-relativistic bulk motions. Figure~\ref{fig:PDFtau}(a) shows the extent by which velocity gradients alone can enhance the optical depth, by assuming $\delta_{\rho_{\rm HI}} = \delta_{\rho_{\rm H}}$. The PDF at large optical depth increases by roughly an order of magnitude when we decrease the spin temperature by an order of magnitude. 
In the pessimistic case ($T_s / T_{\rm CMB} = 0.1$), as many as 0.1\% of the total cells have an optical depth of order unity. 

In Figure~\ref{fig:PDFtau}(b), we plot the same PDF for the actual $\delta_{\rho_{\rm HI}}$ from the reionization simulation. Due to the ``inside-out'' character of reionization, the overdense regions that can have velocity gradients close to $-1$ ionize first and one can expect the effect to be much less. Figure~\ref{fig:PDFtau}(b) shows that for the case $T_s / T_{\rm CMB} = 0.1$, only a fraction of up to $10^{-4}$ of the total number cells approach an optical depth of 1. For the case $T_s / T_{\rm CMB} = 100$ this fraction becomes as low as $10^{-7}$. We therefore conclude that we can safely use the optically thin approximation when calculating the 21cm radiation from the IGM. 

However, we should note that the optically thin approximation may break down to a larger extent in one of the two following scenarios.
\begin{enumerate}
\item 21cm radiation from a halo or in spherical collapse at the turn-around point may be mostly optically thick, because $\delta_{\partial_r v}\sim -1$ there. The breakdown of the optically thin approximation may be more prominent when the 21cm line is in absorption against the CMB. 
\item When the 21cm radiation is computed directly from high-resolution particle data (and not from gridded data as above), a larger fraction of particles can be optically thick, since the particle density is higher in overdense regions. 
\end{enumerate}
The breakdown of the optically thin approximation merits more investigation beyond the scope of this paper where we focus on 21cm radiation from the IGM and this approximation is mostly valid. We defer further analysis to future work.

\subsubsection{When the Mapping from Frequency to Position Along LOS is Multi-valued}
\label{sec:multievent}

In the case of multiple 21cm transitions along the ray path, we label the transition events by $i = 1,\ldots,N$ in sequence along the forward ray path. The optical depth starts from $\tau_0 \equiv 0$ (on the far side of the gas element), to $\tau_i$ (after the ray passes through event $i$), and to $\tau_N =  \tau_{\nu_{\rm obs}}$. We define the differential optical depth $\Delta \tau_i \equiv \tau_i - \tau_{i-1} = \int_{{\rm across}\,i} \kappa_{\nu'} d\xi'$ which can be evaluated using equation~(\ref{eqn:opt-depth}) with the transition location $\xi_r \to \xi_i$. To carry out the integration in equation~(\ref{eqn:radiativetransfer}), we split the integral into a sum of $N$ sub-integrals each over only one transition event, 
i.e., $\int_0^{\tau_{\nu_{\rm obs}}} \ldots  d\tau'_{\nu'} = \sum_{i=1}^{N} \int_{\tau_{i-1}}^{\tau_i} \ldots  d\tau'_{\nu'}$. 
Using the same trick as in \S~\ref{sec:brightT}, we find 
$\int_0^{\tau_{\nu_{\rm obs}}} S_{\nu'}(\xi')  e^{-(\tau_{\nu_{\rm obs}}-\tau'_{\nu'})} d\tau'_{\nu'} = \sum_{i=1}^{N} S_{\nu'}(\xi_i) \Delta \tau_i$ in the {\it optically thin} limit. Using the fact that $\tau_{\nu_{\rm obs}} = \sum_{i=1}^{N} \Delta \tau_i$, we find that 
$\delta T_b(\nu_{\rm obs}) = \sum_{i=1}^{N} \left.\delta T_b(\nu_{\rm obs})\right|_{\xi_i}$, i.e.\ the observed differential brightness temperature is the sum of contributions from all transitions, where each contribution can be evaluated using the equation for a single transition (eq.\ref{eqn:brightnessT}) with ${\bf r}$ the position of each transition.

\subsection{The Distortion of Apparent Location and Brightness Temperature by Peculiar Velocity}
\label{sec:doppler}

In \S~\ref{sec:rederive} we derived the equation for the observed 21cm brightness temperature, evaluating physical properties at the {\it actual} location ${\bf r}$ of the emitting neutral hydrogen atoms. However, observers can only determine the position of the source from the {\it observed} 21cm line frequency, i.e.\ in observer redshift space.
To make theoretical predictions, it is therefore necessary to express the observed 21cm brightness temperature in observer redshift-space coordinates. This subsection deals with solving this issue. 

\subsubsection{Distinguishing the Two Distortion Effects by Peculiar Velocity}
\label{sec:distinguishing}
We should emphasize first that, although the effects of peculiar velocity on observed 21cm brightness temperature and on apparent location of sources are both due to the Doppler shift of the line frequency, the underlying mechanisms do differ. For the former, peculiar velocity distinguishes the emitter space from the FRW space, both of which are physical reference frames, and translates the difference between these two frames, through the transformation of the line profile, to the optical depth that affects the brightness temperature measured by observers today in FRW space. It is a ``real'' effect in the sense that peculiar velocities change the observed brightness temperature, regardless of how observers interpret the location of source.
 
For the latter, the observer redshift-space coordinates are simply an artificial coordinate system that could be replaced by the observer real-space coordinates if observers could measure the peculiar velocities of sources and reconstruct the brightness temperature map in the sources' actual location. This is ``artificial'' in the sense that it is due to the observers' incomplete information on the location of the sources.

The observed brightness temperature we derived in equation~(\ref{eqn:dtb1}) is evaluated in terms of quantities measured in real space. We can rewrite equation~(\ref{eqn:dtb1}) as
\begin{equation}\label{bt-real}
\delta T_b^r ({\bf r}) = \widehat{\delta T}_b (z_{\rm cos}) \, \frac{1+\delta^r_{\rho_{\rm HI}}({\bf r})}{\left|1+\delta^r_{\partial_r v}({\bf r})\right|}\,\left[1 -\frac{ T_{\rm CMB}(a_r)}{T_s^{r,{\rm eff}}({\bf r})}\right]\,, 
\end{equation}
using the superscript ``r'' for real space explicitly. (See our convention of superscripts ``r'' and ``s'' in \S~\ref{sec:term-ref-frame}.)
By definition, the brightness temperature calculated from redshift-space quantities, $\delta T_b^s ({\bf s})$, is equal to $\delta T_b^r ({\bf r})$. So in principle, we can combine the two effects of peculiar velocity and find an expression for the brightness temperature using redshift-space quantities from
\begin{equation}
\delta T_b^s ({\bf s}) = \delta T_b^r ({\bf r}({\bf s}))\,,
\label{eqn:bt-preserved}
\end{equation}
where ${\bf r}({\bf s})$ is the inverse of the real-to-redshift-space mapping ${\bf r} \to {\bf s} = {\bf r} + \frac{(1+z_{\rm obs})}{H(z_{\rm obs})} v_\parallel({\bf r}) \, \hat{r}$. We show below that this relation can be simplified for the 21cm brightness temperature. 
We restrict the discussion to calculating the 21cm brightness temperature in the optically thin approximation here, since this is mostly valid in the IGM.

\subsubsection{21cm Brightness Temperature in Observer Redshift Space: Mathematical Approach}

We present the derivation of the equation for the 21cm brightness temperature in terms of redshift-space quantities in two ways. First in this subsection in a mathematical way, and in the next subsection in a more heuristic physical way. To simplify matters, we for now assume that $T_s \gg T_{\rm CMB}$ and generalize our result to an arbitrary $T_s$ in \S~\ref{sec:spin-reload}. 

Analogous to redshift space distortion in galaxy surveys, where the number of galaxies is preserved between real- to redshift-space, the number of {\it emitting neutral hydrogen atoms} is preserved in the 21cm signal, i.e.\ $n_{\rm HI}^s({\bf s}) \delta V^s({\bf s}) (1+z_{\rm cos})^{-3} = n_{\rm HI}^r({\bf r}) \delta V^r({\bf r})(1+z_{\rm cos})^{-3} $. From the real-to-redshift-space mapping, ${\bf s} = {\bf r} + \frac{(1+z_{\rm obs})}{H(z_{\rm obs})} v_\parallel({\bf r}) \, \hat{r}$, it is easy to find the relation between comoving volume elements in both frames $\delta V^s({\bf s}) = \delta V^r({\bf r}) \left|1+\delta^r_{\partial_r v}({\bf r})\right|$. Therefore, the number density measured in redshift space is 
\begin{equation}\label{eqn:densityrelation}
n_{\rm HI}^s({\bf s}) = \frac{n_{\rm HI}^r({\bf r})}{\left| 1+\delta^r_{\partial_r v}({\bf r})\right|}\,. 
\end{equation}
The mean number density must be preserved too, when averaged over a volume large enough to contain all gas of interest. In terms of fluctuations $\delta^s_{\rho_{\rm HI}}({\bf s}) = \frac{n^s_{\rm HI}({\bf s}) - \bar{n}_{\rm HI}(z_{\rm cos})}{\bar{n}_{\rm HI}(z_{\rm cos})}$, where $\bar{n}_{\rm HI}$ is the mean (physical) HI number density, we have 
$1+\delta^s_{\rho_{\rm HI}}({\bf s}) = \frac{1+\delta^r_{\rho_{\rm HI}}({\bf r})}{\left| 1+\delta^r_{\partial_r v}({\bf r})\right|}$, and hence in the $T_s \gg T_{\rm CMB}$ limit, 
\begin{equation}\label{eqn:Tbredshiftspace}
\delta T_b^s ({\bf s}) 
= \frac{\widehat{\delta T}_b (z_{\rm cos})}{\bar{n}_{\rm HI}(z_{\rm cos})} n^s_{\rm HI}({\bf s})
= \widehat{\delta T}_b (z_{\rm cos}) \, \left[1+\delta^s_{\rho_{\rm HI}}({\bf s})\right]\,. 
\end{equation}
This means that in the high $T_s$ limit, the observed 21cm brightness temperature is directly proportional to the number density of neutral hydrogen atoms measured in observer redshift space. In other words, 21cm tomography maps exactly the neutral hydrogen distribution in redshift-space. This is the result we already found in \S~\ref{sec:21cm-limit}, but now more rigorously derived.  

In case of multiple transitions along the ray path, the brightness temperature is the sum of contributions from all transition events, as discussed in \S~\ref{sec:multievent}. Since these transitions correspond to the same observed frequency and therefore the same redshift-space location, equation~(\ref{eqn:Tbredshiftspace}) still holds for the multi-transition case, since by definition the HI density in redshift-space is the linear addition of HI mass from all such transition spots per unit redshift-space volume. 

\subsubsection{21cm Brightness Temperature in Observer Redshift Space: Physical Approach}
\label{sec:physapproch}

Now we rederive equation~(\ref{eqn:Tbredshiftspace}) by considering the physical meaning of brightness temperature. The
21cm brightness temperature is simply proportional to the specific intensity, i.e.\ $\delta T_{b}(\nu_{\rm obs}) = \frac{c^2}{2 k_B \nu_{\rm obs}^2} \delta I_{\nu_{\rm obs}}$ where $\delta I_{\nu}$ is the differential specific intensity relative to CMB, and equal to the energy received from distant gas per unit observation time per unit transverse collection area per solid angle spanned by sources per unit observed frequency interval. The solid angle is proportional to the transverse area of the source, the observed frequency interval is proportional to the LOS distance interval in {\it redshift} space, and hence the energy received from a patch of sky near $\nu_{\rm obs}$ per unit time per unit collection area is proportional to the brightness temperature times the {\it redshift}-space volume element. I.e., $d^2\Omega = dA^s_\perp / d_A^2(z_{\rm obs})$, $d\nu_{\rm obs} = |ds_\parallel |/y(z_{\rm obs})$, and $dE / dt\, dA_{\rm coll} = C(z_{\rm obs})\, \delta T_{b}(\nu_{\rm obs}) \delta V^s$, where $d_A(z_{\rm obs})$ is the comoving angular diameter distance\footnote{Here $d_A(z)\equiv {c\over H_0}|\Omega_k|^{-1/2}S\left[|\Omega_k|^{1/2}\int_0^z \frac{dz'}{E(z')}\right]$, where $E(z)\equiv \frac{H(z)}{H_0}$ is the relative cosmic expansion rate, and
the function $S(x)$ equals $\sin (x)$ if $\Omega_k<0$, $x$ if $\Omega_k=0$, and $\sinh x$ if $\Omega_k>0$. 
Strictly speaking, it should be $d_A(z_{\rm cos})$ that differs from $d_A(z_{\rm obs})$ by $v_\parallel (1+z_{\rm obs})/H(z_{\rm obs})$. Since $d_A$ is large at high redshift, this difference is negligible.\label{footnote:dA}}, 
$y(z_{\rm obs}) = \lambda_0 (1+z_{\rm obs})^2/H(z_{\rm obs})$, $dA^s_\perp$ is the comoving transverse area in redshift space, $ds_\parallel$ is the comoving radial interval in redshift space, $C(z_{\rm obs}) \equiv 2 k_B \nu_{\rm obs}^2/c^2 d_A^2(z_{\rm obs}) y(z_{\rm obs}) $, and $\delta V^s = dA^s_\perp\,|ds_\parallel| $ is the comoving redshift-space volume element. 

Consider a small region (e.g., a cell or a pixel) of the sky at the telescope's resolution scale. The detector simply smears subcell brightness temperature information by summing energies received from all unresolved subcells. For each subcell, $\delta T_b \,\delta V^s  = \frac{\widehat{\delta T}_b (z_{\rm cos})}{\bar{n}_{\rm HI}(z_{\rm cos})} \frac{ n_{\rm HI}({\bf r})}{ \left|1+\delta^r_{\partial_r v}({\bf r})\right|}\times\delta V^r({\bf r}) \left|1+\delta^r_{\partial_r v}({\bf r})\right| = (1+z_{\rm cos})^3 \frac{\widehat{\delta T}_b (z_{\rm cos})}{\bar{n}_{\rm HI}(z_{\rm cos})} \delta N_{\rm HI}$, where $\delta N_{\rm HI}({\bf r}) =  (1+z_{\rm cos})^{-3}\,n_{\rm HI}({\bf r}) \,\delta V^r({\bf r}) $ is the number of emitting neutral hydrogen atoms from the subcell at $\bf r$. Ignoring the difference of observed frequency and redshift between the subcells, the brightness temperature of the cell is 
$\delta T_{b}(\nu_{\rm obs})  =  \frac{1}{C(z_{\rm obs})\Delta V^s} \sum \left[\frac{dE}{dt\, dA_{\rm coll}}\right]_{\rm sub} 
= \frac{1}{\Delta V^s} \sum \left[ \delta T_{b} \,\delta V^s\right]_{\rm sub} 
= \widehat{\delta T}_b (z_{\rm cos}) \frac{n^s_{\rm HI,cell}}{\bar{n}_{\rm HI}} = \widehat{\delta T}_b (z_{\rm cos}) \, \left[1+\delta^s_{\rho_{\rm HI}}({\bf s})\right]$
(i.e.\ eq.~\ref{eqn:Tbredshiftspace}), where $\Delta V^s$ is the total redshift-space volume of the cell, and $n^s_{\rm HI, cell} = (1+z_{\rm cos})^3 \left(\sum \delta N_{\rm HI,sub}\right)/\Delta V^s = (1+z_{\rm cos})^3 \Delta N_{\rm HI}/\Delta V^s $ is the {\it cell-wise} (physical) HI number density in redshift space.

\subsubsection{Spin Temperature Reloaded} 
\label{sec:spin-reload}

In this subsection we generalize our calculation to the case of arbitrary spin temperature. Following the same algebra as in \S~\ref{sec:physapproch}, for each unresolved subcell, $\delta T_b \,\delta V^s  = (1+z_{\rm cos})^3 \,\frac{\widehat{\delta T}_b (z_{\rm cos})}{\bar{n}_{\rm HI}(z_{\rm cos})} \delta N_{\rm HI} \,\left[1 -\frac{ T_{\rm CMB}(z_{\rm cos})}{T_s^{r,{\rm eff}}({\bf r})}\right] $. Then the brightness temperature of a cell is 
\begin{equation}
\delta T_{b}(\nu_{\rm obs}) = \frac{\widehat{\delta T}_b (z_{\rm cos})}{\bar{n}_{\rm HI}(z_{\rm cos})} \left< n_{\rm HI} \,\left[1 -\frac{ T_{\rm CMB}}{T_s^{\rm eff}}\right] \right>^s_{\rm cell}
\label{eqn:finite_TS1}
\end{equation}
where 
\begin{eqnarray}
\left< n_{\rm HI} \,\left[1 -\frac{ T_{\rm CMB}}{T_s^{\rm eff}}\right] \right>^s_{\rm cell} &=& \frac{1}{\Delta V^s} \sum_{\rm subcells} \biggl\{ n^s_{\rm HI}({\bf s}) \nonumber \\ 
& & \!\!\!\!\!\!\!\!\!\!\!\!\!\!\!\! \times \,\left[1 -\frac{ T_{\rm CMB}(z_{\rm cos})}{T_s^{s,{\rm eff}}({\bf s})}\right] \delta V^s \biggr\}_{\rm sub} 
\label{eqn:finite_TS2}
\end{eqnarray}
is the redshift-space-volume-weighted cell-wise average of $n_{\rm HI} \,\left[1 -\frac{ T_{\rm CMB}}{T_s^{\rm eff}}\right]$, or in other words, the cell-wise total of $\delta N_{\rm HI}\left[1 -\frac{ T_{\rm CMB}}{T_s^{\rm eff}}\right]$ per unit proper redshift-space volume.
Here we implicitly assume that spin temperature is preserved from real- to redshift-space, i.e., $T_s^{s,{\rm eff}}({\bf s}) = T_s^{r,{\rm eff}}({\bf r})$. 

\subsubsection{Breakdown of the Analogy to Galaxy Surveys}

From our results it is clear that the analogy to galaxy redshift surveys breaks down due to two effects: finite optical depth and finite spin temperature, as mentioned before in Sec.~\ref{sec:21cm-limit}. \footnote{There is a third, more technical, difference between galaxy redshift surveys and 21cm surveys. In principle, the apparent location shift from real- to redshift-space results in the difference in the comoving transverse area and, hence, affects the redshift-space volume, in addition to the effect due to the change in the comoving LOS distance interval. This additional effect is non-negligible for galaxy redshift surveys at low redshifts, but small for high-redshift 21cm surveys (as discussed in Footnote~\ref{footnote:dA}). 
We thank Antony Lewis (2011, private communication) for pointing this out to us. 
}

For the first case, when the IGM is optically thick to 21cm radiation, i.e., $\tau_{\nu_{\rm obs}} \gtrsim 1$, the brightness temperature is not linear in $\tau_{\nu_{\rm obs}}$ (see eq.~\ref{eqn:opt-thick-formula}), and the optical depth itself is affected by peculiar velocity through its dependence on spatial derivatives that are higher order than $dv_\parallel/d\,r_\parallel$ (see eqs.~\ref{eqn:opt-sing} and \ref{eqn:opt-beta}). 
Consequently, the brightness temperature is no longer proportional to 
the neutral atom density in redshift space. 

For the second case, e.g. at high redshifts where $T_s \gg T_{\rm CMB}$ is not satisfied \footnote{It is generally assumed that sufficiently late after the formation of the first stars, the spin temperature is well above the CMB temperature. This assumes, e.g., that the IGM is heated but only weakly ionized, as by the X-rays expected from early galaxies and mini-quasars (e.g. \citealt{Chen04}). It also assumes that the first stars produce a strong enough Ly$\alpha$-pumping background to couple $T_s$ to the kinetic temperature of the gas through the Wouthuysen-Field effect (e.g. \citealt{Ciardi03}). However, the length of the transition period from $T_s \lesssim T_{\rm CMB}$ in the Dark Ages to $T_s \gg T_{\rm CMB}$ during the later stages of the EOR is an unsettled topic (see, e.g., \citealt{Baek10}). }, 
neutral atoms in the same redshift-space volume element contribute unequally to the brightness temperature due to their spatial variation in level population, i.e., emitters can have different luminosity. 
Thus the brightness temperature is no longer proportional only to the neutral atom density in redshift space. 
When the mapping from real- to redshift-space is single-valued, the proportionality between observed brightness temperature and neutral atom density in redshift-space is spoiled by the spatially-varying correction factor, $1-T_{\rm CMB}/T_s^{r,\,\rm eff}({\bf r})$, according to equations~(\ref{eqn:finite_TS1}) and (\ref{eqn:finite_TS2}). However, in the more general case in which the mapping may be multi-valued, this correction factor is an average over the different real-space streams that contribute to the same redshift-space element, weighted by their different redshift-space neutral atom densities. 

\subsection{Redshift-space Distortion on 21cm Power Spectrum}
\label{sec:linearRSD}

The 21cm redshift-space-distorted power spectrum in the linear approximation was explored in \citet{Barkana05}, who showed that the linear 21cm power spectrum is distorted in a form analogous to the linear redshift space distortion in galaxy surveys. The authors computed the power spectrum of linearized  peculiar-velocity-corrected 21cm brightness temperature, nevertheless, in {\it real} space, i.e.\ they linearized gas density, neutral fraction, and particularly the velocity gradient correction $1/(1+\delta^r_{\partial_r v}({\bf r})) \approx 1- \delta^r_{\partial_r v}({\bf r})$ by assuming $\delta^r_{\partial_r v} \ll 1$, and computed the Fourier transform of the brightness temperature evaluated in real space. 
The observable power spectrum, however, is in redshift space. Although the expression of power spectrum derived in \citet{Barkana05} can give correct values on large scales, this approach is conceptually incomplete. In addition, the assumption of $\delta^r_{\partial_r v} \ll 1$ may break down on small scales. 
A further complication is that \citet{Barkana05} assume that the product of neutral fraction fluctuation and the gas density fluctuation, $\delta_{x_{\rm HI}} \delta_{\rho_{\rm H}}$, can be neglected, which can be invalid and cause the power spectrum to become inaccurate with a fractional error at the 200\% level on small scales when the universe is 50\% ionized \citep{Lidz07}. 

In this section, we present a reformulation for computing the 21cm power spectrum in observer redshift space, taking into account both distortions in brightness temperature and in apparent location, and give the general equation for the linear redshift-space-distorted power spectrum without assuming either $\delta^r_{\partial_r v}\ll 1$ or $\delta_{x_{\rm HI}} \delta_{\rho_{\rm H}}\ll 1$. 

\subsubsection{Fully nonlinear power spectrum with finite optical depth}

Consider a slice $\delta T_b^s ({\bf s})$ of a 3D data cube near $z_{\rm cos}$, in redshift space. The brightness temperature in Fourier redshift space is $\widetilde{\delta T_b^s} ({\bf k}) \equiv \int d^3s \,\, e^{-i {\bf k \cdot s}} \,\,\delta T_b^s ({\bf s})$. Since predictions of power spectra from theoretical modeling are made in real space, we should relate this to real-space quantities. The redshift- and real-space coordinates are related by equation~(\ref{eqn:red-to-real-coord}), and so the volume elements are related by $d^3 s = d^3 r  \left|1+\delta^r_{\partial_r v}({\bf r})\right|$. The observed brightness temperature is preserved (see eq.~\ref{eqn:bt-preserved}), and, in the general case of finite optical depth, evaluated using equation~(\ref{eqn:opt-thick-formula}) with optical depth using equation~(\ref{eqn:opt-dep-rewrit}). The exact Fourier transform of brightness temperature in redshift-space is,  
\begin{eqnarray}
& & \widetilde{\delta T_b^s} ({\bf k}) = \int d^3r \,\,e^{-i {\bf k \cdot r}}
\cdot \exp{\biggl[-i \left(\frac{1+z_{\rm cos}}{H(z_{\rm cos})}\right)k_\parallel v_\parallel ({\bf r})\biggr]} \nonumber \\
& & \!\!\!\!\!\!\!\!\!\!\!\!\times T_{\rm CMB,0}\left| 1+\delta_{\partial_r v}({\bf r})\right| \,\left[\alpha({\bf r})(1-\frac{v_\parallel}{c})-1\right]\left[ 1- e^{-\tau_{\nu_{\rm obs}}} \right],
\label{eqn:nonlinear_dtb_finite_opt}
\end{eqnarray}
where $k_\parallel = {\bf k}\cdot \hat{r}$. Note that $\tau_{\nu_{\rm obs}}$ is an implicit function of $\bf r$, too. The fully-nonlinear power spectrum can be calculated by its definition $ \left< \widetilde{\delta T_b^{s}}^* ({\bf k}) \widetilde{\delta T_b^{s}} ({\bf k}') \right> \equiv (2\pi)^3 P_{\Delta T}^{s} ({\bf k}) \delta^{(3)}({\bf k} - {\bf k}') $.

\subsubsection{Nonlinear power spectrum in the optically-thin approximation}

In the optically thin limit, we can use the approximation $1- e^{-\tau_{\nu_{\rm obs}}} = \tau_{\nu_{\rm obs}}$. As before, the velocity gradient corrections for the optical depth and the redshift-space volume element cancel in equation~(\ref{eqn:nonlinear_dtb_finite_opt}), and we find that the fully nonlinear Fourier transform of brightness temperature in redshift-space in the optically thin limit is given by
\begin{eqnarray}
& & \widetilde{\delta T_b^s} ({\bf k}) =
\widehat{\delta T}_b (z_{\rm cos}) \,\int d^3r \,\,e^{-i {\bf k \cdot r}} 
\left[1+\delta^r_{\rho_{\rm HI}}({\bf r})\right] \nonumber \\
& & \!\!\!\!\!\!\!\!\times \exp{\biggl[-i \left(\frac{1+z_{\rm cos}}{H(z_{\rm cos})}\right)k_\parallel v_\parallel ({\bf r})\biggr]} 
\biggl[1 -\frac{ T_{\rm CMB}(z_{\rm cos})}{T_s^{r,{\rm eff}}({\bf r})}\biggr]\,.
\label{eqn:bt-red-general}
\end{eqnarray}

\subsubsection{Quasi-linear $\mu_{\bf k}$-decomposition Scheme} 
\label{eqn:lin-muk-decomp}

We work out a ``quasi-linear'' case in this subsection. In this we only take the density and velocity fluctuations to be linear, but the reionization fluctuations are allowed to be nonlinear. This means that we do not assume $\delta_{x_{\rm HI}} \delta_{\rho_{\rm H}}\ll 1$ and thus our approach is more general than that of \cite{Barkana05}. We therefore choose not to call it ``linear theory'', but instead introduce the new name {\it quasi-linear $\mu_{\bf k}$-decomposition scheme}. 

On large scales corresponding to small enough $k$ so that $\left(\frac{1+z_{\rm cos}}{H(z_{\rm cos})}\right)k_\parallel v_\parallel \ll 1$, we can linearize the exponential and keep the linear term in $v$. We also linearize the spin-temperature-dependent term 
\begin{equation}
\eta^r({\bf r}) \equiv \left[1 -\frac{ T_{\rm CMB}(z_{\rm cos})}{T_s^{r,{\rm eff}}({\bf r})}\right]
\end{equation}
by defining its fluctuations as $\delta_\eta^r({\bf r}) = \left[\eta^r({\bf r}) - \bar{\eta}(z_{\rm cos})\right]/\bar{\eta}(z_{\rm cos})$ where $\bar{\eta}(z_{\rm cos})$ is the mean value of $\eta$. We keep only the linear terms in velocity, {\it neutral} density fluctuations, and $\eta$-fluctuations, and find 
$\widetilde{\delta T_b^{s,{\rm qlin}}} ({\bf k}) = 
\widehat{\delta T}_b (z_{\rm cos}) \bar{\eta}(z_{\rm cos}) 
\biggl[ - i \left(\frac{1+z_{\rm cos}}{H(z_{\rm cos})}\right)k_\parallel \widetilde{v^r_\parallel}({\bf k}) 
+ \widetilde{\delta^r_{\rho_{\rm HI}}}({\bf k}) 
+ \widetilde{\delta_\eta^r}({\bf k}) \biggr] $.  
Here $\widetilde{a^r}({\bf k}) \equiv \int d^3r \,\, e^{-i {\bf k \cdot r}} a^r({\bf r})$ is the Fourier transform of the quantity $a^r({\bf r})$ in real space. On large scales, the velocity field is linear, 
$\widetilde{v^r_\parallel}({\bf k}) = i\left(\frac{H(z_{\rm cos})}{1+z_{\rm cos}}\right)\widetilde{\delta^r_{\rho_{\rm H}}}({\bf k}) \frac{\mu_{\bf k}}{k}$, 
where $\mu_{\bf k} = k_\parallel/k$, $k=|{\bf k}|$, and $\widetilde{\delta^r_{\rho_{\rm H}}}({\bf k}) $ is the total hydrogen density fluctuation in Fourier real-space. So we find 
\begin{equation}
\widetilde{\delta T_b^{s,{\rm qlin}}} ({\bf k}) = 
\widehat{\delta T}_b (z_{\rm cos}) \bar{\eta}(z_{\rm cos}) 
\left[ \widetilde{\delta^r_{\rho_{\rm H}}}({\bf k}) \mu_{\bf k}^2 
+ \widetilde{\delta^r_{\rho_{\rm HI}}}({\bf k}) 
+ \widetilde{\delta_\eta^r}({\bf k}) \right]\,.
\label{eqn:lin-scheme-F}
\end{equation}
The power spectrum in the quasi-linear $\mu_{\bf k}$-decomposition scheme in redshift space, defined as $ \left< \widetilde{\delta T_b^{s,{\rm qlin}}}^* ({\bf k}) \widetilde{\delta T_b^{s,{\rm qlin}}} ({\bf k}') \right> \equiv (2\pi)^3 P_{\Delta T}^{s,{\rm qlin}} ({\bf k}) \delta^{(3)}({\bf k} - {\bf k}') $, is 
\begin{equation}
P_{\Delta T}^{s,{\rm qlin}} ({\bf k}) = P_{\mu^0}(k) + P_{\mu^2}(k) \mu_{\bf k}^2 +
P_{\mu^4}(k) \mu_{\bf k}^4 \,,
\label{eqn:lin-scheme}
\end{equation}
where the moments of $\mu_{\bf k}$-polynomial expansion are 
\begin{eqnarray}
P_{\mu^0}(k) &=& \left(\widehat{\delta T}_b \bar{\eta}\right)^2 \left[ P^r_{\delta_{\rho_{\rm HI}},\delta_{\rho_{\rm HI}}}(k) + P^r_{\delta_\eta,\delta_\eta}(k) \right. \nonumber\\
& & \left.+ 2 P^r_{\delta_{\rho_{\rm HI}},\delta_\eta}(k)\right]\,,\\
P_{\mu^2}(k) &=& 2 \,\left(\widehat{\delta T}_b \bar{\eta} \right)^2 \left[ P^r_{\delta_{\rho_{\rm HI}},\delta_{\rho_{\rm H}}}(k) + P^r_{\delta_\eta,\delta_{\rho_{\rm H}}}(k) \right] \,,\\
P_{\mu^4}(k) &=& \left(\widehat{\delta T}_b \bar{\eta}\right)^2 P^r_{\delta_{\rho_{\rm H}},\delta_{\rho_{\rm H}}}(k)\,,\label{eqn:linmu4}
\end{eqnarray}
where all quantities here depend implicitly on the redshift $z_{\rm cos}$. 
Here $P^r_{a,a}$ denotes the auto power spectrum of the quantity $a^r({\bf r})$, and $P^r_{a,b}$ is the cross power spectrum between fields $a^r({\bf r})$ and $b^r({\bf r})$, both in real space. Note that, strictly speaking, the power spectra involving $\delta_\eta$ are not statistically isotropic due to the distortion by peculiar velocity as in equation~(\ref{eqn:dist-spin}). Since the correction is of order $v/c$, we ignore it here. 
When $T_s \gg T_{\rm CMB}$, $\eta = 1$ and $\delta_\eta = 0$, and the power spectrum in quasi-linear $\mu_{\bf k}$-decomposition scheme reduces to equation~(\ref{eqn:Kaiser3D}). 

Although we derived the scheme by assuming linear density and velocity fluctuations, when using it on simulation data, we normally use the non-linear density fluctuations given by the simulation. 

As pointed out above, each moment of the $\mu_{\bf k}$-decomposition can contain higher-order auto- and cross-correlations involving density and ionization fluctuations, 
because $\delta^r_{\rho_{\rm HI}} = \delta^r_{\rho_{\rm H}} + \delta^r_{x_{\rm HI}} + \delta^r_{\rho_{\rm H}} \, \delta^r_{x_{\rm HI}}$. 
To see this explicitly, for example, in the simple case $T_s \gg T_{\rm CMB}$ in which $\eta = 1$ and $\delta_\eta = 0$, we can rewrite the moments as follows. 
\begin{eqnarray}
P_{\mu^0}(k) &=& \widehat{\delta T}_b^2 P^r_{\delta_{\rho_{\rm HI}},\delta_{\rho_{\rm HI}}}(k) \nonumber \\ 
&=& \widehat{\delta T}_b^2 \left[ 
P^r_{\delta_{x_{\rm HI}},\delta_{x_{\rm HI}}}(k) + 
2 P^r_{\delta_{x_{\rm HI}},\delta_{\rho_{\rm H}}}(k) \right. \nonumber \\ 
& & + P^r_{\delta_{\rho_{\rm H}},\delta_{\rho_{\rm H}}}(k) + 2 P^r_{\delta_{x_{\rm HI}}\delta_{\rho_{\rm H}},\delta_{x_{\rm HI}}}(k) 
 \nonumber \\ 
& & 
\left. + 2 P^r_{\delta_{x_{\rm HI}}\delta_{\rho_{\rm H}},\delta_{\rho_{\rm H}}}(k)
+ P^r_{\delta_{x_{\rm HI}}\delta_{\rho_{\rm H}},\delta_{x_{\rm HI}}\delta_{\rho_{\rm H}}}(k) \right] \label{eqn:qlin-mu0-expansion}\\
P_{\mu^2}(k) &=& 2 \,\widehat{\delta T}_b^2 P^r_{\delta_{\rho_{\rm HI}},\delta_{\rho_{\rm H}}}(k) \nonumber \\ 
&=& 2\, \widehat{\delta T}_b^2 \left[ 
P^r_{\delta_{\rho_{\rm H}},\delta_{\rho_{\rm H}}}(k) 
+ P^r_{\delta_{x_{\rm HI}},\delta_{\rho_{\rm H}}}(k) \right. \nonumber \\ 
& & \left. + P^r_{\delta_{x_{\rm HI}}\delta_{\rho_{\rm H}},\delta_{\rho_{\rm H}}}(k) \right] \,,\\
P_{\mu^4}(k) &=& \widehat{\delta T}_b^2 P^r_{\delta_{\rho_{\rm H}},\delta_{\rho_{\rm H}}}(k)\,.\label{eqn:qlin-mu4-expansion}
\end{eqnarray}

However, the quasi-linear $\mu_{\bf k}$-decomposition scheme neglects the nonlinear coupling of peculiar velocity and ionization fluctuations, which we will investigate in future work \citep{Shapiro11}.

\subsubsection{Linear Theory} 

\cite{Barkana05} linearizes both density and ionization fluctuations, and  discards all three- and four-point correlations in the expansion of moments, i.e. in the simple case $T_s \gg T_{\rm CMB}$, equations~(\ref{eqn:qlin-mu0-expansion}-\ref{eqn:qlin-mu4-expansion}) reduce to 
\begin{eqnarray}
P_{\mu^0}(k) &=& \widehat{\delta T}_b^2 \left[ 
P^r_{\delta_{x_{\rm HI}},\delta_{x_{\rm HI}}}(k) + 
2 P^r_{\delta_{x_{\rm HI}},\delta_{\rho_{\rm H}}}(k) \right. \nonumber \\ 
& & \left. + P^r_{\delta_{\rho_{\rm H}},\delta_{\rho_{\rm H}}}(k) \right] \,,
\label{eqn:BL05-mu0}\\
P_{\mu^2}(k) &=& 2\, \widehat{\delta T}_b^2 \left[ 
P^r_{\delta_{\rho_{\rm H}},\delta_{\rho_{\rm H}}}(k) 
+ P^r_{\delta_{x_{\rm HI}},\delta_{\rho_{\rm H}}}(k) \right] \,,\\
P_{\mu^4}(k) &=& \widehat{\delta T}_b^2 P^r_{\delta_{\rho_{\rm H}},\delta_{\rho_{\rm H}}}(k)\,.\label{eqn:BL05-mu4}
\end{eqnarray}

\cite{Lidz07} demonstrated that, if peculiar velocity is not taken into account, i.e. only zeroth moment is concerned, the neglect of higher-order correlations can result in significant errors in 21cm power spectrum. 
They also pointed out that, for the same reason, 21cm redshift-space power spectrum computed using the linear theory of \cite{Barkana05} can have large errors, but they did not provide any detail or analysis of computing the non-linear power spectrum, nor did they propose an analytic solution that incorporates all of the relevant higher order terms. 

In our paper, in addition to investigating the fully nonlinear power spectrum, we propose the quasi-linear $\mu_{\bf k}$-decomposition scheme as a solution that can as well separate the cosmological density fluctuations from the ionization fluctuations just as the linear theory \citep{Barkana05} does, but account for higher order correlations due to nonlinear ionization fluctuations. 

\section{Computational Schemes to Predict Brightness Temperature in Redshift Space}
\label{sec:schemes}

\subsection{Exact Steps in the Case of Finite Optical Depth}

Analytical models and semi-numerical or numerical simulations provide us with {\it real}-space data. In order to make predictions for the observed 21cm power spectrum, we need to calculate the fully nonlinear 21cm brightness temperature accurately and efficiently in {\it redshift} space, accounting for all effects of peculiar velocities.

As explained in \S~\ref{sec:distinguishing}, the effects of peculiar velocity can be separated into an effect on the observed brightness temperature and one on the apparent location of the 21cm emission source. So in principle, in order to compute the signal in redshift space, the brightness temperature should (1) first be corrected by the velocity gradient, evaluated in real space, using the exact formula of 21cm brightness temperature (eq.~\ref{eqn:opt-thick-formula}) with finite optical depth (eq.~\ref{eqn:opt-dep-rewrit}), and (2) then shifted to the apparent location corresponding to the Doppler frequency shift, with the volume element re-sized according to the velocity gradient, and (3) finally resampled onto a regular grid in redshift space. Power spectra calculated this way should be equivalent to those using equation~(\ref{eqn:nonlinear_dtb_finite_opt}). This process is in general computationally cumbersome.

\subsection{\textit{Real-to-Redshift-Space-Mapping (RRM)} Schemes}

Since the optically-thick cells are very rare in the IGM, as we have shown in \S~\ref{sec:opt-thin-revis}, we may evaluate brightness temperature in the optically-thin approximation (eq.~\ref{eqn:dtb1}). In doing this, although brightness temperature in an optically-thick cell would become artificially divergent in real-space, its net contribution to the brightness temperature in redshift-space is still finite and proportional to the total number of neutral atoms in that cell, because the redshift-space volume element of this cell is compressed accordingly. This has been well discussed in \S~\ref{sec:doppler}. We can exploit the proportionality between the 21cm brightness temperature and the neutral atom number density both measured in redshift space. Inspired by common wisdom in large-scale structure simulations, we propose two computational schemes based on mapping the neutral atom density from real- to redshift-space, and then computing the 21cm brightness temperature in redshift space using equation~(\ref{eqn:Tbredshiftspace}).
We also assume $T_s \gg T_{\rm CMB}$ in this section, but our schemes can be readily generalized to the arbitrary $T_s$ case.  

Strictly speaking, these two schemes are accurate only when the optically-thick cells are rare enough, because neutral atoms in those cells should be ``self-shielded'' to 21cm radiation. We will revisit in detail the accuracy of power spectrum in the optically-thin approximation in \S~\ref{sec:finite-opt-comparison}. 

\begin{figure*}
\begin{center}
  \includegraphics[height=0.6\textheight]{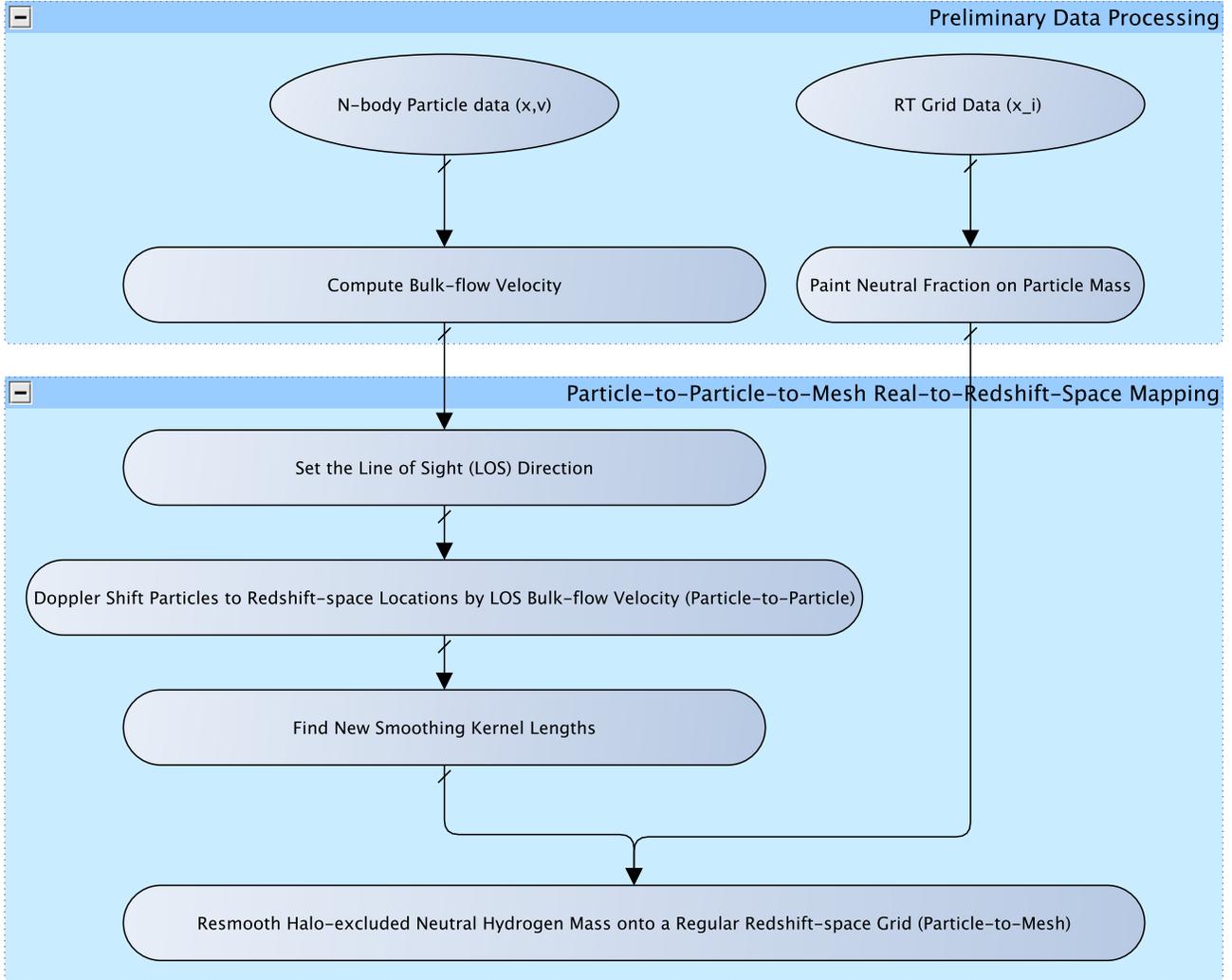} 
\end{center}
\caption{Flowchart of the PPM-RRM scheme. 
}
\label{fig:flowchart-PPM-RRM}
\end{figure*}

\begin{figure*}
\begin{center}
  \includegraphics[height=0.6\textheight]{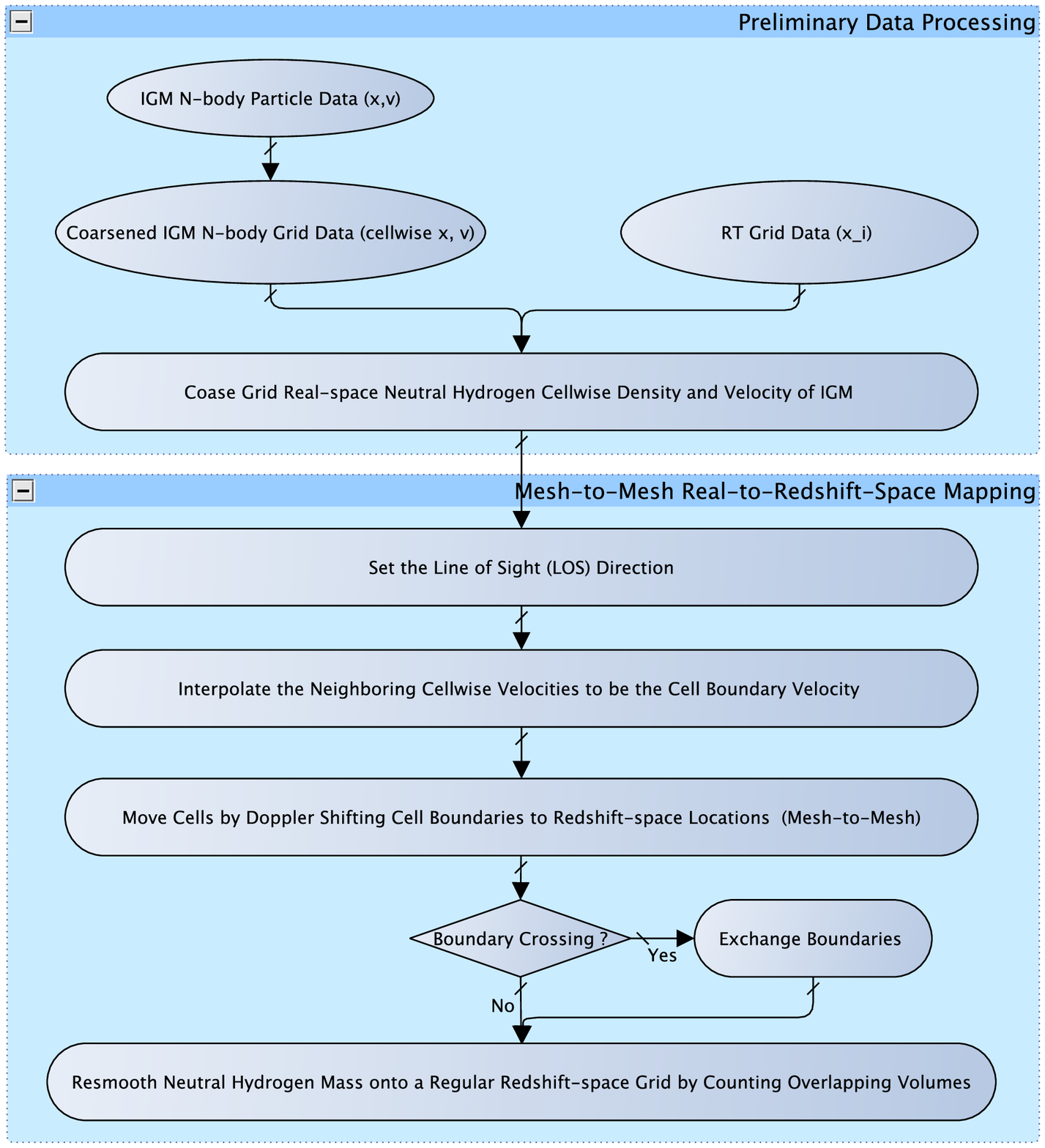} 
\end{center}
\caption{Flowchart of the MM-RRM scheme. 
}
\label{fig:flowchart-MM-RRM}
\end{figure*}

\subsubsection{Particle-to-Particle-to-Mesh(PPM)-RRM scheme}
\label{sec:PPM-RRM}

Most numerical simulations of reionization are processed as follows. First one runs a large-scale N-body simulation, from which one obtains gridded density fields and the collapsed halo information such as location and mass. The reionization simulation is then run on these gridded density fields using the halos as sources of radiation. Since the RT grid resolution is typically coarser than the N-body particle resolution, the most accurate 3D map of the neutral atom distribution in redshift space that can be possibly achieved from a given reionization simulation is made by taking advantage of the high-resolution N-body particle information. We propose the {\it Particle-to-Particle-to-Mesh Real-to-Redshift-Space-Mapping} (``PPM-RRM'') scheme as follows: 
\begin{itemize}
\item We compute the bulk-flow velocity of the IGM at the position of particles directly from N-body particle data using an adaptive-kernel, SPH-like approach. The SPH-smoothed bulk velocity assigned to each particle is the smoothed momentum density divided by the smoothed mass density, evaluated at the particle location. \footnote{If a hydrodynamical simulation is coupled to N-body cold dark matter (CDM) simulation, then the gas particle velocity can be directly used. But since our simulations are dark matter only, we approximate the gas bulk-flow velocity as the SPH-smoothed velocity at the particle location (see Appendix~\ref{app:sph}). One cannot use the particle velocities directly because those can be multi-streaming. In all this we assume that the gas traces the dark matter exactly, which is a good approximation on large scales.}

\item We assign each particle the neutral fraction from the RT grid cell that it is located in. 

\item For a given LOS direction, we Doppler-shift the N-body particles to their apparent locations according to the LOS bulk-flow velocity, 
in accordance to equation~(\ref{eqn:red-to-real-coord}). 

\item We compute new smoothing kernel lengths using the new particle positions in redshift-space. 

\item We use those kernel lengths to smooth the particle data (i.e.\ H~I   mass) onto a regular, redshift-space grid (see the discussion of grid resolution below). In this step, we exclude particles contained in halos\footnote{
We simulate the reionization of the IGM, and therefore compute the 21cm brightness temperature only from the IGM, so excluding particles in halos. }.

\item From this latter, gridded density field, we compute the H~I density fluctuations in redshift-space, and from this the 21cm brightness temperature measured in redshift space using equation~(\ref{eqn:Tbredshiftspace}). 

\end{itemize} 
Some details of the particle smoothing algorithm are discussed in Appendix~\ref{app:sph}. We use adaptive kernels rather than fixed-kernels so as to better resolve the small scale spatial variations in overdense regions.

The high wavenumber modes in the power spectrum can be inaccurate due to sampling effects when calculating the power spectrum using the fast Fourier transform (FFT). Instead of correcting the power spectrum using the method proposed by \cite{Jing05}, we partly avoid the sampling effect by gridding the particle data onto a redshift-space grid at four times higher resolution than the RT grid, but only keeping the modes in the power spectrum with $k\le \pi/ \Delta L$ ($\Delta L$ is the RT grid spacing), i.e.\ one-quarter of the Nyquist wavenumber for a grid with the resolution $\Delta L/4$. The reason for this and a summary of the sampling effect are discussed in more detail in \S~\ref{sec:aliasing}. 

The PPM-RRM prescription can be summarized as ``P$_{\rm r}\to$ P$_{\rm s}\to$ M$_{\rm s}$(4$\times$RT)'' where ``P'' means particle data, ``M'' means mesh data, subscript ``r'' means real-space, ``s'' means redshift-space, and ``4$\times$RT'' indicates that the grid resolution is 4 times finer than RT grid resolution. Figure~\ref{fig:flowchart-PPM-RRM} shows the flow chart for the PPM-RRM scheme.

\subsubsection{Mesh-to-Mesh(MM)-RRM scheme}
\label{sec:MM-RRM}

Manipulating N-body particle data is accurate but computationally costly (see Table~\ref{tab:accu-eff} below). Since the N-body particle data typically already have been smoothed onto a regular, real-space grid in order to simulate the
radiative transfer, we propose an alternative scheme, the {\it Mesh-to-Mesh Real-to-Redshift-Space-Mapping} (``MM-RRM'') scheme. \citet{Mellema06b} were actually the first to use the MM-RRM scheme to produce brightness temperature spectra and maps along the LOS (their Figs.~4, 9 and 10), but did not provide a detailed description of the method in their paper. 
This scheme saves computational resource by using the real-space grid data such as cell-wise mass density, velocity, and ionization fraction, but gives consistent results (depending on the grid resolution, to be tested in \S~\ref{sec:MMRRM-test}).  
The MM-RRM scheme works as follows:
\begin{itemize}

\item As the preliminary step, we grid the N-body particle data in the IGM (i.e.\ particles in the halo excluded) onto a regular, real-space grid with a resolution $n$ times finer than the RT resolution, using our adaptive kernel SPH-like smoothing. This provides us with cell-wise density and velocity fields.

\item We assign each cell the neutral fraction from the RT grid that this fine cell belongs to. 

\item We assume the cell-wise velocity to be the velocity at the cell center, and compute the LOS velocity at the boundary between two LOS-neighboring cells by linear interpolation.

\item We shift the cell boundaries to their apparent locations according to their LOS velocity, in accordance with equation~(\ref{eqn:red-to-real-coord}), whereby the real-space cell can get stretched or compressed in redshift space. In high density cells the boundaries of a cell can cross each other in redshift space, an effect known as the {\it finger of God}. When this happens, we switch the cell's crossing boundaries so that the cell size is always positive. 

\item We regrid the neutral hydrogen mass from the real-space grid onto a regular, redshift-space grid at the same resolution, by counting the overlapping volumes; e.g., if the LOS is along the $x$-axis, a real-space cell $(i,j,k)$ with the size $\Delta x$ stretches to the length $\Delta x'_i$ in redshift-space, with a portion of this length, $\Delta L_{i,i'}$, overlapping the cell $(i',j,k)$ in the regridded, redshift-space, mesh, then all real-space cells $(i,j,k)$ contribute to the neutral hydrogen density of the redshift-space cell $(i',j,k)$, according to 
\begin{equation}
\rho_{\rm HI}^s (i',j,k) = \sum_i F_{i,i'}\, \rho_{\rm HI}^r (i,j,k)\,,
\end{equation}
where $F_{i,i'}$ is the fractional overlap of the real-space volume $i$ with the redshift-space volume $i'$, i.e. $F_{i,i'}=\Delta L_{i,i'}/ \Delta x'_i$ (the indices $j$ and $k$ are not relevant here because we move all cells along the $x$-axis). 

\item We compute the HI density fluctuations in redshift-space, and from this the 21cm brightness temperature using equation~(\ref{eqn:Tbredshiftspace}). This is done at at $n$ times higher resolution than the RT grid, but when calculating the power spectrum we only keep modes with $k\le \pi/ \Delta L$ ($\Delta L$ is the RT grid spacing). 

\end{itemize}
The MM-RRM scheme can be summarized as ``[P$_{\rm r}\to$ M$_{\rm r}$($n\times$RT)]$\to$ M$_{\rm s}$($n\times$RT)'', where the operation inside the square bracket is the prerequisite step. In \S~\ref{sec:MMRRM-test} we will experiment with different resolution factors $n$ to find the optimal resolution. Figure~\ref{fig:flowchart-MM-RRM} shows the flow chart for the MM-RRM scheme.

\subsubsection{The Redshift-space-distorted Lightcone Effect}

Both the PPM-RRM and MM-RRM schemes deal with simulation data from a finite volume at a fixed cosmic time, implicitly assuming that the cosmic evolution of both neutral fractions and density perturbations are negligible during the light travel time across the simulation box, $t_\mathrm{cross}$.
For the typical simulation volume sizes (100-200 Mpc) one does not expect much evolution in the density field during  $t_\mathrm{cross}$. However, the
neutral fractions may evolve much more rapidly during some periods of the EoR. If $(d\ln x_i/dt) \delta t \gtrsim 1$, then we must take into account this so-called lightcone effect \citep{Barkana06} and couple it to peculiar velocity. This implies first time-interpolating the particle data to the appropriate look back time and the corresponding real-space location and then shifting the particles to their apparent location according to its interpolated LOS peculiar velocity, and finally mapping these time interpolated particles onto a regular redshift-space grid on the lightcone. The full version of the lightcone PPM-RRM scheme is beyond the scope of this paper\footnote{\citet{Mellema06b} {\it did} apply such a time interpolation of {\it grid} data, both on the neutral fraction and density fields.} and we postpone an investigation of this effect to a future paper in this series.

\subsection{Simulations}
\label{sec:sim}

For our reionization simulation we use a new large-scale, high-resolution N-body simulation of the $\Lambda$CDM universe (performed with the CubeP$^3$M code, \citealt{Iliev08b}) in a comoving volume of $L_{\rm box} = 114$\,Mpc/$h$ on each side using $3072^3$ (29 billion) particles. To find the halos, we use the spherical overdensity method and require them to consist of at least 20 N-body particles; this implies a minimum halo mass of $10^8\,M_\odot$. 

Assuming that the gas traces the CDM particles exactly, we grid the density on a $256^3$ grid using SPH-like smoothing with an adaptive kernel. The halo lists and density fields are then processed with the radiative transfer code C$^2$Ray \citep{Mellema06a}. Each halo releases $f_\gamma$ ionizing photons per baryon per $\Delta t = 11.5$\,Myrs, with $f_\gamma=150$ ($f_\gamma=10$) for halos below $10^9\,M_\odot$ (above $10^9\,M_\odot$), respectively. To incorporate feedback from reionization, halos less massive than $10^9\,M_\odot$ located in ionized regions are not producing any photons.

The simulations were run on the University of Texas Sun Constellation
Linux Cluster {\it Ranger}, one of the largest computational resources
in the world. Both codes are massively parallel, using 512 compute
nodes, each with one Quad-Core 64-bit processor. We refer the readers to
\citet{Friedrich11} and \citet{Iliev11} for more details of this simulation which in those papers is labelled as ``163Mpc\_g8.7\_130S''.

The simulations used the following set of cosmological parameters $\Omega_\Lambda=0.73, \Omega_{\rm M}=0.27, \Omega_{\rm b}=0.044, h=0.7, \sigma_8=0.8, n_\mathrm{s}=0.96$ where $H_0 = 100h$\,km\,s$^{-1}$\,Mpc$^{-1}$, consistent with the {\it WMAP} seven-year results \citep{Komatsu11}.

\begin{figure}
\begin{center}
  \includegraphics[height=0.35\textheight]{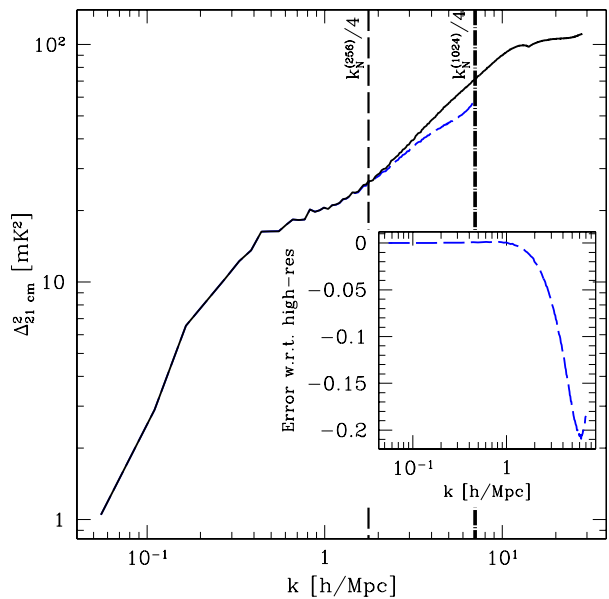} 
\end{center}
\caption{Aliasing effect in the PPM-RRM scheme: 
21cm redshift-space 1D power spectrum at $z=9.457$ (50\% ionized), when the particle data is smoothed onto a regular, redshift-space, grid with the RT grid resolution ($256^3$, long-dashed, blue), or four times finer ($1024^3$, solid, black). The vertical lines are at $ k = k_N^{(256)}/4 = 1.75$\,h/Mpc (thin long-dashed) and $k=k_N^{(1024)}/4 =k_N^{(256)} = 7$\,h/Mpc (thick dot-long-dashed), respectively. The fractional error plotted in the inset is with respect to the power from the $1024^3$ grid.
}
\label{fig:PPM-aliasing}
\end{figure}

\subsection{Sampling Effects}
\label{sec:aliasing}

Measuring power spectra using a fast Fourier transform (FFT) of gridded data suffers from the so-called {\it sampling effect}. This effect is due to the mass assignment of particle data or continuous fields to a chosen grid. In cosmology, it was first extensively discussed for power spectrum measurements of density fields in large scale structure (see, e.g., \citealt{Jing05}, \citealt{Cui08}, and references therein). The mass assignment is equivalent to convolving the true density field with a window function and sampling this convolved density field with a finite number of grid points. The power spectrum of the convolved field is a {\it biased} one, i.e., \citep{Jing05} 
\begin{equation}
P^{f}({\bf k}) = \sum_{\bf n} \left| \tilde{W}({\bf k}+2k_N {\bf n})\right|^2 P({\bf k}+2k_N {\bf n}) + P_{\rm shot}\,,
\label{eqn:sampling}
\end{equation}
where $P^{f}({\bf k})$ and $P({\bf k})$ are power spectra of the convolved and true field, respectively, $\tilde{W}({\bf k})$ is the Fourier transform of the window function, $P_{\rm shot}$ is the shot noise, and the summation is over all three-dimensional integer vectors $\bf n$. 
The sampling effects include three aspects that can affect the true power spectrum measurement \citep{Cui08}.
\begin{itemize}

\item Smoothing effect: the Fourier window function $|\tilde{W}({\bf k})|^2$ falls off sharply from $|\tilde{W}({\bf 0})|^2 = 1$, e.g., for a Cloud-In-Cell (CIC) window function, $|\tilde{W}|^2 = 0.90$ at $k=k_N/4$, but $|\tilde{W}|^2 = 0.66$ at $k=k_N/2$, 
where $k_N = \pi/a$ is the Nyquist wavenumber for some grid spacing $a$. 

\item Anisotropy effect: the Fourier window function is not isotropic for a given $k$, and the anisotropy is significant for $k\sim k_N$.

\item Aliasing effect: higher wavenumber modes (${\bf n}\ne 0$) contaminate the true mode at ${\bf k}$, preventing us from relating $P^f({\bf k})$ and $P({\bf k})$ straightforwardly. For a FFT, $(-k_N, k_N)$ is the range in ${\bf k}$-space that a finite resolution grid can probe. Thus those high-wavenumber modes that contaminate are due to modes of the unresolved field below the grid resolution. 

\end{itemize}

The smoothing effect and anisotropy effect can easily be corrected for, e.g.\ by just deconvolving $P^f({\bf k})$ with the normalization $|\tilde{W}({\bf k})|^2$. Correcting the aliasing effect is more difficult, and may be done using the iterative method proposed and tested for the density power spectrum by \citet{Jing05}. Instead, we can be less ambitious and define a ``comfort'' zone ($k\le$ some critical value) where the FFT power spectrum has negligible errors. This can be done because at low enough $k$, all these sampling effects should be insignificant. The test problems in both \cite{Jing05} and \cite{Cui08} seem to agree that the raw density power spectra for different window functions agree at $k\lesssim k_N/4$. Here we test this on the 21cm power spectrum. 
In Figure~\ref{fig:PPM-aliasing} we compare two power spectra, both calculated with the PPM-RRM scheme but differing in the resolution chosen for gridding the redshift-space particle data, $256^3$ and $1024^3$, respectively. As can be seen in the figure, both power spectra agree for $k\lesssim k_N^{(256)}/4$, where $k_N^{(256)}$ is the Nyquist wavenumber of the $256^3$ grid. We therefore conclude that if we use the $256^3$ grid, we can trust the results for $k \le k_N^{(256)}/4$.

However, this comparison also shows that we can use our high-resolution N-body data to try to capture the modes between $k=k_N^{(256)}/4$ and $k=k_N^{(256)}$. By sampling the Doppler-redshifted particle data onto a grid with a resolution of 4$\times$RT = $1024^3$ we can minimize the smoothing and anisotropy effects. We also minimize the aliasing effect due to the gridded density and velocity data.
The aliasing effect due to the finite resolution of the ionization fraction field obviously cannot be corrected for this way. However, this effect may be quite small due to the nature of the ionization fraction field. Recall that the aliasing effect is due to the contamination from high-wavenumber modes unresolved by the grid resolution. 
For blackbody type sources, 
the edges of ionized regions are sharp, i.e.\ the ionization fraction is very close to 1 inside and very nearly 0 outside ionized regions. Therefore only cells that contain boundaries of ionized regions have unresolved subcell information. The fraction of boundary cells for an ionized region of $N$ cells in each dimension is $\sim N^2/N^3 = 1/N$. The peak of the H~II bubble size distribution can be $\sim 10$ Mpc, corresponding to $\sim 22$ RT cells across a bubble (see, e.g.\ \citealt{Friedrich11}). For such bubbles only $\sim 4\%$ of the cells contribute to the aliasing effect. Only if there are many small bubbles of size less than an RT cell, would the ionization field introduce a substantial aliasing effect.

Given this argument it would seem prudent to choose the smoothed grid resolution to be four times smaller than the RT resolution, as this minimizes the sampling effects for the 21cm power spectra. We therefore adopt this approach. The modes between $k_N/4$ and $k_N$ (where $k_N$ here corresponds to RT grid resolution) may still be affected by the aliasing effect due to the finite RT grid resolution, but we expect this to be a minor effect.

\subsection{Tests of PPM-RRM Scheme}
\label{sec:testofPPM}

Since the PPM-RRM (4$\times$RT) scheme retains the particle data the longest by mapping them directly into redshift space, it can be expected to be more accurate than the MM-RRM scheme. We therefore first present tests for the PPM-RRM scheme in this section and in the next section compare the results of the two schemes.

\subsubsection{Conservation of Mass}

The mean total (and neutral) hydrogen density is conserved between real- and redshift-space, because (i) the total (and neutral) hydrogen atom number is conserved, and (ii) the the total space is conserved for a volume large enough (peculiar velocity vanishes for large distances) or a periodic box, since $\int \delta V^r \delta^r_{\partial_r v}({\bf r})$ is a total derivative.

For the simulation box, the total volume is automatically conserved. We can therefore check whether our schemes conserve mass by checking the conservation of mean hydrogen density and HI density. Conservation of the mean density could be violated if a scheme would undercount particles after shifting particles to redshift-space. 
\footnote{To parallel-process N-body particle data using Message Passing Interface (MPI) software, the simulation volume is divided into cubic partitions and particles are assigned to the partition within which they are located. Each partition is processed independently by a given node in the parallel computer. The mapping described here of particle locations from real- to redshift-space can move a particle out of its original (real-space) partition into another, even to one which is not a neighbor partition. In that case, the number of partition pairs that must share particle data, to exchange particles, can be large and, hence, computationally inefficient. Fortunately, we find that the size of each partition in our N-body simulations (which we also use for our real- to redshift-space mapping) is larger than the maximum Doppler shift of particles in comoving coordinates, so only neighboring partitions need exchange particle data.} 

Our code passes this test by showing that the fractional differences of the mean total (and neutral) hydrogen density between real-space and redshift-spaces with three distinct LOS directions are zero, i.e.\ smaller than machine error. 

\begin{figure*}
\begin{center}
  \includegraphics[height=0.60\textheight]{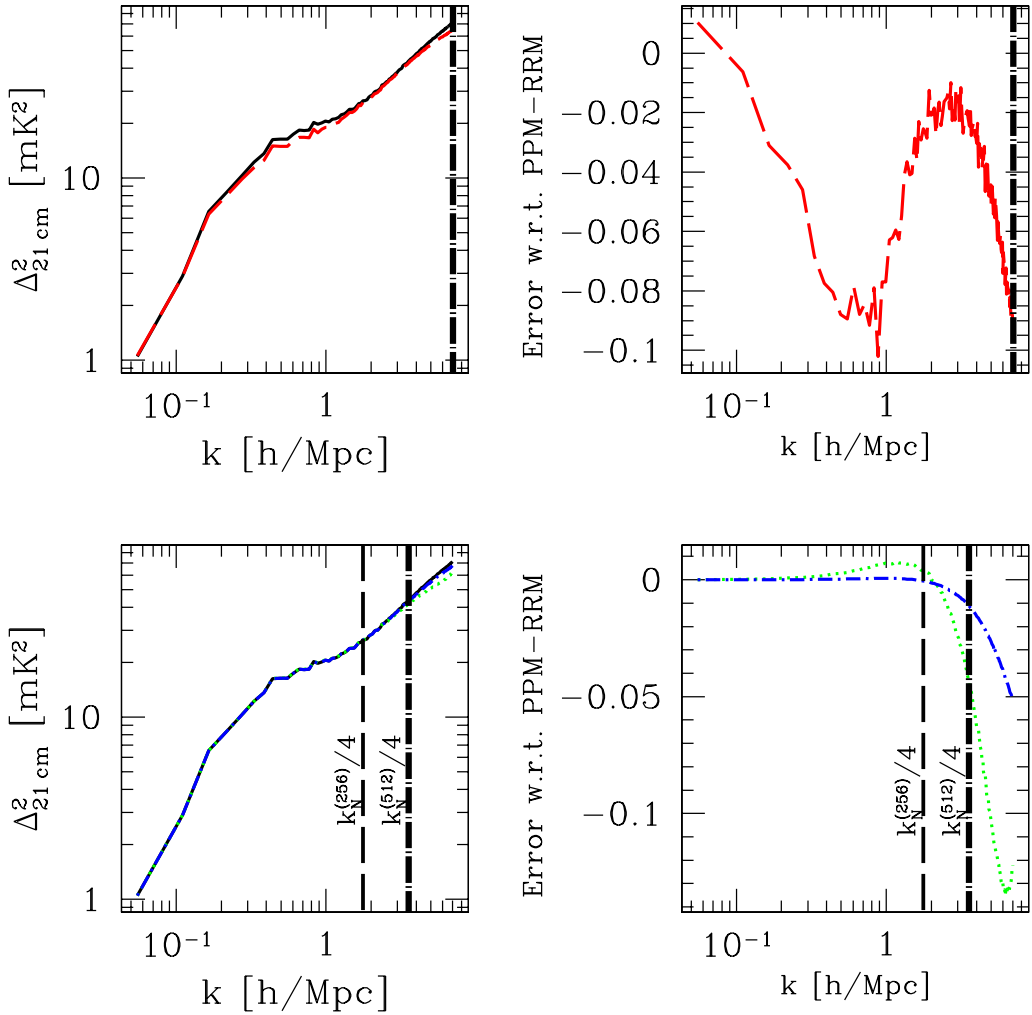} 
\end{center}
\caption{Tests of the PPM-RRM scheme for the 21cm redshift-space 1D power spectrum at $z=9.457$ (50\% ionized). 
Top panels: large scale test of the PPM-RRM scheme (solid, black) against the quasi-linear $\mu_{\bf k}$-decomposition scheme (long-dashed, red), both computed from a $1024^3$ grid, and Fourier modes kept only at $k \le k_N^{(1024)}/4 = k_N^{(256)}= 7$\,h/Mpc (the thick vertical lines). The fractional error is that of the quasi-linear $\mu_{\bf k}$-decomposition scheme result with respect to the PPM-RRM result. \newline
Bottom panels: small scale test of the PPM-RRM scheme computed from $1024^3$ grid (solid, black) against the {\it DEMRF} scheme, computed on a $256^3$ grid (dotted, green) and on a $512^3$ grid (dot-short-dashed, blue), respectively. We only keep modes $k\le k_N^{(256)} = 7$\,h/Mpc. The black and blue curves are almost indistinguishable until at the very large $k$ in the left bottom panel. The fractional errors are with respect to the PPM-RRM result. 
The vertical lines at $ k = k_N^{(256)}/4 = 1.75$\,h/Mpc (thin long-dashed) and $ k = k_N^{(512)}/4 = 3.5$\,h/Mpc (thick dot-long-dashed) delimit the comfort zone for the {\it DEMRF} result computed on a $256^3$ and $512^3$ grid, respectively. 
}
\label{fig:PPM-tests}
\end{figure*}

\subsubsection{Large Scale Test}
\label{sec:large-scale-test}

As shown in \S~\ref{sec:linearRSD}, the fully nonlinear power spectrum reduces to the quasi-linear $\mu_{\bf k}$-decomposition power spectrum at large scales. We use
this here to test the PPM-RRM scheme. Figure~\ref{fig:PPM-tests} (top panels) shows the 1D dimensionless \footnote{It still has the unit mK$^2$. It is dimensionless with regard to Fourier space units.} power spectrum $\Delta^2_{\rm 21cm}(k) = k^3 P_{\Delta T}^s(k)/2\pi^2$ calculated with the PPM-RRM scheme. In order to minimize noise, we averaged the power spectra for three distinct LOS directions (namely, along $x$-, $y$- and $z$-axes). Plotted in the same figure is the 1D quasi-linear $\mu_{\bf k}$-decomposition power spectrum calculated directly from the real-space ionization fraction (on the $256^3$ grid) and density and velocity data (on the $1024^3$ grid), using equations~(\ref{eqn:lin-scheme}) - (\ref{eqn:linmu4}). We choose the 50\% ionized epoch for this comparison. Note that even though we use the quasi-linear $\mu_{\bf k}$-decomposition scheme equations to evaluate the power spectrum, we use the fully nonlinear density and ionization fraction fields from the simulation. 

The comparison shows that the nonlinear power spectrum computed from the PPM-RRM scheme agrees with the quasi-linear $\mu_{\bf k}$-decomposition power spectrum at large scales ($k\lesssim 0.3\, h$/Mpc) within 5\%. This confirms that the 21cm brightness temperature data cube constructed by the PPM-RRM scheme captures the correct large-scale fluctuations in redshift space as dictated by the quasi-linear $\mu_{\bf k}$-decomposition scheme. The nonlinear power spectrum deviates from the quasi-linear $\mu_{\bf k}$-decomposition power spectrum at intermediate scales ($0.3 \lesssim k \lesssim 2\, h$/Mpc) at the level of $\sim 10\%$, and even larger deviations can be found at smaller scales. In the second paper of this series \citep{Shapiro11}, we will investigate in detail the cause of this departure from linearity and how it affects the use of 21cm observations for cosmology.

\subsubsection{Test Down to Small Scales}
\label{sec:DEMRF}

Similar to the quasi-linear $\mu_{\bf k}$-decomposition scheme test that employs the real-space grid data to compute the redshift-space statistics, we can compute the redshift-space power spectrum at all scales, in principle, by evaluating the integral in equation~(\ref{eqn:bt-red-general}). The integration can be carried out by a fast Fourier transform of the data cube $F({\bf r}) = \exp{\left[-i \left(\frac{1+z_{\rm cos}}{H(z_{\rm cos})}\right)k_\parallel v_\parallel ({\bf r})\right]}\cdot 
\left[1+\delta^r_{\rho_{\rm HI}}({\bf r})\right] $ (assuming $T_s \gg T_{\rm CMB}$)  for any given $k_\parallel$, and then picking up only those modes with the LOS component $k_\parallel$, i.e., $ \widetilde{\delta T_b^s} ({\bf k}) = \widehat{\delta T}_b (z_{\rm cos})\,\widetilde{F}({\bf k}) $ only if ${\bf k}\cdot \hat{n} = k_\parallel$. We can construct the whole Fourier data cube by making such FFT evaluation for each ${\bf k}$-space plane of constant $k_\parallel \ge 0$, exploiting the symmetry $\widetilde{\delta T_b^s} (-{\bf k}) = \widetilde{\delta T_b^s}^* ({\bf k}) $, with $k_\parallel$ discretized in units of $2\pi/L_{\rm box}$. 

Note that in order for the discrete Fourier transform to be a good approximation to the continuous Fourier transform, the particle data should in principle be  smoothed to compute the cell-wise average $\left< F({\bf r}) \right>_{\rm cell}$ and in particular $\left< \exp{\left[-i \left(\frac{1+z_{\rm cos}}{H(z_{\rm cos})}\right)k_\parallel v_\parallel ({\bf r})\right]}\right>_{\rm cell}$ directly. However, to take advantage of existing cell-wise density and velocity data on the grid, we evaluate $\left< \exp{\left[-i \left(\frac{1+z_{\rm cos}}{H(z_{\rm cos})}\right)k_\parallel v_\parallel ({\bf r})\right]} \right>_{\rm cell} \to \exp{\left[-i \left(\frac{1+z_{\rm cos}}{H(z_{\rm cos})}\right)k_\parallel \left< v_\parallel ({\bf r}) \right>_{\rm cell} \right]}$. We compute the 1D power spectrum from the Fourier modes, averaged over three independent LOS directions. We name this method of evaluating the power spectrum the {\it Direct Evaluation by Multiple Real-space FFTs} (DEMRF) scheme. 

Obviously, the DEMRF scheme is accurate only when the cell size is not too small so that $\left< v_\parallel^n ({\bf r}) \right>_{\rm cell} \approx \left< v_\parallel ({\bf r}) \right>_{\rm cell}^n $ for any $n > 1$ in the Taylor expansion of $\left< \exp{\left[-i \left(\frac{1+z_{\rm cos}}{H(z_{\rm cos})}\right)k_\parallel v_\parallel ({\bf r})\right]}\right>_{\rm cell}$. On the other hand, if the grid is too coarse, the high-$k$ powers are subject to the sampling effect and become inaccurate. We experiment on the trade-off by trying out the DEMRF scheme on grid data with different resolutions ($256^3$, $512^3$ and $1024^3$), and find that for a box of size 114 Mpc/$h$, the $1024^3$ grid is too fine and fails to make sensible results due to the subcell nonlinearity. We plot the DEMRF result computed from $256^3$ and $512^3$ grids in Figure~\ref{fig:PPM-tests} (bottom panels), and find that within the comfort zone for each grid (1.75 Mpc/$h$ and 3.5 Mpc/$h$, respectively), the PPM-RRM result agrees with the DEMRF results within 1\%. 

The three tests presented thus show that the PPM-RRM (4$\times$RT) scheme is accurate on both large and small scales. We can now use this to test our other scheme.
\begin{figure}
\begin{center}
  \includegraphics[height=0.35\textheight]{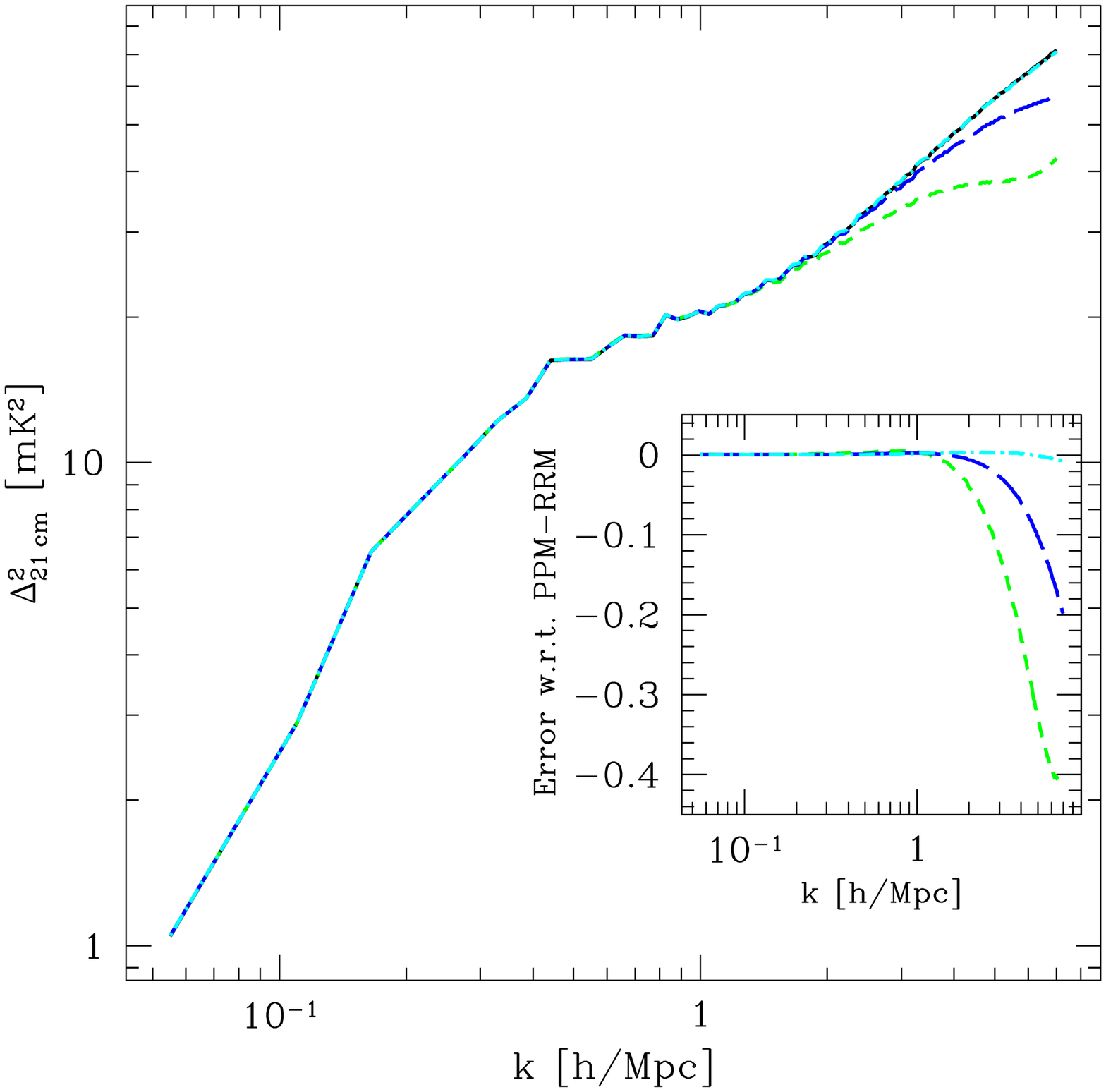} 
\end{center}
\caption{Test of the MM-RRM scheme: 
21cm redshift-space 1D power spectrum at $z=9.457$ (50\% ionized). We experiment with grids of the RT grid resolution (short-dashed, green), 2 times (long-dashed, blue), and 4 times higher resolution (dot-short-dashed, cyan). The benchmark is the result from the PPM-RRM (4$\times$RT) scheme (solid, black). The power spectra of the MM-RRM (4$\times$RT) scheme and the PPM-RRM (4$\times$RT) scheme are indistinguishable on all scales shown. The fractional errors of the MM-RRM results with respect to the PPM-RRM result are shown in the inset. 
}
\label{fig:PPMvsMM}
\end{figure}

\subsection{Test of MM-RRM Scheme}
\label{sec:MMRRM-test}

The MM-RRM scheme is expected to be less accurate than the PPM-RRM scheme since it grids the particle data before moving to redshift space and inevitably small scale information is lost in the process. For example, the gas density within an RT cell is assumed to be uniform, so that the resized cell in redshift-space can be uniformly regridded by counting overlapping volumes. This assumption obviously ignores the subcell clumpiness. Second, the scheme treats the velocity of cell boundary as the linear interpolation between cell-wise velocities of two neighboring cells and thus ignores small-scale velocity fluctuations at the inter-cell scale. Third, the treatment of cell boundary crossing is approximate and a careful treatment should require particle data to mimic the {\it finger of God} effect. 
However, the scales that are affected depend on the resolution chosen, and if one can choose a grid with fine enough resolution, there is a hope that the MM-RRM scheme can yield as accurate power spectra at $k\le k_N$ (corresponding to RT grid resolution) as the PPM-RRM scheme does. 

We experiment with the resolution of the MM-RRM scheme by choosing $n=1,$ 2 or 4 in the pipeline ``[P$_{\rm r}\to$ M$_{\rm r}$($n\times$RT)]$\to$ M$_{\rm s}$($n\times$RT)'' (where $n\times$RT means the grid resolution $n$ times finer than RT grid resolution). We compute the 21cm power spectrum for each of these three resolutions and plot them for the modes $k\le k_N^{(256)}$ in Figure~\ref{fig:PPMvsMM}. As above we average over three LOS directions. We use the PPM-RRM(4$\times$RT) result as a benchmark. All MM-RRM results agree with the PPM-RRM result down to the scale $k\lesssim 1\,h$/Mpc, while at high $k$ the MM-RRM(1$\times$RT) and (2$\times$RT) results deviate from the benchmark by up to 40\% and 20\%, respectively. Fortunately, the MM-RRM(4$\times$RT) result agrees with the benchmark within 1\% error on all scales down to $k \le k_N^{(256)}$. We therefore
conclude that MM-RRM(4$\times$RT) gives as accurate results as the PPM-RRM(4$\times$RT) scheme for $k\le k_N$. 

\begin{table*}
\begin{center}
\begin{minipage}{1\linewidth}
\caption{Usability, accuracy and efficiency of various computational schemes for the redshift-space brightness temperature. Our simulation is in a box with 114\,Mpc/$h$ on each side, has $3072^3$ N-body particles, and evolves reionization on a $256^3$ RT grid. }
\label{tab:accu-eff}
\begin{tabular}{@{}p{1.1cm}p{1.4cm}p{1.4cm}p{1.4cm}p{1cm}p{1cm}p{1cm}p{1.4cm}p{1.4cm}p{2.6cm}}
 \hline\hline
  &  &  \multicolumn{2}{c}{PPM-RRM}  &  \multicolumn{3}{c}{MM-RRM} &  \multicolumn{2}{c}{{\it DEMRF}} &  Quasi-linear \mbox{$\mu_{\bf k}$-decomposition}   \\
 \cline{3-10}
  &  & 1$\times$RT & 4$\times$RT & 1$\times$RT & 2$\times$RT & 4$\times$RT & 1$\times$RT & 2$\times$RT & 4$\times$RT \\
 \hline
\multicolumn{2}{c}{Input data type}  & 
    \multicolumn{2}{p{3cm}}{N-body particle $({\bf x},\,{\bf v})$ ($3072^3$ particles), and RT grid $x_i$ ($256^3$ grid size)}  & 
    \multicolumn{3}{p{3.5cm}}{Cell-wise $({\bf x},\,{\bf v})$ in $256^3$, $512^3$, and $1024^3$ grid size ($1\times$, $2\times$ and $4\times$RT, respectively), and RT grid $x_i$ ($256^3$ grid size)} & 
    \multicolumn{2}{p{3cm}}{Cell-wise $({\bf x},\,{\bf v})$ in $256^3$ and $512^3$ grids ($1\times$ and $2\times$, respectively), and RT grid $x_i$ ($256^3$ grid)} & 
    Cell-wise $({\bf x},\,{\bf v})$ in $1024^3$ grid size, and RT grid $x_i$ ($256^3$ grid size), or real-space power spectra directly \\
 \hline
\multicolumn{2}{c}{Output data type} & 
    \multicolumn{2}{p{3cm}}{HI density in redshift-space grid in $256^3$ and $1024^3$ grid size ($1\times$ and $4\times$RT, respectively)} & 
    \multicolumn{3}{p{3.5cm}}{HI density in redshift-space grid in $256^3$, $512^3$, and $1024^3$ grid size ($1\times$, $2\times$ and $4\times$RT, respectively)} & \multicolumn{2}{p{3cm}}{Power spectrum only 
    \footnote{In principle, a brightness temperature data cube in redshift space can be constructed by taking the inverse Fourier transform of the ${\bf k}$-space brightness temperature evaluated using equations~(\ref{eqn:bt-red-general}) and (\ref{eqn:lin-scheme-F}) for the DEMRF scheme and quasi-linear $\mu_{\bf k}$-decomposition scheme, respectively. However, aliasing effects from multiple forward and backward FFTs can introduce errors. It is beyond the scope of this paper to test these effects.}} & 
    Power spectrum only   \\
  \hline
\multicolumn{2}{c}{Usability} & \multicolumn{2}{p{3cm}}{Numerical simulations} 
  & \multicolumn{3}{p{3.2cm}}{Numerical or semi-numerical simulations} 
& \multicolumn{2}{p{3cm}}{Numerical or semi-numerical simulations}  
& Analytical modeling (no realization), numerical or semi-numerical simulations  \\
  \hline
  \multicolumn{2}{c}{Well defined} & \multicolumn{2}{c}{Yes} & \multicolumn{3}{p{3cm}}{Inaccurate assumptions on small scales} & \multicolumn{2}{p{3cm}}{Unable to use on a grid too fine (see \S~\ref{sec:DEMRF})} & Yes  \\
 \hline \hline    
 
Error \footnote{All errors here are with respect to the results from the PPM-RRM (4$\times$RT) scheme, which is the most accurate.} in 1D  & at $k~\le~2\,h$/Mpc & 
    $\lesssim 2\%$ & benchmark &  $\lesssim 4\%$ & $\lesssim 2\%$ & $0\%$  & $\lesssim 1\%$ & $0\%$ & $\lesssim 10\%$\\  \cline{2-10}
power spectrum   &  at $ 2\!\!\!~<~\!\!\!k~\!\!\!<~\!\!\!7\,h$/Mpc &    
 \vskip0.001cm  $\lesssim 20\%$ & \vskip0.001cm benchmark & \vskip0.001cm  $\lesssim 40\%$  & \vskip0.001cm $\lesssim 20\%$ & \vskip0.001cm $\lesssim 1\%$  & \vskip0.001cm  $\lesssim 14\%$ & \vskip0.001cm $\lesssim 5\%$ & \vskip0.001cm $\lesssim 10\%$ \\
  \hline\hline

SUs & Preliminary\footnote{Preliminary SUs for the MM-RRM scheme, quasi-linear $\mu_{\bf k}$-decomposition scheme, and DEMRF scheme refers to the SUs used to smooth particle density and velocity data onto a regular, real-space, grid.}
 & 0 & 0  & 350  & 358 &  375 &  350 &  358 &  375 	\\ 
  \cline{2-10}
(=cores   & Processing   & 2048 & 2127 & 0.1 & 0.7 & 8.5 & 52 & 887 & 5.3\\
  \cline{2-10}  
$\times$hours)    & Total      &  2048 & 2127 & 350 & 359 & 384 & 402 & 1245 & 380\\
  \hline \hline
\end{tabular}
\end{minipage}
\end{center}
\end{table*}

\subsection{Computational Accuracy and Efficiency}
\label{sec:comparison}

So far we have discussed four viable schemes to compute 21cm brightness temperatures in redshift space: the PPM-RRM scheme (\S~\ref{sec:PPM-RRM}), the MM-RRM scheme (\S~\ref{sec:MM-RRM}), the quasi-linear $\mu_{\bf k}$-decomposition  scheme (\S~\ref{eqn:lin-muk-decomp}), and the DEMRF scheme (\S~\ref{sec:DEMRF}). To facilitate the usage of these schemes, we compare their usability, accuracy and efficiency in Table~\ref{tab:accu-eff}. 

With N-body particle data (numerical simulation), the PPM-RRM scheme has no ambiguity in finding new particle locations in redshift space. When particle data is re-smoothed onto a redshift-space grid four times finer than RT grid resolution, PPM-RRM (4$\times$RT) can accurately compute the power spectrum down to the RT resolution scale. However, the scheme is very computationally expensive and difficult to code, so we recommend to use it only as a development tool and for benchmarking, not for production work.

The MM-RRM (4$\times$RT) scheme is the perfect tool for production work. It requires only 1/6 of total computational effort of the PPM-RRM (4$\times$RT) scheme (including preliminary calculations), and the results are just as accurate. Using the fine (4$\times$RT), instead of coarse (RT) grid does not really add to the total computational effort. Note also that it can be directly used for semi-numerical simulations that evolve density on grids and do not use particles.

The DEMRF scheme is a nicely posed scheme since it is just a mathematical integration. However, in practise if we wish to substitute the cell-wise average $\left< \exp{\left[-i \left(\frac{1+z_{\rm cos}}{H(z_{\rm cos})}\right)k_\parallel v_\parallel ({\bf r})\right]}\right>_{\rm cell}$ with $\exp{\left[-i \left(\frac{1+z_{\rm cos}}{H(z_{\rm cos})}\right)k_\parallel \left< v_\parallel ({\bf r}) \right>_{\rm cell} \right]}$ using cell-wise velocity, this scheme loses accuracy at the cell size $\sim 114/1024\approx 0.11\,{\rm Mpc}/h$. Moreover, the DEMRF (2$\times$RT) scheme is neither the most accurate nor the most efficient, so it is not to be recommended for production work. However, it is useful for validating the results from the PPM-RRM and MM-RRM schemes.

In the case of no realization, one can employ the quasi-linear $\mu_{\bf k}$-decomposition scheme which yields the redshift-space power spectrum with moderate accuracy and the least computational effort. It also has as a useful feature that it can proceed with only real-space statistics as input, making it an ideal tool for pure analytical modeling. 

The upshot is that we recommend the MM-RRM (4$\times$RT) scheme for practical usage, and the PPM-RRM (4$\times$RT) for development.

\section{How Much do Rare Optically-thick Cells Affect the Accuracy of Power Spectrum in the Optically-thin Approximation?}
\label{sec:finite-opt-comparison}

In the optically thin limit, computation of 21cm redshift-space brightness temperature can be simplified by taking advantage of the proportionality of brightness temperature and neutral atom density, both in redshift-space. However, we have shown in Figure~\ref{fig:PDFtau} that there is a nonzero, albeit small, chance to find large 21cm optical depth in the IGM. So in principle, the observed 21cm power spectrum in the redshift-space that takes the finite optical depth into account can be different from the result in the optically-thin approximation. The difference depends on the population of optically-thick cells.  
In this section, we revisit the accuracy of the optically-thin approximation with regard to the 21cm power spectrum. 

We use the DEMRF method to calculate two power spectra in redshift-space: one with finite optical depth in equation~(\ref{eqn:nonlinear_dtb_finite_opt}), and one in the optically-thin approximation in equation~(\ref{eqn:bt-red-general}). We have demonstrated in \S~\ref{sec:DEMRF} that, in the optically thin limit, the power spectrum using the DEMRF scheme agrees with the PPM-RRM result in the comfort zone ($k \le k_N/4$). Here, we smooth the density, velocity and velocity gradient fields on the fine ($512^3$) grid with two times better resolution than RT grid ($256^3$), and focus on the region $k \le k_N^{(512)}/4 = 3.5$ Mpc$/h$. We use SPH-like smoothing of our N-body particle data to compute the velocity gradient on the grid. Details of this technique are discussed in Appendix~\ref{app:sph}. 

The 21cm optical depth depends on the spin temperature which, however, is beyond the scope of our cosmological radiative transfer simulations. For the purpose of demonstration, we assume $\alpha = T_s/T_{\rm CMB}(z_{\rm cos})$ is a spatial constant, and investigate the cases $\alpha = 100$, 10, and 0.1 (the $\alpha = 1$ case has no 21cm radiation contrast to CMB). 
In the optically-thin approximation, the power spectrum with finite (but constant) spin temperature is just the power spectrum with high spin temperature ($T_s \gg T_{\rm CMB}$), i.e. the result in \S~\ref{sec:DEMRF}, scaled by the factor $(1-\frac{1}{\alpha})^2$. 

\begin{figure*}
\begin{center}
  \includegraphics[height=0.35\textheight]{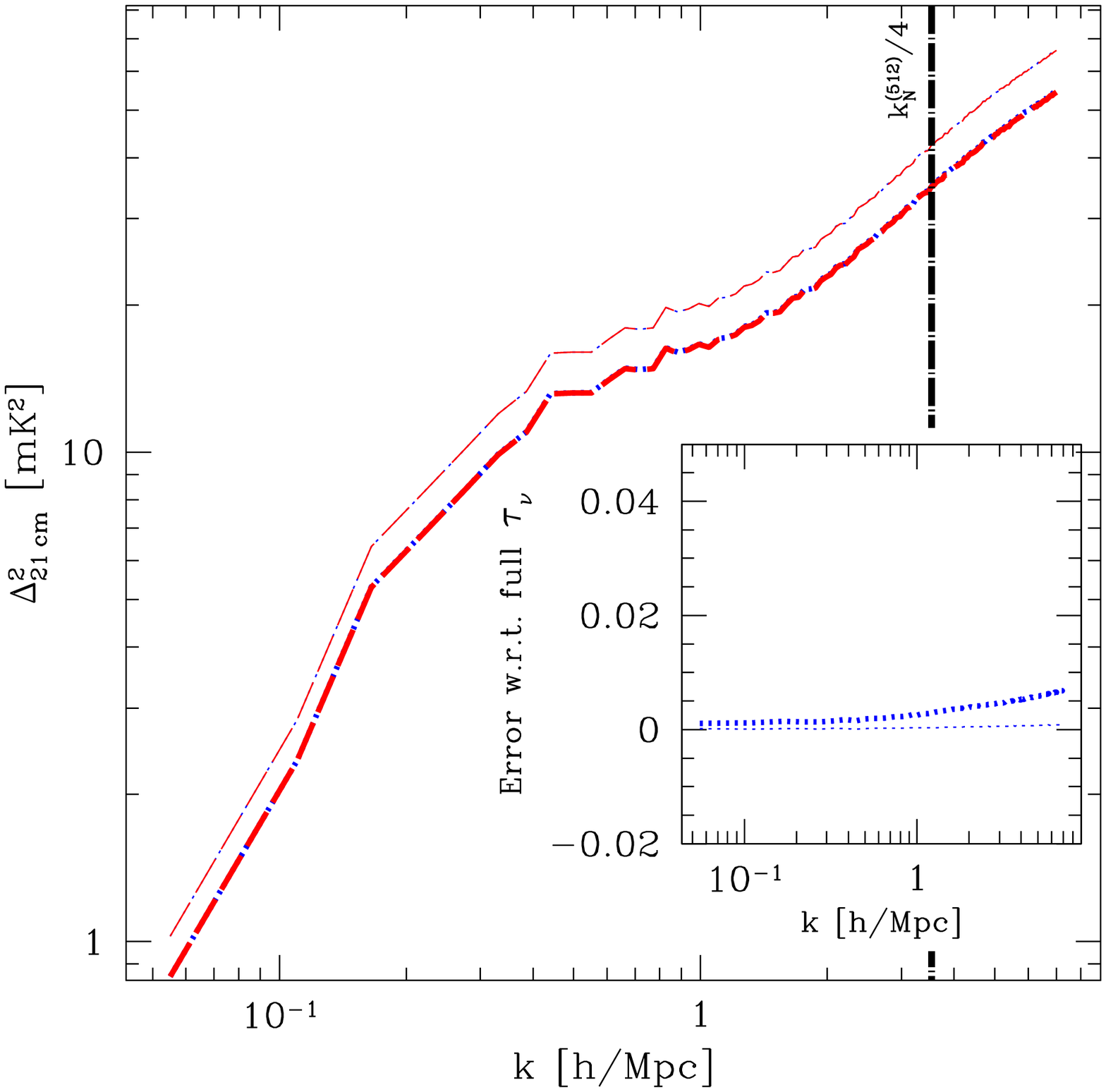} 
  \includegraphics[height=0.35\textheight]{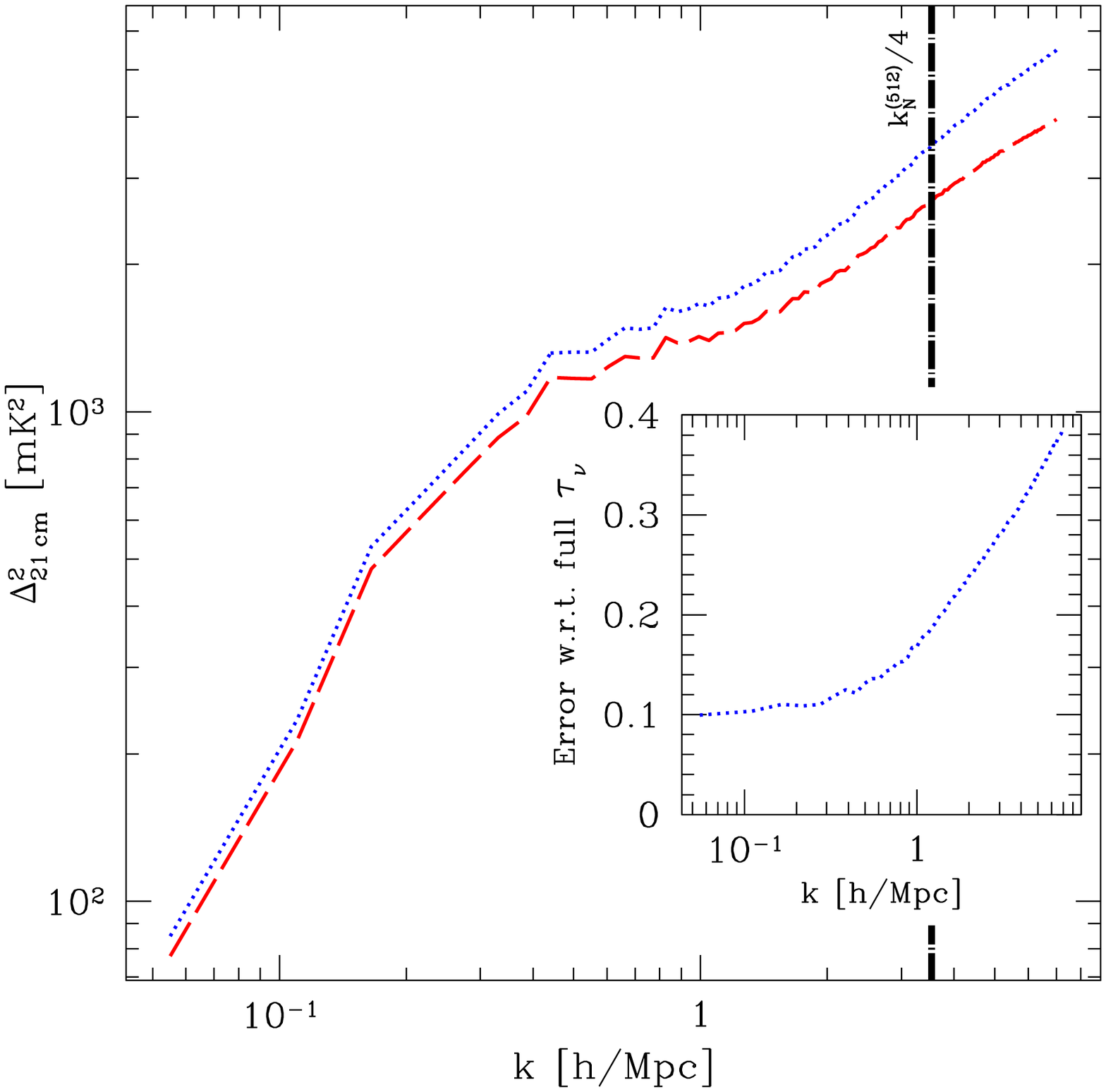} 
\end{center}
\caption{Power spectra of 21cm brightness temperature in redshift-space calculated in the optically-thin approximation (dotted, blue), and the results that take finite optical depth into account (long-dashed, red), both using the {\it DEMRF} scheme on a grid ($512^3$) two times finer than the RT grid. 
(Left) When $T_s$ is high, i.e. $\alpha = T_s/T_{\rm CMB} = 100 $ (thin lines) and $10$ (thick lines). In each set, two curves (finite optical depth vs. optically-thin approximation) overlap almost exactly. 
(Right) When $T_s$ is low, i.e. $\alpha = 0.1$. All results use the RT simulation data at 50\% ionized epoch ($z=9.457$). The fractional errors of the optically-thin approximation are with respect to the results with finite optical depth. The vertical lines at $ k = k_N^{(512)}/4 = 3.5$\,h/Mpc delimit the comfort zone for the {\it DEMRF} results computed on the $512^3$ grid. 
}
\label{fig:fullvsDEMRF}
\end{figure*}

In Figure~\ref{fig:fullvsDEMRF} (left panel), we find that the power spectra in the optically-thin approximation are so highly accurate, in the $\alpha = 100$ and 10 cases, that the two curves (finite optical depth vs. optically-thin approximation) are almost indistinguishable. However, Figure~\ref{fig:fullvsDEMRF} (right panel) shows that, in the low $T_s$ case ($\alpha = 0.1$), the optically-thin approximation can result in an error of $\sim 10\%$ in the power spectrum on large scales, and $\gtrsim 30\%$ on small scales. The large-scale error is due to the offset in the global mean signal, because the optically-thin approximation overestimates the brightness temperature (i.e. $\delta T_b \propto \tau_\nu$) in the optically-thick cells, which should otherwise be suppressed in the exact form $\delta T_b \propto [1-\exp{(-\tau_\nu)}]$ when optical depth is large. This decreases the small-scale power spectrum, too, because the 21cm brightness temperature in these overdense regions (where $\tau_\nu \gtrsim 1$) fails to encode the complete statistical information of density and ionization fluctuations. 

Is the optically-thin approximation accurate with regard to 21cm power spectrum? The answer depends on the spin temperature, because 21cm optical depth is inversely proportional to $T_s$. As Figure~\ref{fig:PDFtau} shows, the low $T_s$ case has much higher chance to find optically-thick cells than the high $T_s$ case, i.e. roughly an order of magnitude smaller in $T_s$, an order of magnitude larger in the probability of $\tau_\nu \gtrsim 1$. This is consistent with our results that the optically-thick cells are too rare to virtually affect the power spectrum when $T_s/T_{\rm CMB} \ge 10$, but they are non-negligible when $T_s$ is lower than $T_{\rm CMB}$. 

The upshot is that the power spectrum in the redshift-space calculated in the optically-thin approximation, e.g. using the PPM-RRM or MM-RRM scheme, is accurate with respect to the result that takes finite optical depth into account, {\it only} when $T_s$ is high ($T_s/T_{\rm CMB} \ge 10$). The low $T_s$ case merits further careful investigation which we defer to future work. 

\section{How Accurate Is Linear Theory?}
\label{sec:BL05}

The linear theory formula for 21cm redshift-space power spectrum \citep{Barkana05} has been widely employed in the literature (e.g. \citealt{Santos06,Zahn07,Mao:2008ug,Adshead11}), but it is derived under two assumptions that may both break down. First, the ionization fluctuations are assumed to be linear. This is only valid on scales much larger than the size of the H~II region which can be rather large ($\sim 10$ Mpc, see, e.g. \citealt{Friedrich11}). Second, the matter density and velocity fluctuations are assumed to be linear, i.e.\ the velocity is dictated by the density through the linear relation, $\widetilde{v^r_\parallel}({\bf k}) = i\left(\frac{H(z_{\rm cos})}{1+z_{\rm cos}}\right)\widetilde{\delta^r_{\rho_{\rm H}}}({\bf k}) \frac{\mu_{\bf k}}{k}$. This relation is also inaccurate on small scales. 
Is linear theory spoiled by the breakdown of these approximations? For simplicity, we restrict our discussion in this section to the simple case $T_s \gg T_{\rm CMB}$. 

\begin{figure}
\begin{center}
  \includegraphics[height=0.35\textheight]{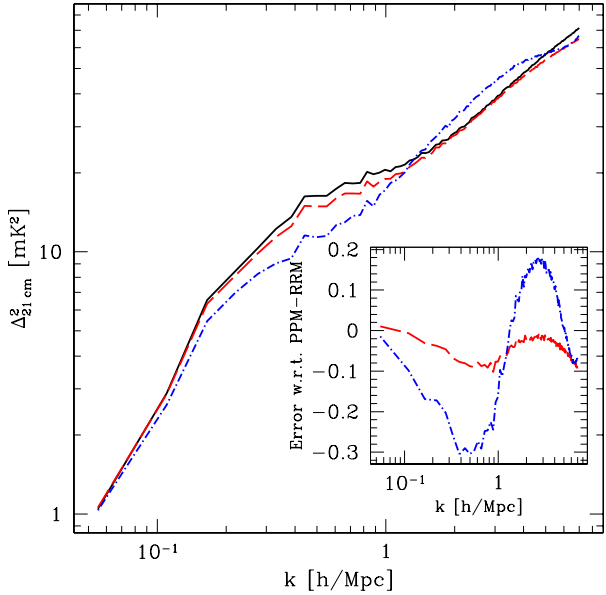} 
\end{center}
\caption{Test of the linear theory: 21cm redshift-space 1D power spectrum at $z=9.457$ (50\% ionized), calculated using the linear theory of \citet{Barkana05}  (dot-short-dashed, blue), the quasi-linear $\mu_{\bf k}$-decomposition scheme (long-dashed, red), and the PPM-RRM (4$\times$RT) scheme (solid, black), respectively. }
\label{fig:BL05-test}
\end{figure}

We compute the 21cm redshift-space 1D power spectrum in the {\it linear theory} by angle-averaging equation~(\ref{eqn:lin-scheme}) with moments in equations~(\ref{eqn:BL05-mu0}) - (\ref{eqn:BL05-mu4}), (\citealt{McQuinn:2005hk,Zahn07,Lidz07})
\begin{eqnarray}
P_{\Delta T}^{s,{\rm lin, 1D}} (k) &=& \widehat{\delta T}_b^2 \left[ 
P^r_{\delta_{x_{\rm HI}},\delta_{x_{\rm HI}}}(k) + 
\frac{8}{3} P^r_{\delta_{x_{\rm HI}},\delta_{\rho_{\rm H}}}(k) \right.\nonumber \\
& & \left. + \frac{28}{15} P^r_{\delta_{\rho_{\rm H}},\delta_{\rho_{\rm H}}}(k) 
\right]\,.
\end{eqnarray}
We compare it with the angle-averaged fully nonlinear power spectrum in redshift-space, computed using the PPM-RRM ($4\times$RT) scheme. In Figure~\ref{fig:BL05-test}, we find that the linear theory power spectrum departs from the fully nonlinear result with $\lesssim 30\%$ error in the intermediate range of $k \sim 0.1 - 1\,h/$Mpc, at the 50\% ionized epoch. It crosses the nonlinear result at $k \sim 1\,h/$Mpc, and deviates more from the latter at smaller scales.  This is in qualitative agreement \footnote{The deviation increases monotonically at $k > 1\,h/$Mpc in \cite{Lidz07}, while there seems to be a turn-around at large $k$ in our Figure~\ref{fig:BL05-test}. This turnaround is not real because it is the resolution effect. We are forced to compute the linear theory power spectrum on the RT grid (the grid for ionization fraction fields). The aliasing effect suppresses the linear theory result at $k>k_N^{(256)}/4=1.76\,h/$Mpc in our simulation, while \cite{Lidz07} result is free of aliasing effect at $k\lesssim 10\,h/$Mpc by adopting an RT grid of much higher resolution.}
with a similar comparison in \cite{Lidz07} (their Fig.~10, but they did not provide any detail of how they computed the nonlinear power spectrum in redshift-space). 

\cite{Lidz07} pointed out that such a large error in linear theory may result from the neglect of higher order auto- and cross-correlations involving density and ionization fluctuations, i.e. the breakdown of the first assumption we mentioned above, but they did not provide a solution that incorporates all of relevant higher order terms in redshift-space power spectrum. Here we propose in \S~\ref{eqn:lin-muk-decomp} the {\it quasi-linear $\mu_{\bf k}$-decomposition} scheme as such a solution that not only incorporates these higher order corrections, but can decompose 21cm redshift-space power spectrum in polynomials of $\mu_{\bf k}$, just as the linear theory does. How accurate is this new scheme? Figure~\ref{fig:BL05-test} also shows that the angle-average power spectrum of the quasi-linear $\mu_{\bf k}$-decomposition scheme (calculated using eq.~\ref{eqn:Kaiser-ave.}) agrees with the fully nonlinear result to $\sim 10\%$ accuracy at $k \sim 0.3 - 2\,h/$Mpc, but with increasing errors at smaller scales. 
We will defer the detailed investigation of the errors in this scheme associated with the neglect of additional nonlinearity to \cite{Shapiro11}. 

The large errors in the linear theory for redshift-space distortion suggest that it is a simple, but by no means accurate, tool to predict 21cm power spectrum. One should either employ the {\it quasi-linear $\mu_{\bf k}$-decomposition} scheme for improved (but not perfect) accuracy, or follow the numerical schemes we proposed above (PPM-RRM and MM-RRM) to obtain fully nonlinear results.

\section{How Accurate is the ``$\bmath{\nabla \upsilon}$-limited'' Prescription?}
\label{sec:vecgrad-lim}

\subsection{The ``$\bmath{\nabla v}$-limited'' prescription vs. the avoidance of the divergence problem in observer space}

\citet{Santos10} treated the effects of peculiar velocity on the 21cm brightness temperature by evaluating an equation equivalent to our equation~(\ref{eqn:dtb1}) at each point in a {\it real-space} grid at a given time. They found that the 21cm brightness temperature diverges in some overdense regions where $\delta_{\partial_r v} \rightarrow -1$. As such, the power spectrum computed from the Fourier transform of this 21cm brightness temperature evaluated in {\it real-space} diverges, too. They deal with this divergence problem by replacing the actual value calculated for $\delta_{\partial_r v}$ from their real-space grid data whenever it is close to -1, by a fixed minimum value larger than $-1$ (e.g.,  $-0.7$ in their paper), so as to cap the divergence and obtain finite results for both brightness temperature and its power spectrum. This approach was also adopted by the {\it 21cmFAST} code \citep{Mesinger11} (with the cap $-0.5$). 

Before analyzing the accuracy of the $\nabla v$-limited prescription, we would like to explain why the divergence encountered for $\delta_{\partial_r v} \rightarrow -1$ is a mathematical, but not a physical one. As we shall show, the appearance of the divergence is avoided naturally for physical observables in observer redshift-space. 

The first part of this explanation was already considered in \S~\ref{sec:singular-revis}. Equation~(\ref{eqn:dtb1}) was derived under the assumption of low optical depth. However, the locations at which $\delta_{\partial_r v}$ approaches -1 are not optically thin. The 21cm brightness temperature must be evaluated using equation (\ref{eqn:opt-thick-formula}), instead, at these locations, to take finite optical depth into account. When this is done, the brightness temperature does not diverge for $\delta_{\partial_r v} \rightarrow -1$.

However, even in the optically-thin approximation, it is unnecessary to apply a cap to the velocity gradient in order to prevent divergence in the physical observables, as long as we account properly for redshift-space distortion. 
The approach in which equation (\ref{eqn:dtb1}) is applied to real-space grid data does not fully account for the remapping of real- to redshift-space locations of 21cm sources. 
While this remapping cannot remove the divergence of 21cm brightness temperature at those locations at which $\delta_{\partial_r v} \rightarrow -1$, the power spectrum in redshift-space is guaranteed to be finite. The reason is simply that real-space regions for which $\delta_{\partial_r v} \rightarrow -1$ become infinitesimally small in redshift-space, since $d^3s = d^3r \left|1+\delta^r_{\partial_r v}({\bf r})\right|$. The Fourier transform of brightness temperature in {\it redshift-space}, 
$\widetilde{\delta T_b^s} ({\bf k}) \equiv \int d^3s \,\, e^{-i {\bf k \cdot s}} \,\,\delta T_b^s ({\bf s})$, is a finite integration, and so is the power spectrum computed from it, because the divergent factor in $\delta T_b^s ({\bf s}) = \delta T_b^r ({\bf r}) \propto 1/\left|1+\delta^r_{\partial_r v}\right| $ is exactly cancelled by its inverse in the volume element $d^3s$ . 
In addition, in the optically-thin approximation, while, strictly speaking, the 21cm brightness temperature still diverges at those locations at which $\delta_{\partial_r v}$ approaches -1, it, too, becomes finite when smoothed over finite band- and beam-width in observer redshift space (see also \S~\ref{sec:physapproch}). 
This, of course, makes perfect sense since the 21cm emitted photons produced by a finite region of space, in the optically-thin limit, are proportional to the number of neutral hydrogen atoms in that region, which is always finite and is conserved by the mapping from real- to redshift-space. 
Since what observers actually measure are this pixelized brightness temperature and the power spectrum, full account of redshift-space distortion gives a physical result for these observables without resorting to artificial caps. 

\begin{figure*}
\begin{center}
  \includegraphics[height=0.35\textheight]{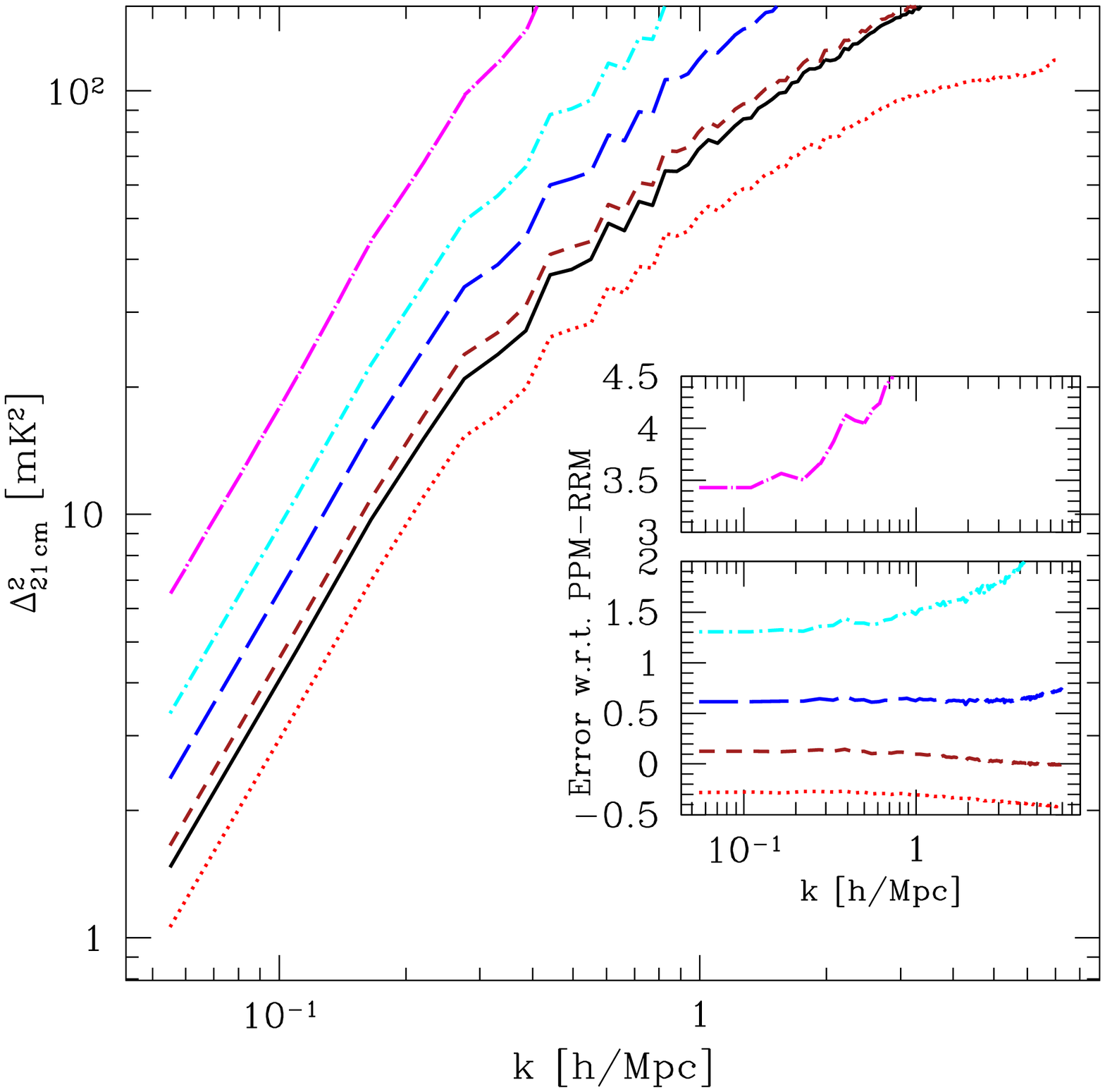} 
  \includegraphics[height=0.35\textheight]{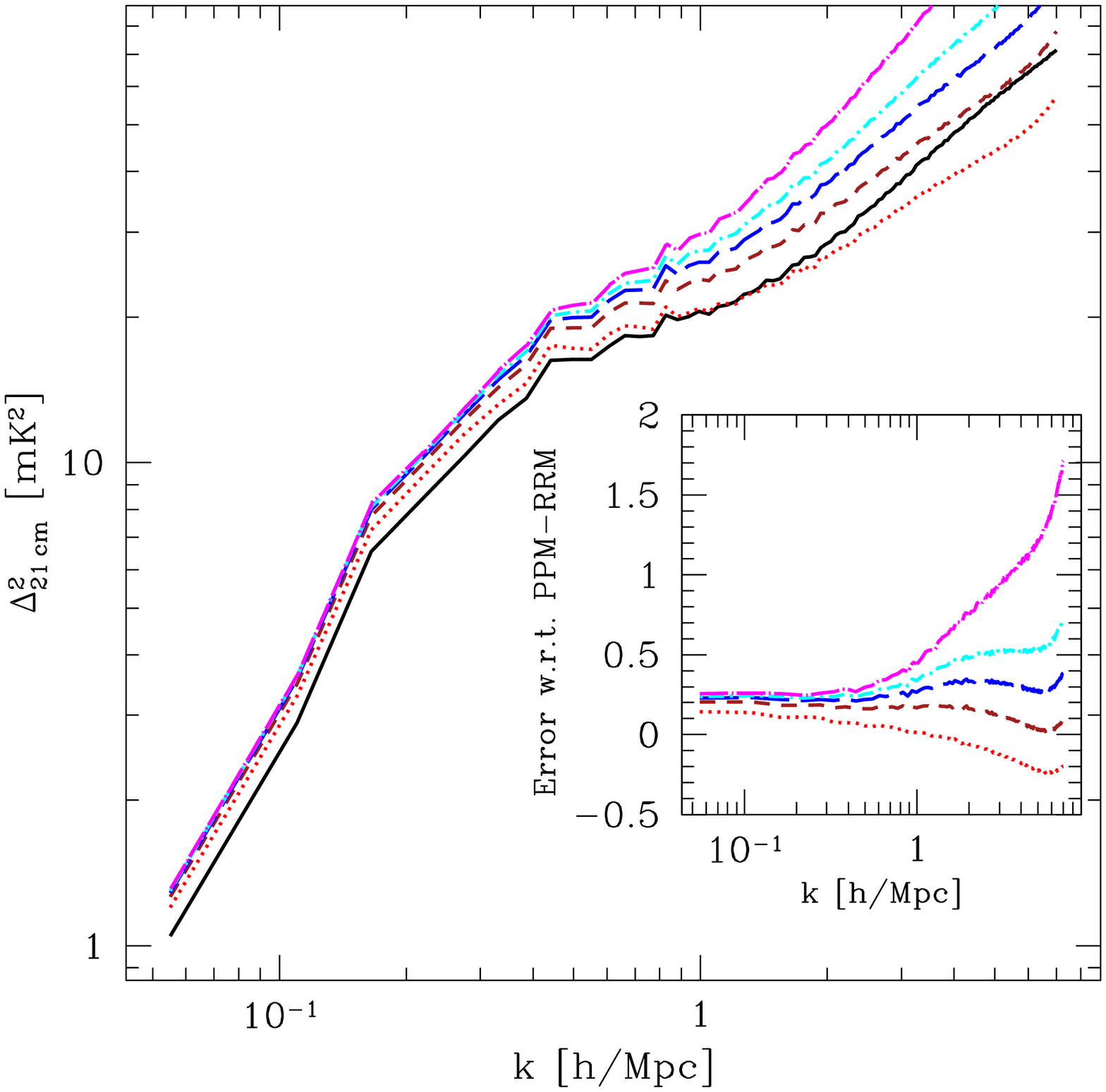} 
  \includegraphics[height=0.35\textheight]{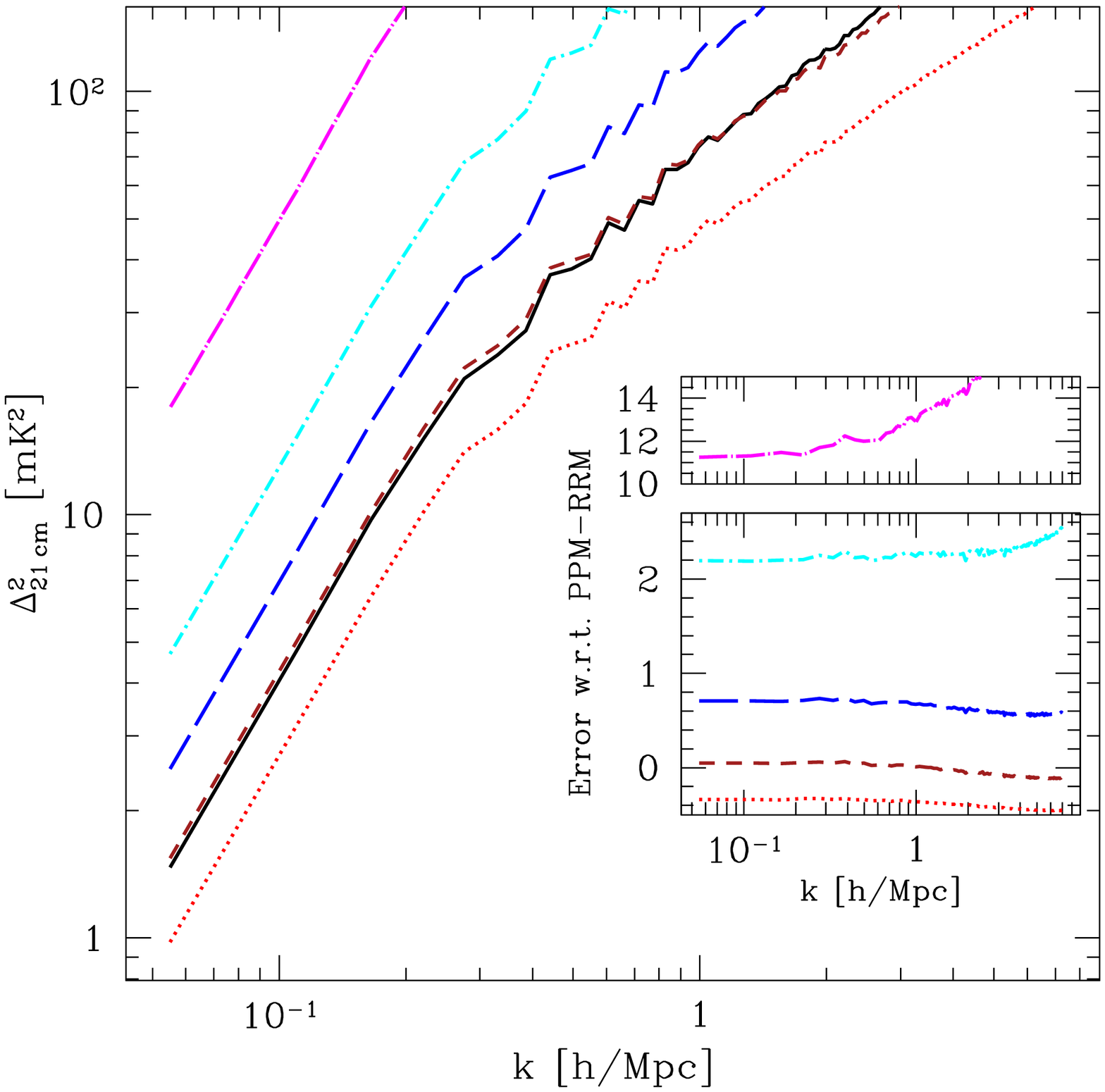} 
  \includegraphics[height=0.35\textheight]{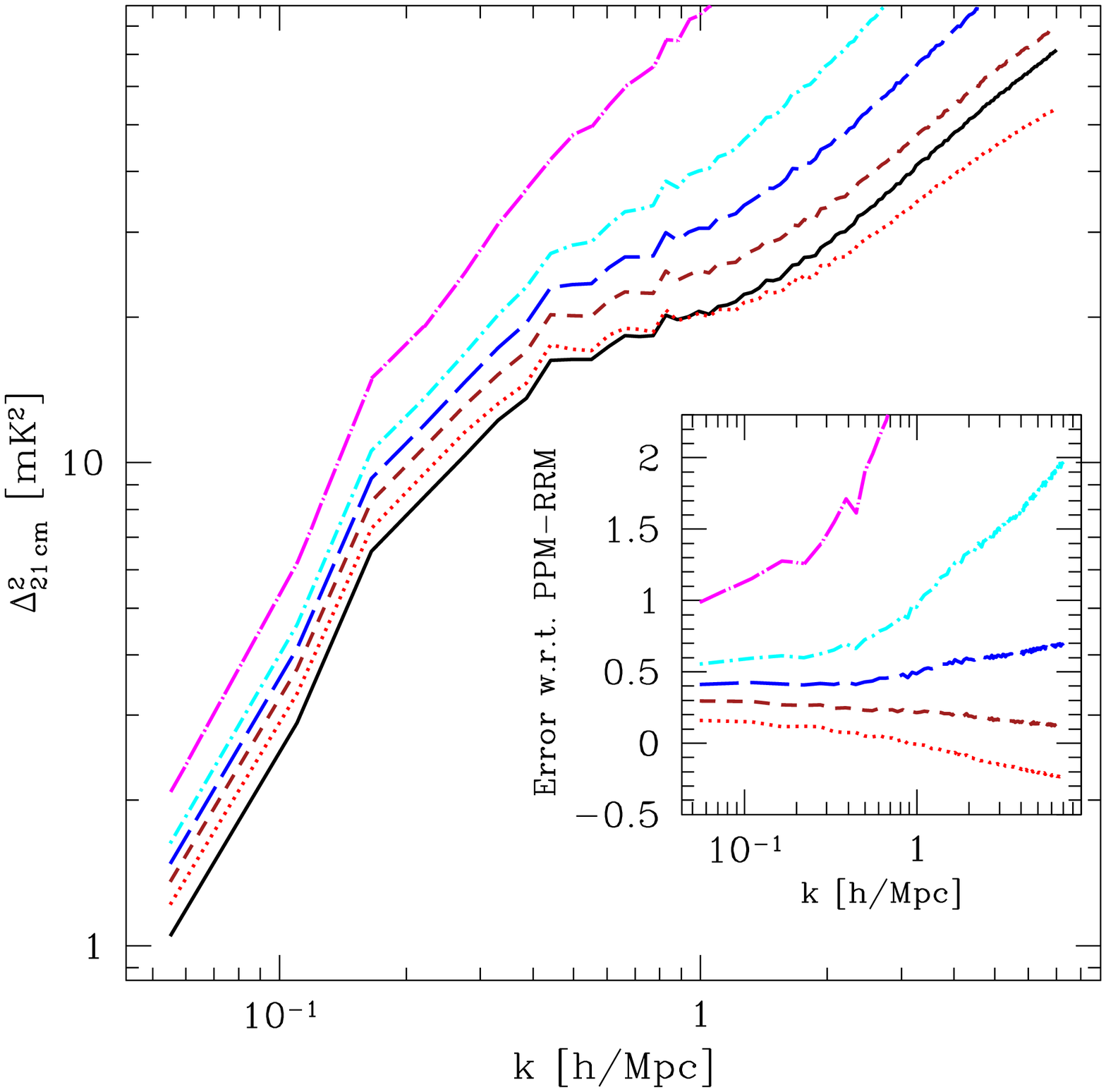} 
\end{center}
\caption{Power spectra of 21cm brightness temperature at 50\% ionized epoch ($z=9.457$). We show the results of the $\nabla v$-limited prescription with the cap $\left<\delta_{\partial_r v}\right>_{\rm cell} \ge -\lambda$ at 
$\lambda = $0.1 (dotted, red), 0.3 (short-dashed, brown), 
0.5 (long-dashed, blue), 0.7 (dot-short-dashed, cyan), and 0.9 (dot-long-dashed, magenta), compared to the results of the PPM-RRM (4$\times$RT) scheme (solid, black). The fractional errors of the $\nabla v$-limited prescription are with respect to the PPM-RRM (4$\times$RT) result, plotted in the inset. Upper panels: using velocity gradient field smoothed on the RT grid resolution; lower panels: using velocity gradient field smoothed on the fine grid (4 times finer than RT grid resolution). 
Left panels: assuming a fully neutral Universe ($x_{\rm HI} = 1$); right panels: using the actual reionization fluctuations from the simulation. }
\label{fig:PPMvsgradv}
\end{figure*}

\subsection{Evaluating the accuracy of the ``$\bmath{\nabla v}$-limited'' prescription} 

Although based on a conceptual artifact (truncation of an unphysical divergence) and providing an incomplete fix (calculating the power spectrum in real- instead of redshift-space), the ``$\nabla v$-limited'' prescription may still provide a practical solution to the problem of the diverging brightness temperature.  \citet{Santos10} argued that although an ad-hoc solution, imposing this limit only affects a very small number of cells, and thus has no influence on global statistics such as the power spectrum. Since our methods avoid the divergence problem, we are now able to test this assertion. Furthermore, to be a practical solution to the problem, the results should not depend too much on the choice for the cap on the velocity gradient. Here we test these two issues by comparing the results of the $\nabla v$-limited prescription to those from our PPM-RRM (4$\times$RT) scheme. For simplicity, we restrict our discussion in this section to the simple case $T_s \gg T_{\rm CMB}$.

In order to find gridded values for the velocity gradient, we use SPH-like smoothing of our N-body particle data to compute the velocity gradient. Details of this technique are discussed in Appendix~\ref{app:sph}. We implement the $\nabla v$-limited prescription by 
replacing the actual value of $\left<\delta_{\partial_r v}\right>_{\rm cell}$ by the cap value of $-\lambda$ whenever $\left<\delta_{\partial_r v}\right>_{\rm cell} < -\lambda$, 
for a range of cap values $\lambda = $ 0.1, 0.3, 0.5, 0.7, and 0.9, 
evaluating and Fourier transforming the brightness temperature in real space. We then average the power spectra over three LOS directions. In this section, we assume the limit of high spin temperature, $T_s \gg T_{\rm CMB}$. 

To most clearly show the effects of the $\nabla v$-limited prescription we first take our volume to be fully neutral, by setting $x_{\rm HI}=1$ everywhere. Figure~\ref{fig:PPMvsgradv} (top left panel) shows the power spectra from the $\nabla v$-limited prescription for five different values of the upper limit $\lambda$ as well as the power spectrum calculated with the PPM-RRM scheme. Here we use the smoothed velocity gradient field on the RT grid resolution. 
We find that different values for $\lambda$ yield rather different power spectra even on large scales, and 
none of the previously proposed values of caps ($\lambda = 0.5$ or 0.7) is consistent with the PPM-RRM result.

This is of course the most extreme case since in a fully neutral medium all locations with $\delta_{\partial_r v} \rightarrow -1$ contribute. Since these regions are preferably located in high density areas, which typically reionize earlier, one can expect that the effect is much less severe when considering a neutral fraction distribution obtained from a reionization calculation. Figure~\ref{fig:PPMvsgradv} (top right panel) shows this indeed to be case. 
On large scales, different limits in the $\nabla v$-limited prescription yield converging power spectra which, however, have an offset of 
$\sim$ 20\% from the power spectrum of the PPM-RRM scheme. This offset is due to the enhancement in the mean brightness temperature averaged in real space, i.e.\ although the distribution of $\delta_{\partial_r v}({\bf r})$ has zero mean, the distribution of $\left(1+\delta_{\rho_{\rm HI}}({\bf r})\right)/\left|1+\delta_{\partial_r v}({\bf r})\right|$ does not have the (volume-weighted) mean of unity \footnote{This can be compared to the mean in redshift space, where the averaging is over redshift-space volume elements $d^3s = d^3r\, |1+\delta^r_{\partial_r v}({\bf r})| $, equivalent to averaging in real-space weighted by $|1+\delta^r_{\partial_r v}({\bf r})| $, and therefore the distribution of $\left(1+\delta_{\rho_{\rm HI}}({\bf r})\right)/\left|1+\delta_{\partial_r v}({\bf r})\right|$ has 1 as the mean in redshift space.\label{ftnt:redshift-mean}} due to the nonlinear function $1/\left|1+\delta_{\partial_r v}\right|$. Although converged on large scales, on small scales the results of the $\nabla v$-limited prescription still depend on the cap value chosen and can have inaccuracy as large as 
$\lesssim 40\%$ for $\lambda = 0.5$, or $\lesssim 50\%$ for $\lambda = 0.7$, both at $k \lesssim 2 h/$Mpc, and more divergent for larger caps 
(as exemplified by $\lambda = 0.9$). 
However, these inaccuracies are substantially smaller than the ones found for the fully neutral case. 

These inaccuracies, of course, depend on the grid resolution of the velocity gradient field. We redo the above analyses, using a fine grid (four times finer than the RT grid), as shown in Figure~\ref{fig:PPMvsgradv} (bottom panels). We find that the errors in the results of the $\nabla v$-limited prescription are significantly amplified. This is because the velocity and its gradient become more nonlinear on smaller scales. Hence, a larger population of cells are ``clipped'' on fine grids than on coarse grids. 

The reasons why the $\nabla v$-limited prescription does not work well are twofold. First, this prescription deals with data cubes in real-space coordinates. Consequently, their Fourier transform and power spectra are real-space quantities, unlike in the redshift-space as this prescription claimed to do. 
Second, even though the $\nabla v$-limited prescription was invented to circumvent the unphysical divergence in 21cm brightness temperature in real-space regions that are optically thick to 21cm radiation, the cells that are affected by imposing a cap on velocity gradient are actually much more numerous than the optically-thick cells; e.g., at 50\% ionized epoch in our simulation, we find that only a fraction of $\sim 10^{-7}$ amongst all cells are optically thick in the best case ($T_s/T_{\rm CMB}=100$), or $\sim 0.01\%$ in the worst case ($T_s/T_{\rm CMB}=0.1$), (see \S~\ref{sec:opt-thin-revis}), 
but the $\nabla v$-limited prescription affects all those cells for which $\left|\delta_{\partial_r v}\right|$ exceeds the cap, which can be a much larger fraction of the cells than that of the optically thick cells. A fraction $\sim 1\%$ of the cells have $\delta_{\partial_r v} \le -0.7$, while $\sim 3\%$ have $\delta_{\partial_r v} \le -0.5$. 
(These specific fractions can depend on the grid resolution of the simulation. The numbers here are counted using velocity gradient field on RT grid resolution.) In other words, the $\nabla v$-limited prescription affects a lot more cells than necessary. 

Even though the $\nabla v$-limited prescription cannot provide the most accurate treatment of the effect of peculiar velocity, it still serves as a simple and useful tool to estimate the 21cm power spectrum, if a smaller cap is chosen than those previously proposed. We optimize this prescription by comparing its results using $\lambda = 0.1$ and 0.3 with our PPM-RRM result, and find that with the actual reionization fluctuations, the $\nabla v$-limited prescription with $\lambda = 0.1$ approximates the PPM-RRM result with the least errors $\lesssim 20\%$, while, if we assume a fully neutral universe, $\lambda = 0.3$ is the optimal cap, with errors $\lesssim 10\%$. The optimal value of the cap depends on the grid resolution, and perhaps on the redshift and the ionization fraction, as well.

\section{Conclusions}
\label{sec:conclusion}

\begin{itemize}

\item We have demonstrated that the neglect of peculiar velocity introduces a substantial error in 21cm brightness temperature spectra from the EOR and noticeable anisotropy in the 21cm power spectrum. We did this in three different ways: first, we compared the 3D power spectra computed uncorrected for peculiar velocity (UPV scheme), from the quasi-linear $\mu_{\bf k}$-decomposition scheme, and from a particle-based numerical scheme (PPM-RRM); second, we compared the 21cm brightness temperature spectra computed from the UPV scheme and the PPM-RRM scheme, along 5 different sightlines; lastly, we compared the angle-averaged 21cm power spectra computed from the quasi-linear $\mu_{\bf k}$-decomposition scheme and the UPV scheme, respectively. The non-trivial difference between results with and without peculiar velocity correction motivates our thorough investigation of the effect of peculiar velocity on 21cm signal as well as a more careful treatment of this effect on reionization simulation data than previously made. 

\item We clarify that peculiar velocity distorts the mapping of 21cm brightness temperature not only by shifting the apparent location in redshift-space, but also by modifying the brightness temperature itself in real-space. We show that the combined effect, which we call ``21cm redshift-space distortion'', establishes, in the limit of low optical depth and high spin temperature, the exact proportionality between observed 21cm brightness temperature and the neutral hydrogen density as measured in redshift-space. This proportionality makes it possible to infer the three-dimensional distribution of neutral hydrogen density using 21cm brightness temperature measurements. 

\item We show that this proportionality between 21cm observed brightness temperature and the redshift-space neutral hydrogen density, however, can break down when $\tau_{\rm 21cm} \gtrsim 1$ and/or $T_s \lesssim T_{\rm CMB}$. For the first case, we check the optically thin approximation, and demonstrate that this widely-assumed approximation is mostly valid in the IGM, but it can be invalid in some cases, e.g. in virialized halos where the peculiar velocity gradient can be large enough to cancel the Hubble flow. For the second case, we show that the proportionality mentioned above is spoiled by the spatially-varying $T_s$-dependent factor $1-T_{\rm CMB}/T_s^{r,\,\rm eff}({\bf r})$. This $T_s^{r,\,\rm eff}({\bf r})$ includes a correction to $T_s$ of the order $v/c$  due to peculiar velocity.

\item The unphysical divergence in 21cm brightness temperature results from the neglect of finite optical depth, which eliminates the divergence. 
We show that, in the optically thick limit, the optical depth can depend upon higher order spatial derivatives of peculiar velocity than $dv_\parallel/dr_\parallel$. 

\item  We derive the fully nonlinear Fourier transform of 21cm brightness temperature fluctuations, with finite optical depth, as measured in {\it redshift-space}, in terms of the density, velocity and its gradient, ionization fraction, and spin temperature fields in real space, following the combined effect of 21cm redshift-space distortion. We further simplify it in the optically-thin approximation. We further show that, when redshift-space distortion is properly accounted for, however, the observed power spectrum in redshift-space remains finite {\it even} in the optically-thin approximation.

\item We investigate the effect of finite 21cm optical depth. The 21cm power spectrum in redshift-space calculated in the optically-thin approximation is accurate with respect to the results which take finite optical depth into account, {\it only} when spin temperature is high relative to the CMB temperature ($T_s/T_{\rm CMB} \ge 10$). 

\item We clarify that it is the bulk velocity of the gas but not the thermal velocity that is responsible for the velocity correction to the optical depth and 21cm brightness temperature. This is done by showing that the latter constitutes only a negligible contribution to the correction, compared to the former, when $\tau_{\rm 21cm} \lesssim 1$.

\item To make a careful treatment of the peculiar velocity effect on 21cm brightness temperature when using reionization simulation data, we propose and test two numerical schemes that compute the 21cm brightness temperature as measured in a redshift-space grid from real-space simulation data, 
in the limit of high spin temperature. 
Both schemes take advantage of the mapping from real- to redshift-space, one particle-based (PPM-RRM), and one grid-based (MM-RRM). We show that the MM-RRM scheme can be optimized to achieve the same high accuracy in the angle-averaged power spectrum as the PPM-RRM scheme, while being much more computationally efficient than the latter. If the RT grid resolution (on a mesh with Nyquist wavenumber $k_{N,\rm RT}$, the mesh on which the ionization fluctuation field is determined) is coarser than the resolution of the density and peculiar velocity fields, we optimize the grid-based MM-RRM resolution by including all modes with $k\le k_{N,\rm RT}$ in a grid with Nyquist wavenumber $4k_{N,\rm RT}$ which uses the finer-resolution density and velocity data, together with the coarser-resolution ionized fractions. This reduces the aliasing errors which would otherwise spoil the results for $k > k_{N,\rm RT}/4$ if all data were coarsened to the RT-grid resolution. We show that this optimized MM-RRM scheme can compute the angle-averaged 21cm power spectrum within $\lesssim 1\%$ error with respect to the PPM-RRM(4$\times$RT) results, for all modes $k\le k_{N,\rm RT}$.

\item We examine the {\it linear theory} formula widely employed to compute the 21cm redshift-space power spectrum \citep{Barkana05}, and find large inaccuracy ($\sim 30\%$) at the intermediate range $k \sim 0.1 - 1\,h/$Mpc at the 50\% ionized epoch, in the high spin temperature regime. This suggests that linear theory cannot work as an accurate tool to predict the 21cm power spectrum in redshift-space.

\item We develop the ``quasi-linear $\mu_{\bf k}$-decomposition scheme'' which can decompose 21cm power spectrum in polynomials of $\mu_{\bf k}$, just as the linear theory does, but it incorporates relevant higher order correlations of ionization and density fluctuations.  We find that the fully nonlinear 21cm 1D power spectrum deviates from the prediction of quasi-linear $\mu_{\bf k}$-decomposition scheme by roughly 10\% at the 50\% ionized epoch (see \S~\ref{sec:large-scale-test}). The nonlinearity may introduce larger deviations when the 3D power spectrum is decomposed to extract only the $P_{\mu^4}(k)$ for cosmology. It is important to understand the nature of this nonlinear effect, and estimate its impact on 21cm cosmology. We will address these issues in the second paper of this series \citep{Shapiro11}. 

\item Our careful treatment of brightness temperature fluctuations in redshift space avoids the divergences that appear in the real-space evaluation when peculiar-velocity gradients are large. Such large gradients are a natural result of nonlinear structure formation on small scales. We find that previous attempts to escape these divergences by numerically ``clipping'' the velocity gradients whenever they exceed some threshold (referred to here as the ``$\nabla v$-limited prescription'') introduce a non-negligible inaccuracy in the 21cm power spectra on {\it all} scales, including scales much larger than that of the nonlinearity. 
We show that the errors associated with this prescription, however, can be reduced if the value of the cap is properly chosen (e.g. $\lambda\sim 0.1$ yields an error $\sim 15\%$ at $k\sim 0.1\,h/$Mpc), but this error grows with increasing spatial resolution of the grid, and may depend on redshifts and ionization fraction, too. 

\item The upshot is that we provide an integrated understanding of how peculiar velocity affects 21cm tomography, and also an accurate and efficient numerical algorithm (MM-RRM) for practical numerical application.

\end{itemize}


\section*{Acknowledgments} 

The authors wish to thank Marcelo Alvarez, Kanan Datta, Leon Koopmans, Antony Lewis, Matthew McQuinn, Ue-Li Pen, Mario Santos, Max Tegmark, and Jun Zhang for useful discussions. YM is deeply indebted to Eiichiro Komatsu and Donghui Jeong for enlightening discussions on the redshift space distortion in galaxy redshift surveys. YM and PRS would like to acknowledge the hospitality of the Aspen Center for Physics where part of this work was accomplished. The authors acknowledge the Texas Advanced Computing Center
(TACC\footnote{\url{http://www.tacc.utexas.edu}}) at The University of Texas at Austin for providing HPC resources, under NSF TeraGrid grants TG-AST0900005 and TG-080028N and TACC internal allocation grant ``A-asoz'', as well as the Swedish National Infrastructure for Computing (SNIC) resources at HPC2N 
(Ume\aa, Sweden), which have contributed to the research results reported within this paper. This work was supported in part by Swiss National Science Foundation grant 200021-116696/1, NSF grants AST-0708176 and AST-1009799, NASA grants NNX07AH09G, NNG04G177G and NNX11AE09G, Chandra grant SAO TM8-9009X, and Swedish Research Council grant 2009-4088. 
ITI was supported by The Southeast Physics Network (SEPNet) and the Science and Technology Facilities Council grant numbers ST/F002858/1 and ST/I000976/1. KA is supported in part by Basic Science Research Program
through the National Research Foundation of Korea (NRF)
funded by the Ministry of Education, Science and Technology 
(MEST; 2009-0068141,2009-0076868) and by KICOS
through K20702020016-07E0200-01610 funded by MOST.


\begin{appendix}

\section{SPH-like Smoothing with Adaptive kernel}\label{app:sph}

In this section, we briefly describe the SPH-like technique to smooth N-body particle data onto a grid. We refer readers to \cite{Shapiro96} for a comprehensive discussion of smooth particle hydrodynamics with an adaptive kernel.

Assume that the continuous density and velocity fields are represented by $N_p$ particles with mass $m_i$, location ${\bf r}_i$, and velocity ${\bf v}_i$ ($i=1,\ldots,N_p$). We define a particle's kernel $h_i$ to be the distance between the particle $i$ and its 32$^{\rm nd}$ nearest neighbor particle. We take the ``scatter'' approach to smooth particle data (see Fig.~2 of \citealt{Shapiro96} for an illustration of the {\it scatter} vs. {\it gather} approaches), i.e., a field point at ${\bf r}$ is influenced by a particle $i$ if this particle's {\it own} influence zone covers this field point (e.g., in the case of isotropic kernel, $|{\bf r} - {\bf r}_i |\le h_i$). 

We employ the triangular kernel function with adaptive kernel size $h$, $W({\bf r};h) = f_h(x) f_h(y) f_h(z)$, centered at the particle location to smooth its data. The function $f_h(x)$ is triangular-shaped with width $2h$, i.e., 
\begin{equation}
f_h(x) = \biggl\{   
  \begin{tabular}{lcl} 
    $-\frac{x}{h^2} + \frac{1}{h}$ & , & $0 \le x \le h$ \\
    $\frac{x}{h^2} + \frac{1}{h}$  & , & $-h \le x < 0$ \\
    0 & , & otherwise 
  \end{tabular}  
\end{equation}

\subsection*{Smoothed Fields at a Point}

The smoothed mass density and momentum density fields are defined, respectively, by 
\begin{eqnarray}
\rho({\bf r}) &=& \sum_i m_i W({\bf r}-{\bf r}_i;h_i)\,,\\
{\bf \mathcal{P}}({\bf r}) &=& \sum_i m_i {\bf v}_i W({\bf r}-{\bf r}_i;h_i)\,.
\end{eqnarray}
To preserve momentum, the continuous velocity field is defined by 
\begin{equation}
{\bf v}({\bf r}) = {\bf \mathcal{P}}({\bf r}) / \rho({\bf r}) \,.
\label{eqn:smooth-v}
\end{equation}
We identify the bulk-flow velocity of the IGM at a particle's position ${\bf r}_i$ to be the smooth field ${\bf v}({\bf r}_i)$ evaluated at ${\bf r}_i$. 

\subsection*{Smoothed Fields of a Cell}

To smooth particle data onto a regular grid, we use the following approach to compute the cell-wise mass density, 
\begin{eqnarray}
\left< \rho \right>_{\rm cell} & =& \frac{1}{V_{\rm cell}} \int_{\rm cell} \rho({\bf r}) d^3r \nonumber \\
& =& \frac{1}{V_{\rm cell}} \sum_i m_i \int_{\rm cell} W({\bf r}-{\bf r}_i;h_i)d^3r\,, \label{eqn:rhoint}
\end{eqnarray}
where the integral $\int_{\rm cell} W({\bf r}-{\bf r}_i;h_i)d^3r$ can be evaluated analytically, and is only a function of $h_i$ and the relative location between the particle $i$ and the cell boundaries. Similarly, the cell-wise momentum density is 
\begin{eqnarray}
\left< {\bf \mathcal{P}} \right>_{\rm cell} & =& \frac{1}{V_{\rm cell}} \int_{\rm cell} {\bf \mathcal{P}}({\bf r}) d^3r \nonumber \\
& =& \frac{1}{V_{\rm cell}} \sum_i m_i {\bf v}_i \int_{\rm cell} W({\bf r}-{\bf r}_i;h_i)d^3r \,.
\end{eqnarray}
The cell-wise velocity is defined in a momentum-preserving way, 
\begin{equation}
\left< {\bf v} \right>_{\rm cell} = \left< {\bf \mathcal{P}} \right>_{\rm cell} / \left< \rho \right>_{\rm cell}\,.
\end{equation}

\begin{figure}
\begin{center}
  \includegraphics[height=0.2\textheight]{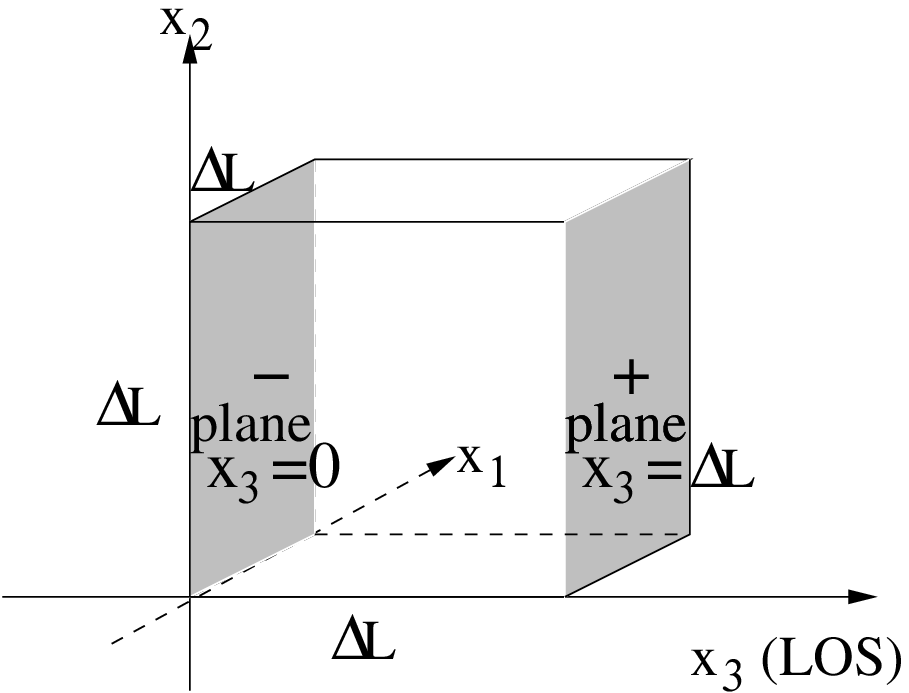} 
\end{center}
\caption{Cartoon of computing $\left< dv_\parallel / dr_\parallel \right>_{\rm cell}$ for a cubical cell of size $\Delta L$ on each side. In this cartoon, we assume the LOS along the $x_3$-axis, then the ``$+$ plane'' (``$-$ plane'') is the $x_1$-$x_2$ plane with $x_3=\Delta L$ ($x=0$).}
\label{fig:dvdr}
\end{figure}

\subsection*{$\mathbf{dv_\parallel/dr_\parallel}$ of a Cell}
\label{sec:dvdrmethod}

We compute the cell-wise velocity gradient $\left< dv_\parallel / dr_\parallel \right>_{\rm cell}$ in the following way (assuming the LOS is along one of the principal axes of a cubical cell), 
\begin{eqnarray}
\left< dv_\parallel / dr_\parallel \right>_{\rm cell} & = & 
  \frac{1}{V_{\rm cell}} \int_{\rm cell} \frac{dv_\parallel}{dr_\parallel}({\bf r})\,d^3r \nonumber \\
 & = & \frac{1}{\Delta L}\left[\left< v_\parallel \right>_{+\,\rm plane} - \left< v_\parallel \right>_{-\,\rm plane}\right]\,,
\label{eqn:dvdr}
\end{eqnarray} 
where $\Delta L$ is the size of the cubical cell, ``+ plane'' (``- plane'') is the cell wall perpendicular to the LOS with larger (smaller) location along the $r_\parallel$-axis, and $\left< v_\parallel \right>_{+\,\rm plane} $ is the velocity mean on the ``+'' cell wall, i.e.\ $\left< v_\parallel \right>_{+\,\rm plane} = \frac{1}{(\Delta L)^2} \int_{+\,\rm plane} d^2 r_\perp\, v_\parallel(\vec{S}_\perp, +\,\rm plane)$. Unfortunately, we cannot apply the same smoothing as in equation~(\ref{eqn:rhoint}) to compute the velocity average, because the velocity defined in equation~(\ref{eqn:smooth-v}) involves a summation in the denominator. 

To circumvent this, we approximate the smoothed velocity averaging on a cell wall by the momentum-preserving velocity, i.e.
\begin{equation}
\left< v_\parallel \right>_{\rm plane} \to \left< {\mathcal{P}_\parallel} \right>_{\rm plane} / \left< \rho \right>_{\rm plane}\,,
\end{equation}
where the r.h.s. is the center-of-mass velocity of a thin layer on the cell wall. The surface mass density of the cell wall is 
\begin{eqnarray}
\left< \rho \right>_{\rm plane} & =& \frac{1}{(\Delta L)^2} \int_{\rm plane} \rho({\bf r}) d^2r_\perp \nonumber \\
& =& \frac{1}{(\Delta L)^2} \sum_i m_i \int_{\rm plane} W({\bf r}-{\bf r}_i;h_i)d^2r_\perp\,,
\end{eqnarray}
where the integral can be evaluated analytically, 
\begin{eqnarray}
\int_{\rm plane} W({\bf r}-{\bf r}_i;h_i)d^2r_\perp &=& \left[\int_{x_{1,c}-\Delta L/2}^{x_{1,c}+\Delta L/2} f_{h_i}(x_1 - x_{1,i}) dx_1 \right] \nonumber\\
& & \!\!\!\!\!\!\!\!\!\!\!\!\!\!\!\!\!\!\!\!\!\!\!\!\!\!\!\!\!\!\!\!\!\!\!\!\!\!\!\!\!\!\!\!\!\!\!\!\!\!\!\!\!\!\!\!\!\!\!\!\!\!\!\!\!\!\!\!\!\!\!\!\!\!\!\!\!\!\!\!
\times \left[\int_{x_{2,c}-\Delta L/2}^{x_{2,c}+\Delta L/2} f_{h_i}(x_2 - x_{2,i}) dx_2 \right]\times f_{h_i}(x_{3, {\rm plane}} - x_{3,i})\,.
\end{eqnarray}
Here we take $x_1$ and $x_2$ to be axes in the cell wall perpendicular to the LOS axis $x_3$, $x_{3, {\rm plane}}$ is the LOS coordinate of the cell wall, $x_{1,c}$ and $x_{2,c}$ are the transverse coordinates of the center of the cell, and ${\bf x}_i$ are the three-dimensional coordinates of the particle $i$. 

Similarly, we use
\begin{eqnarray}
\left< {\mathcal{P}_\parallel} \right>_{\rm plane} & =& \frac{1}{(\Delta L)^2} \int_{\rm plane} {\mathcal{P}_\parallel}({\bf r}) d^2r_\perp \nonumber \\
& =& \frac{1}{(\Delta L)^2} \sum_i m_i v_{i,\parallel} \int_{\rm plane} W({\bf r}-{\bf r}_i;h_i)d^2r_\perp\,. \nonumber 
\end{eqnarray}

\end{appendix}


\label{lastpage}

\end{document}